\documentclass{article}
\usepackage{epubtk}
\usepackage{amssymb}
\usepackage{amsmath}
\usepackage{booktabs}
\usepackage{graphicx}
\usepackage{subeqnarray}
\usepackage{color}
    \definecolor{Blue}{rgb}{0.0,0.0,1.0}
    \definecolor{Red}{rgb}{1.0,0.0,0.0}
    \definecolor{Green}{rgb}{0.0,1.0,0.0}
    \DeclareMathAlphabet{\mathpzc}{OT1}{pzc}{m}{it}
\newcommand{\der}[2]{\ensuremath{\frac{{\rm d} #1}{{\rm d} #2}}}

\newcommand{\pder}[2]{\ensuremath{\frac{\partial #1}{\partial #2}}}

\def\<{{\langle}}
\def\>{{\rangle}}

\newcommand{\unit}[1]{\mathrm{\ #1}}
\newcounter{destination_saved}

\begin{document}

\title{Foundations of Black Hole Accretion Disk Theory}

\author{%
\epubtkAuthorData{Marek A.\ Abramowicz}{%
Physics Department, G\"oteborg University \\
SE-412-96 G\"oteborg, Sweden\\
and\\
N.\ Copernicus Astronomical Center\\
Bartycka 18, PL-00-716 Warszawa, Poland}{%
Marek.Abramowicz@physics.gu.se}{%
http://fy.chalmers.se/~marek }%
\\
~\\
\epubtkAuthorData{P.\ Chris Fragile}{%
Department of Physics \& Astronomy, College of Charleston \\
Charleston, SC 29424, USA}{%
fragilep@cofc.edu}{%
http://fragilep.people.cofc.edu/ }%
}

\date{}
\maketitle

\begin{abstract}
This review covers the main aspects of black hole accretion disk
theory. We begin with the view that one of the main goals of the
theory is to better understand the nature of black holes themselves.
In this light we discuss how accretion disks might reveal some of the
unique signatures of strong gravity: the event horizon, the innermost
stable circular orbit, and the ergosphere.  We then review, from a
first-principles perspective, the physical processes at play in
accretion disks.  This leads us to the four primary accretion disk
models that we review: Polish doughnuts (thick disks),
Shakura--Sunyaev (thin) disks, slim disks, and advection-dominated
accretion flows (ADAFs).  After presenting the models we discuss
issues of stability, oscillations, and jets.  Following our review of
the analytic work, we take a parallel approach in reviewing numerical
studies of black hole accretion disks.  We finish with a few select
applications that highlight particular astrophysical applications:
measurements of black hole mass and spin, black hole vs.\ neutron star
accretion disks, black hole accretion disk spectral states, and
quasi-periodic oscillations (QPOs).

%
\end{abstract}

\epubtkKeywords{black hole accretion disks}

\newpage

\section{Introduction}
\label{section-introduction}

Because of its firm connection to black holes themselves, black hole accretion disk theory
belongs to the realm of fundamental physics. However, the theory itself employs a
complicated maze of fluid-dynamics results and several phenomenological estimates and
guesses known only to specialists. This Living Review aims to give readers a useful
guide, a ``road map'' if you will, through the unavoidable complexity of the subject.

Below we list fourteen ``Destinations'' on this road map and explain their logical
connections. In our opinion, these Destinations are the key issues, or landmarks, of the
theory being reviewed. The particular road map that we present is, of course, biased by our
own ideas and research histories. However, we are confident that the grand landscape of this
field will, nevertheless, shine through in the end.

We start by pointing out that black holes are one of the most remarkable inventions of the
human mind. Their bizarre properties capture nearly everyone's imagination, from Princeton
string theorists to Hollywood movie makers. Initially, black holes only existed virtually,
as weird theoretical and mathematical ideas. Models of them were studied with interest, but
their real existence was questioned by many, including Albert Einstein himself. This
situation changed in the latter part of the twentieth century, after unambiguous and robust
detections were made of several astrophysical black hole ``candidates'' within our Galaxy
and in many others. Thus far, two%
\epubtkFootnote{An interesting, unanswered question is if there are
  any missing populations of black holes, e.g., ``intermediate'' mass
  black holes in the range $100\,M_{\odot} \lesssim M \lesssim
  10^6\,M_{\odot}$~\cite{madau_01,miller_02a, farrell_09, davis_11} or
  ``primordial'' black holes~\cite{carr_74}.}
classes of black holes have been observed by astronomers. In our
theory-minded Living Review we do not give detailed descriptions of
their observational properties. Instead, we stress their importance by
starting our road map from the two classes of observed black holes:
\newcounter{destination}
\begin{list}{\tt Destination \arabic{destination}:}{\usecounter{destination}}
\setcounter{destination}{\value{destination_saved}}
\item \textit{Quasars} and other similar supermassive objects, which
  are collectively called ``active galactic nuclei'' (or AGN), having
  masses%
\epubtkFootnote{$M_{\odot} = 1.99 \times 10^{33}\unit{[g]}$ denotes the
  mass of the Sun, used in astrophysics as a mass unit.}
in the range $10^6\,M_{\odot} < M < 10^9\,M_{\odot}$. They reside at
centers of our and other galaxies. The ``active'' ones among them are
the most powerful steady energy sources known in the universe. Many
have radiant powers $L$ in excess of their corresponding Eddington
luminosities.%
\epubtkFootnote{Radiant power in astrophysics is traditionally called
  ``luminosity.'' At the Eddington luminosity, $L_{\mathrm{Edd}}
  \equiv 1.2\times 10^{38} M/M_{\odot}\unit{[erg/s]}$, radiation force
  balances the gravity of the central object (with mass $M$). In the
  case of stars, ``super-Eddington'' luminosities, $L >
  L_{\mathrm{Edd}}$, are not possible, as this would mean radiation
  pressure would blow the star apart. 
  }
The high efficiency of quasars, $\eta \equiv L/{\dot M}c^2 > 0.1$, where ${\dot M}$ is the mass supply rate from accretion (``the accretion rate''), is puzzling. Black hole accretion disk theory predicts that $L > L_{\mathrm{Edd}}$ would imply small accretion efficiency $\eta \ll 0.1$. However, the famous ``So{\l}tan argument,'' based on quasar counts, shows that on a long time average, $t \sim t_{\mathrm{Hubble}}$, quasars can have both $L > L_{\mathrm{Edd}}$ and $\eta \sim 0.1$~\cite{soltan_82, raimundo_12}. AGN are described in much greater detail in the authoritative monograph by Krolik~\cite{krolik_99a}.

\item \textit{Microquasars} and similar ``stellar-mass'' black holes, having $M \sim 10\,M_{\odot}$. The term ``microquasars'' was invented by Mirabel~\cite{mirabel_92} to convey that these objects, in many regards, behave like scaled-down versions of quasars. A few tens of them have been found in our Galaxy as members of X-ray binaries~\cite{mcclintock_06}. The natural scaling $\text{(time)}\sim \text{(mass)}$ adds importance to observations of microquasar variability, because the same processes that takes hundreds of years (say) in quasars, takes only minutes in microquasars. Particularly interesting are spectral state changes, which occur on timescales $\sim 1~\text{day}$ (see \cite{remillard_06} for a review), and quasi-periodic oscillations (QPOs), which have timescales as short as $\sim 1~\text{ms}$~\cite{vanderklis_06,mcclintock_06,remillard_06}.
\setcounter{destination_saved}{\value{destination}}
\end{list}
%
Quasars and microquasars are among the most intriguing astrophysical
objects ever discovered, and the goal of black hole accretion disk
theory is, obviously, to explain their observed properties -- but one
hopes for much more! One hopes that observations of quasars and
microquasars, together with their proper theoretical interpretation,
would eventually test the very heart of black hole physics
itself. When this happens, we may meaningfully constrain our knowledge
of the fundamental properties of space and time.%
\epubtkFootnote{Observations of black holes may eventually cast light
  on the quantum gravity structure of the physical
  vacuum~\cite{abramowicz_98}, possibly constraining string
  theory~\cite{damour_08} and the hypothesis of extra
  dimensions~\cite{arkani_99, psaltis_07, johannsen_09}.}

To accomplish this goal, black hole accretion disk theory must find ways to filter out from the observational data the parts that bear clear \textit{black hole signatures} from the remaining \textit{dirty astrophysics} parts. The logical structure of our Living Review attempts to reflect as closely as possible this main task of the theory. Thus, our review starts (in Section~\ref{section-strong-gravity}) with the pure gravity part, concentrating on the three particular signatures of black hole gravity that are crucial to black hole accretion disk theory:
\begin{list}{\tt Destination \arabic{destination}:}{\usecounter{destination}}
\setcounter{destination}{\value{destination_saved}}
\item \textit{Event horizon.} This is a sphere of radius $\sim GM/c^2$ surrounding the black hole singularity, from within which nothing may emerge --- a one-way membrane.  Note that this means that \textit{black holes have no rigid surfaces}. This is a \emph{unique} signature of black holes; other relativistic features may be observable around non-black hole objects, specifically sufficiently compact neutron stars, but the event horizon is a defining property of black holes.
\item \textit{Ergosphere.} This is a region around a rotating black hole where spacetime itself is dragged along in the direction of rotation at a speed greater than the local speed of light in relation to the rest of the universe. In this region, negative energy states are possible, which means that the rotational energy of the black hole can be tapped through various manifestations of the ``Penrose process''~\cite{penrose_69}.
\item \textit{Marginally stable orbit} (also called the ``innermost
  stable circular orbit'' or ISCO). This is the smallest circle ($r =
  r_{\mathrm{ms}}$) along which free particles may stably orbit around
  a black hole. No stable circular motion is possible for $r <
  r_{\mathrm{ms}}$. This is a unique feature of relativity, as in
  Newtonian theory, orbits at all radii are possible.%
\epubtkFootnote{Strictly speaking, this statement is only true for
  nearly spherical gravity sources. Higher order (octopole) moments
  allow for the formation of an ISCO, even in Newtonian
  theory~\cite{amsterdamski_02}.}
\setcounter{destination_saved}{\value{destination}}
\end{list}

As fascinating as these destinations are, they are not easily probed in nature. The
radiation we receive from quasars and microquasars comes not from the black holes
themselves, but instead originates in the accretion disks which surround
them%
\epubtkFootnote{Astrophysical black holes do not themselves
  radiate. The temperature associated with Hawking radiation is $T_H =
  (\hbar c^3)/(8\pi G M k_B)$. For a stellar-mass black hole $T_H \sim
  10^{-8}{\,}[^{\circ}\unit{K}]$. Thus, Hawking radiation is
  completely suppressed by the thermal bath of the
  $3{\,}[^{\circ}\unit{K}]$ cosmic background radiation. For
  supermassive black holes, the Hawking temperature is at least five
  orders of magnitude smaller still.}
(see Figure~\ref{figure-quasar-microquasar}).  In these accretion disks,
angular momentum is gradually removed by some presumably (although not
necessarily~\cite{blandford_82}) dissipative process, causing matter
to spiral down into the black hole, converting its gravitational
energy into heat, and then, by various processes, radiating this
energy.%
\epubtkFootnote{The rate ${\dot M}_*$ at which matter accretes is therefore
regulated by internal torques and radiative processes. \textit{Only}
if ${\dot M}_* = {\dot M}_0 = \text{const}$ everywhere, with ${\dot
  M}_0$ being the outside mass supply, can the accretion process be
stationary. Since the internally determined ${\dot M}_*$ may change
due to instabilities, limit cycles, etc., an occurrence of a really
long-term steady accretion flow should be considered a fine-tuned
eigenstate. Note, too, that in many astrophysical situations ${\dot
  M}_0$ is also genuinely variable.}
The radiation subsequently leaks through the disk, escapes from its
surface, and travels along trajectories curved (in space) by the
strong gravity of the black hole, eventually reaching our
telescopes. Bingo! That is what we are interested in this Living Review.

\epubtkImage{}{%
\begin{figure}[htbp]
  \centerline{\includegraphics[width=0.95\textwidth]{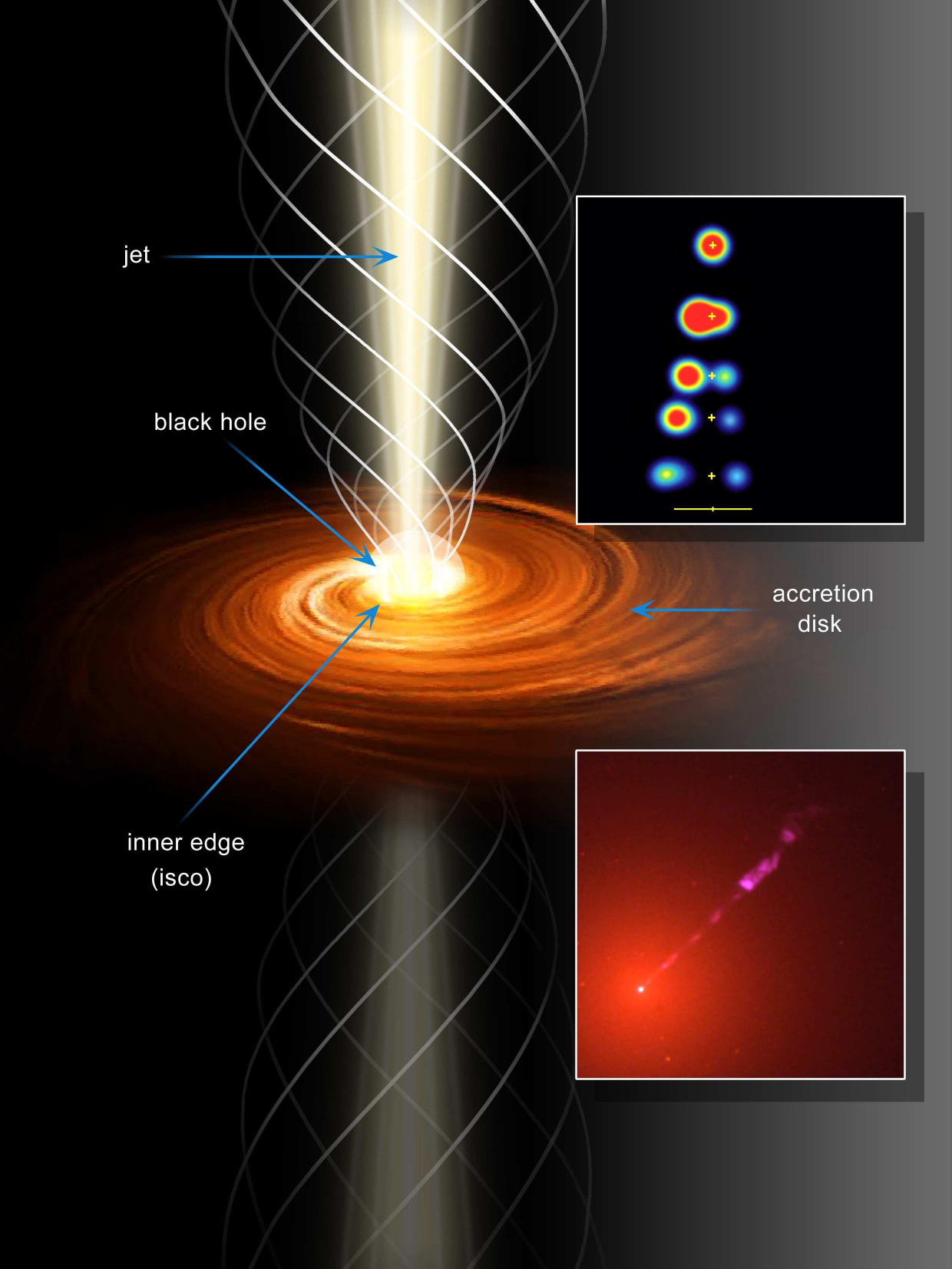}}
 \caption{An artist's rendition of a generic black hole accretion disk and jet. Inset figures include a time sequence of radio images from the jet in microquasar, GRS~1915+105 \cite{mirabel_94} and an optical image of the jet in quasar, M87 (Credit: J.A.~Biretta et al., Hubble Heritage Team (STScI /AURA), NASA).}
\label{figure-quasar-microquasar}
\end{figure}}

The preceding description of how black hole accretion works is well established in general,
yet several rather crucial details remain either not sufficiently understood or are too
complex to be studied even by the most powerful present-day computers. To deal with this
seemingly hopeless situation, several purely phenomenological approaches have been adopted.
Two of them should certainly be marked as Destinations, in and of themselves, on the road
map, because they are among the very basic ingredients of the accretion process:
\begin{list}{\tt Destination \arabic{destination}:}{\usecounter{destination}}
\setcounter{destination}{\value{destination_saved}}
\item \textit{Angular momentum transport and energy dissipation}. The
  quasi-steady accretion of a particle of mass $m$ through a Keplerian
  disk from a large outer radius, $r_{\mathrm{out}}$, to an inner radius,
  $r_{\mathrm{in}}$, requires that the particle give up an amount of energy%
\epubtkFootnote{This makes black hole accretion the most efficient
  energy generation process in the universe, short of
  matter-antimatter annihilation.}
$\sim 0.1\,mc^2$. To do this, the particle must also give up an amount of angular momentum $\sim (GM r_{\mathrm{out}})^{1/2}$. We will see in Section~\ref{subsection-stress-part} that viscous stresses within the fluid can facilitate this (mass transfer in, angular momentum transfer out, and energy dissipation).  However, the stresses can not come from ordinary molecular viscosity, as this is much too weak in astrophysical accretion disks. Instead, the stresses likely come from turbulence that acts like an \emph{effective viscosity}.

\item \textit{Radiative processes and radiative transfer.} These depend on the thermodynamic state of matter (electron density, ion density, temperature), its motion, and the magnetic field, but most importantly on whether the matter is opaque or transparent to radiation, i.e., whether its optical depth is large, $\tau \gg 1$, or small, $\tau \ll 1$. In the general case when the matter is optically thick $\tau > 1$, the accretion disk can be quite luminous and also efficiently cooled by radiation. Accretion disks with $\tau < 1$ are inefficiently cooled and thus less luminous.
\setcounter{destination_saved}{\value{destination}}
\end{list}
%



With these complexities, it is useful to have both analytic theories
and numerical simulations to facilitate progress in the field. Before
getting into details, though, let us simply study the parameter space
available before us. Let us first consider three generic types of
physical processes that must occur in black hole accretion disks:
``dynamical'' processes with a characteristic time-scale,
$t_{\mathrm{dyn}} \sim 1/\Omega$, where $\Omega$ is the orbital
angular velocity; ``thermal'' processes with a characteristic
time-scale, $t_{\mathrm{th}} \sim c_s^2/\nu_* \Omega^2$, where $c_s$ is
the sound speed and $\nu_*$ is the kinematic viscosity; and
``viscous'' processes with a characteristic time-scale,
$t_{\mathrm{vis}} \sim r^2/\nu_*$, where $r$ is the radial distance
from the black hole. One excepts that, typically,%
\epubtkFootnote{Note that there is an analogy with stars, where there
  are three timescales -- dynamical, thermal, and nuclear -- which for
  most of the life of a star obey $t_{\mathrm{dyn}} \ll
  t_{\mathrm{th}} \ll t_{\mathrm{nuc}}$.}
\begin{equation}
t_{\mathrm{dyn}} \ll t_{\mathrm{th}} \ll t_{\mathrm{vis}}.
\label{three-time-scales}
\end{equation}
Because dynamical processes act much faster than thermal or viscous
ones, in the first approximation one may consider only the dynamical
structure of the disk. The disk dynamical structure is governed by
three forces: gravity, which at a given place is fixed by the Kerr
geometry; pressure; and rotation forces. The relative importance of
each of these depend on the particular astrophysical situation ---
their values may either be ``small'' (dynamically unimportant) or
``large'' (dynamically important). Correspondingly, there are four
``genuine'' types of dynamical situations, as described in the upper
half of Table~\ref{table-four-dynamical-types}. The four types of
accretion disks we consider in this Living Review (thin, slim, thick,
ADAF) are located in three of the four regions. The fourth region
corresponds to a ``free-fall'' of dust.%
\epubtkFootnote{The slow rotation cases are sometimes referred to as
  ``Bondi flows''  in honor of Bondi's pioneering works on
  spherically-symmetric (non-rotating) accretion~\cite{bondi_52}.}
It is remarkable that the same types of accretion disks pop-up when one introduces other, \textit{very different}, criteria for dividing the parameter space, such as accretion rate and opacity, as shown in lower half of Table~\ref{table-four-dynamical-types}. These four types of accretion disks are the next four Destinations on our road map.  We characterize them in terms of:  relative thickness, $h = H/R$; dimensionless accretion rate, ${\dot m} = 0.1{\dot M} c^2/L_{\mathrm{Edd}}$; optical depth, $\tau$; importance of advection, $q = Q_{\mathrm{adv}}/Q_{\mathrm{rad}}$, where $Q$ represents an energy flux; importance of radiation pressure, $\beta =P_{\mathrm{gas}}/(P_{\mathrm{gas}} + P_{\mathrm{rad}})$; location of inner edge, $r_{\mathrm{in}}$; and accretion efficiency $\eta$.

%
%
\begin{table}
\caption{\textit{Upper half:} The four \emph{a priori} possible types
  of dynamical states of accretion structures, corresponding to the
  division of the dynamical parameter space into fast-slow rotation
  and large-small pressure. \textit{Lower half:} A different division
  of the parameter space, corresponding to high-low accretion rate and
  large-small opacity.}
\label{table-four-dynamical-types}
\centering
{\small
  \begin{tabular}{c c c}
    \toprule
    ~ & Fast rotation (Disk) & Slow rotation (Bondi)
    \\ \midrule
    Large pressure &
    \textbf{ slim, thick} &
    \textbf{ ADAFs }
    \\ 
    Small pressure &
    \textbf{ thin} &
    \textbf{ free-fall }
    \\ \toprule
    ~ & Accretion rate high & Accretion rate low
    \\ \midrule
    Large opacity &
    \textbf{ slim, thick} &
    \textbf{ thin }
    \\ 
    Small opacity &
    \textbf{ --} &
    \textbf{ ADAF }
    \\ \bottomrule
\end{tabular}}
\end{table}

%

\begin{list}{\tt Destination \arabic{destination}:}{\usecounter{destination}}
\setcounter{destination}{\value{destination_saved}}
\item \textit{Thick Disk.}
\\$h > 1$, ${\dot m} \gg 1$, $\tau \gg 1$, $q \sim 1$, $\beta \ll 1$,
$r_{\mathrm{in}} \sim r_{\mathrm{mb}}$, $\eta \ll 0.1$
\item \textit{Thin Disk.}
\\$h \ll 1$, ${\dot m} < 1$, $\tau \gg 1$, $q = 0$, $\beta \sim 1$,
$r_{\mathrm{in}} = r_{\mathrm{ms}}$, $\eta \sim 0.1$
%
\item \textit{Slim Disk.}
\\$h \sim 1$, ${\dot m} \gtrsim 1$, $\tau \gg 1$, $q \sim 1$, $\beta <
1$, $r_{\mathrm{mb}} < r_{\mathrm{in}} < r_{\mathrm{ms}}$, $\eta < 0.1$

\item \textit{Advection-Dominated Accretion Flow (ADAF).}
\\$h < 1$, ${\dot m} \ll 1$, $\tau \ll 1$, $q \sim 1$, $\beta = 1$,
$r_{\mathrm{mb}} < r_{\mathrm{in}} < r_{\mathrm{ms}}$, $\eta \ll 0.1$
\setcounter{destination_saved}{\value{destination}}
\end{list}
%
Interestingly, the parameter space of each of these types of accretion disks overlaps that of other solutions. For instance, ADAF solutions exist that have the same mass accretion rates as thin disk solutions~\cite{chen_95}. In such cases, it is not clear how nature might choose one over the other.

In Sections~\ref{section-thick-disks}\,--\,\ref{section-ADAFs}, we will demonstrate that analytic (or at least semi-analytic) models exist for each of these types of accretion disks. However, this does not guarantee that they are stable. The issue of stability is our next destination.
\begin{list}{\tt Destination \arabic{destination}:}{\usecounter{destination}}
\setcounter{destination}{\value{destination_saved}}
\item \textit{Stability.} Stability analysis is important because the systematic differential rotation that is one of the defining characteristics of accretion disks is also a potential source of destabilizing energy. On the one hand, this may be essential, as the angular momentum transport and energy dissipation required for accretion may \textit{require} disks to be mildly unstable. On the other hand, if a model is violently unstable, then the basic assumption of a ``steady-state'' would be violated.
\setcounter{destination_saved}{\value{destination}}
\end{list}
Along with stability, we also look at the natural oscillation modes associated with accretion disks. The frequencies of some of these modes are tied to the properties of the black hole space-time, which is how these relate to the fundamental physics issues of interest in this Living Review.
\begin{list}{\tt Destination \arabic{destination}:}{\usecounter{destination}}
\setcounter{destination}{\value{destination_saved}}
\item \textit{Oscillations.} As with any finite distribution of fluid, accretion disks have natural oscillation modes associated with them. If these modes can be excited at appreciable amplitudes, they may be able to modify the observed light curve of the disk in measurable ways. This makes disk oscillations a leading candidate for explaining the quasi-periodic oscillations (QPOs) that we discuss in Section~\ref{section-QPOs}.
\setcounter{destination_saved}{\value{destination}}
\end{list}
Because of the close observational links between black hole accretion disks and jets \cite{fender_04,fender_09}, we include jets as the final Destination of our review.
\begin{list}{\tt Destination \arabic{destination}:}{\usecounter{destination}}
\setcounter{destination}{\value{destination_saved}}
\item \textit{Jets.} Jets are narrow ($\text{opening angle} < 5^\circ$), long ($\text{length} > 10^7$~ly in the case of AGN), and fast ($v > 0.9c$) streams of matter emerging from very compact regions around the black hole, usually in opposite directions, presumably normal to the plane of the accretion disk. Jets can play a significant role in transporting energy and angular momentum away from the accretion disk \cite{blandford_82}. They also play an important role in shaping the black hole's environment far beyond the gravitational reach of the black hole itself, affecting galactic evolution, particle acceleration, and intragalactic ionization.
\setcounter{destination_saved}{\value{destination}}
\end{list}

Going hand-in-glove with analytic models of accretion disks are direct numerical simulations.  Although analytic theories have been extremely successful at explaining many general observational properties of quasars and microquasars, numerical simulations can be critically important in at least two respects: 1) as an extension of analytic work, by treating nonlinear perturbations and higher order coupling terms, and 2) in cases that are highly time variable or contain little symmetry, such that the prospects of finding an analytic solution are poor. There is also an important overlap region where various analytic and numerical methods are applicable and can be used to independently validate results. Because of these close connections between analytic and numerical work, we have a section (Section~\ref{section:numerical_simulations}) dedicated to the discussion and review of direct numerical simulation of black hole accretion disks.

We finish this Living Review with a section (Section~\ref{section:applications}) that tries to make some connections between the concepts discussed in earlier sections and actual observational phenomena.  We emphasize that we are not aiming to provide a comprehensive review of black hole observations, but rather to highlight a very small subset of these that are of particular relevance to our review.

Throughout this review, we adopt the $-+++$ metric signature and often use units where $c = 1 = G$. To make all physical quantities dimensionless, we sometimes also use the mass of the black hole as a unit, $M = 1$. We use the common Einstein summation convention, where repeated indices in a formula imply summation over the range of that index.  We also follow the common convention where Greek (Latin) indices are used for four-(three-)dimensional tensor quantities.

\newpage

\section{Three Destinations in Kerr's Strong Gravity}
\label{section-strong-gravity}

In this section, we briefly describe the three destinations within Kerr's strong gravity that are most relevant to black hole accretion disk theory:
\begin{enumerate}
\item \textbf{Event Horizon:} That radius inside of which escape from the black hole is not possible;
\item \textbf{Ergosphere:} That radius inside of which negative energy states are possible (giving rise to the potentiality of tapping the energy of the black hole).
\item \textbf{Innermost Stable Circular Orbit (ISCO):} That radius inside of which free circular orbital motion is not possible;
\end{enumerate}
Our principal question is: \textit{Could accretion disk theory unambiguously prove the existence of the event horizon, ergosphere, and ISCO using currently available or future observations?}

In realistic astrophysical situations involving astrophysical black holes (in particular quasars and microquasars), the black hole itself is uncharged, and the gravity of accretion disk is practically negligible. This means that the spacetime metric $g_{\mu\nu}$ is given by the Kerr metric, determined by two parameters: total mass $M_*$ and total angular momentum $J_*$. It is convenient to rescale them by
\begin{subeqnarray}
\label{rescaling}
M &=& \frac{G M_*}{c^2} \slabel{rescaling-mass} \\
a &=& \frac{J_*}{M_* c}, \slabel{rescaling-spin}
\end{subeqnarray}
such that both $M$ and $a$ are measured in units of length.

In the standard spherical Boyer--Lindquist coordinates the Kerr metric takes the form~\cite{bardeen_72},
\begin{alignat}{2}
g_{tt} &= -\left(1 - \frac{2Mr}{\varrho^2} \right),
&\quad ~~~g^{tt} &= -\frac{(r^2 + a^2)^2 -
a^2\Delta\sin^2\theta}{\varrho^2\,\Delta}, \nonumber  \\
%
g_{t\phi} &= -\frac{2Mar\sin^2\theta}{\varrho^2},
&\quad ~~~~g^{t\phi}  &= -\frac{2Mar}{\varrho^2\,\Delta}, \nonumber  \\
g_{\phi\phi} &= \left ( r^2 + a^2 + \frac{2Ma^2r\sin^2 \theta}
{\varrho^2} \right )\,\sin^2\theta,
&\quad ~~~~g^{\phi\phi}   &= \frac{\Delta - a^2\,\sin^2
\theta}{\Delta\,\varrho^2\,\sin^2\theta}, \nonumber  \\
%
g_{rr} &= \frac{\varrho^2}{\Delta}, ~~~g_{\theta\theta} =
\varrho^2,
&\quad ~~~~g^{rr} &= \frac{\Delta}{\varrho^2},
~~~~g^{\theta\theta} = \frac{1}{\varrho^2},
\label{kerr-metric}
\end{alignat}
where $\Delta = r^2 - 2\,M\,r + a^2$ and $\varrho^2 = r^2 + a^2\,\cos^2\theta$.

The Kerr metric depends neither on time $t$, nor on the azimuthal angle $\phi$ around the symmetry axis. These two symmetries can be expressed in a coordinate independent way by the two commuting Killing vectors $\eta^\mu = \delta^\mu_{~t}$ and $\xi^\mu = \delta^\mu_{~\phi}$,
\begin{subeqnarray}
\label{killing-equations}
&\nabla_\mu \eta_\lambda + \nabla_\lambda \eta_\mu = 0,
~~\nabla_\mu \xi_\lambda + \nabla_\lambda \xi_\mu = 0,&
\slabel{killing-equations-vectors}
\\
&\eta^\mu \nabla_\mu \xi^\nu = \xi^\mu \nabla_\mu \eta^\nu.&
\label{killing-equations-commuting}
\end{subeqnarray}
%
%
%
%
%
%
%
%
%
%
Here $\nabla_\mu$ denotes the \textit{covariant} derivative,
 \begin{equation}
 \label{covariant-derivative}
 \nabla_\mu \eta_\lambda = \partial_\mu \eta_\lambda -
\Gamma^\nu_{\mu \lambda} \eta_\nu,
~~\Gamma^\nu_{\mu \lambda} =\frac{1}{2}g^{\nu \kappa}\left(
\partial_\lambda g_{\mu \kappa} +
\partial_\mu g_{\lambda \kappa} -
\partial_\kappa g_{\lambda \mu} \right),
\end{equation}
and $\partial_\mu =
\partial/\partial x^\mu$ denotes the standard \textit{partial} derivative.
Formulae for the Kerr metric (\ref{kerr-metric}) and all its non-zero Christoffell symbols
$\Gamma^\nu_{\mu \lambda}$ (\ref{covariant-derivative}), are available from
\cite{tollerud_notebook}.

In Boyer--Lindquist coordinates the $t$ and $\phi$ components
of the Kerr metric can be expressed as scalar products of the Killing vectors,
\begin{equation}
\label{metric-by-killig-vectors}
g_{tt} = \eta^\mu \eta_\mu, ~~
g_{t\phi} = \eta^\mu \xi_\mu, ~~
g_{\phi\phi} = \xi^\mu \xi_\mu.
\end{equation}
From the Killing vectors one can also define the following constants of motion for a particle or photon with four-momentum $p_\mu$
\begin{subeqnarray}
\label{constant-motion}
\text{energy:} && {\cal E} \equiv -\eta^\mu p_\mu = - p_t,
\slabel{constant-motion-energy} \\
\text{angular~momentum:} && {\cal L} \equiv  \xi^\mu p_\mu =
p_\phi,
\slabel{constant-motion-angular} \\
\text{specific~angular~momentum:} &&\ell \equiv \frac{\cal L}{\cal
E} = -\frac{p_\phi}{p_t} = -\frac{u_\phi}{u_t},
\slabel{constant-motion-specific} \\
\text{Carter~constant:} && {\cal C} = K^{\mu\nu}p_\mu\,p_\nu =
(p_\theta)^2 + \frac{p_\phi^2}{\sin^2\theta}.
\slabel{constant-motion-Carter}
\end{subeqnarray}
The Carter constant ${\cal C}$ is connected to the Killing tensor $K_{\mu\nu}$, which exists
in the Kerr metric. Killing tensors obey,
\begin{equation}
\label{killing-tensor}
\frac{1}{6}\left[
\nabla_\kappa K_{\mu \nu} +
\nabla_\mu K_{\nu \kappa} +
\nabla_\nu K_{\kappa \mu} +
\nabla_\kappa K_{\nu \mu} +
\nabla_\nu K_{\mu \kappa} +
\nabla_\mu K_{\kappa \nu}
\right] = 0.
\end{equation}
The other coordinates often used in black hole accretion disk research are the Kerr-Schild
coordinates, in which the metric takes the form,
\begin{eqnarray}
\label{Kerr-Schild-coordinates}
g_{\mu \nu} &=& \eta_{\mu \nu} + f\,k_{\mu}\,k_{\nu},
~~\eta_{\mu \nu} = \mathrm{diag} (-1, 1, 1, 1 ) \nonumber \\
%
%
%
%
f &=& \frac{2Mr^3}{r^4 + a^2z^2}, \nonumber \\
k_t &=& 1, \nonumber \\
k_x &=& \frac{rx + ay}{r^2 + a^2}, \nonumber \\
k_y &=& \frac{ry - ax}{r^2 + a^2}, \nonumber \\
k_z &=& \frac{z}{r},
\end{eqnarray}
where $k_\nu = (k_t, k_x, k_y, k_z)$ is a unit vector, and $r$ is given implicitly by the condition,
\begin{equation}
\label{Kerr-Schild-coordinates-01}
\frac{x^2 + y^2}{r^2 + a^2} + \frac{z^2}{r^2} = 1.
\end{equation}
Note that we have given the Kerr--Schild metric in its Cartesian form to prevent confusion with the spherical-polar Boyer--Lindquist coordinates.  In keeping with this, unless specifically stated otherwise, the indices $\{t,\phi,r,\theta\}$ will always refer to the Boyer--Lindquist coordinates in this review.

\subsection{The event horizon}
\label{section-horizon}

The mathematically precise, general, definition of the event horizon involves topological considerations \cite{misner_73}. Here, we give a definition which is less general, but in the specific case of the Kerr geometry is fully equivalent.

The Boyer--Lindquist coordinates split the Kerr spacetime into a ``time'' coordinate $t$ and a three-dimensional ``space,'' defined as $t = \text{const}$ hypersurfaces. This split may be done in a coordinate independent way, based on the Killing vectors which exist in the Kerr spacetime. Indeed, the family of non-geodesic observers $N^\mu$ with trajectories orthogonal to a family of 3-D spaces $t
= \text{const}$ is defined as,
\begin{subeqnarray}
\label{zamo-observer}
\text{ZAMO:} &&N^\mu \equiv e^{-\Phi}{\tilde \eta}^\mu, ~~ {\tilde
\eta}^\mu = \eta^\mu + \omega\xi^\mu,
\slabel{zamo-observer-vector} \\
\text{frame~dragging:} &&\omega \equiv -\frac{\eta^\nu\xi_\nu}{\xi^\mu
\xi_\mu} = -\frac{g_{t\phi}}{g_{\phi\phi}}.
\slabel{zamo-observer-drag}
\end{subeqnarray}
They are called zero-angular-momentum-observers (ZAMO), because for them, the angular momentum defined by (\ref{constant-motion-angular}) is zero, ${\cal L} = N_\phi = 0$. The ZAMO observers provide the standard of rest in the 3-D space: objects motionless with respect to the ZAMO frame of reference occupy fixed positions in space.

We can also define a gravitational potential in the ZAMO frame:
\begin{eqnarray}
\text{potential:} &&\Phi \equiv -\frac{1}{2}\ln\left[
\frac{\xi^\mu\xi_\mu}{(\eta^\nu\xi_\nu)^2 -
(\eta^\nu\eta_\nu)(\xi^\mu \xi_\mu)}\right] = -\frac{1}{2}\ln
\vert g^{tt} \vert.
\label{zamo-observer-potential}
\end{eqnarray}
The primary reason to call $\Phi$ the gravitational potential is that, in Newton's theory, the observer who stays still in space experiences an acceleration due to ``gravity'' $g_\mu$, which equals the gradient of the gravitational potential. In the Kerr spacetime it is,
\begin{equation}
g_\mu =  (a_\mu)_{\mathrm{ZAMO}} \equiv N^\nu \nabla_\nu N_\mu =
\nabla_\mu \Phi.
\label{gravity-acceleration}
\end{equation}
From (\ref{zamo-observer-vector}) one sees that at the surface
$1/g^{tt} = 0$, the vector ${\tilde \eta}^\mu$ is null, ${\tilde
  \eta}^\mu {\tilde \eta}_\mu = 0$. Therefore, the ZAMO observers who
provide the standard of rest, \textit{move on that surface with the
  speed of light}. In order to stand still in this location, one must
move radially out with the speed of light.%
\epubtkFootnote{\textit{``Well, in our country,``} said Alice, still
  panting a little, \textit{``you'd generally get to somewhere else if
    you run very fast for a long time, as we've been doing.''}
  \textit{``A slow sort of country!''} said the Queen. \textit{``Now,
    here, you see, it takes all the running you can do, to keep in the
    same place.''}}
As it is clear from (\ref{kerr-metric}), $1/g^{tt} = 0$ is equivalent to $\Delta = 0$. The last equation has a double solution,
\begin{equation}
\label{horizon-radius-Kerr}
r = r_H = r_G \left( 1 \pm \sqrt{1 - \frac{a}{M}} \right).
\end{equation}
Note, that for $r < r_+$ the ZAMO ``observers'' are spacelike: standing still at a given radial location implies moving along a \textit{spacelike} trajectory ---  i.e., faster than light. All trajectories that move radially out are also spacelike. Thus, the outer root $r = r_+$ of Eq.~(\ref{horizon-radius-Kerr}) defines the Kerr black hole event horizon: \textit{a null surface that surrounds a region from which nothing may escape}. Outside the outer horizon (i.e., for $r > r_+$) the normalization of $N^\mu$ is non-singular, and therefore the gravitational potential (\ref{zamo-observer-potential}) is a non-singular, well-defined quantity.

\subsubsection{Detecting the event horizon}
\label{section-event-horizon}

One may think of two general classes of astrophysical observations that could provide evidence for a black hole horizon. Arguments in the first class are indirect; they are based on estimating a dimensionless ``compactness parameter''
\begin{equation}
\label{compactness-parameter} {\mathfrak C} = \frac{(\text{size~of~the~object})}{(G/c^2)(\text{mass~of~the~object})}.
\end{equation}
Arguments in the second class are more direct. They are based (in principle) on showing that some amount of radiation emitted by the source is lost inside the horizon.

\textit{Evidence based on estimating the compactness parameter:} A source for which observations indicate ${\mathfrak C} \approx 1$ may be suspected of having an event horizon. Values ${\mathfrak C} \approx 1$ have indeed been found in several astronomical sources. In order to know ${\mathfrak C}$, one must know mass and size of the source. The mass measurement is usually a \textit{direct} one, because it may be based on an application of Kepler's laws. In a few cases the mass measurement is remarkably accurate. For example, in the case of Sgr~A*, the supermassive black hole in the center of our Galaxy, the mass is measured to be $M = (4.3 \pm 0.5)\times10^6\,M_{\odot}$~\cite{gillessen_09}.

Until recently, estimates of size were always indirect, and generally not accurate.  They are usually based on time variability or spectral considerations. For the former, the measurement rests on the logic that if the shortest observed variability time-scale is $\Delta t$, then the size of the source cannot be larger than $R = c\Delta t$.  For the latter, the argument goes like this: If the total radiative power $L$ and the radiative flux $F$ can be independently measured for a black-body source, then its size can be estimated from $L = 4\pi R^2 F$. Keep in mind that one must know the distance to the source in order to measure $L$. The flux can be estimated from $F = aT^4$, where $T$ is the temperature corresponding to the peak in the observed intensity versus frequency electromagnetic spectrum.

It is hoped that in the near future, the next generation of high-tech
radio telescopes will be able to measure directly the size of ``the
light circle'', which is uniquely related to the horizon size (see
Figure~\ref{figure-SgrA-silhouette}). For Sgr~A*, at a distance of
$8.28\pm0.44\mathrm{\ kpc}$~\cite{gillessen_09}, the event horizon
corresponds to an angular size of $\sim 10~\mu{\rm}as$ in the sky,
making it an ideal target for near-future microarcsecond very long
base interferometric techniques~\cite{doeleman_08,
  eisenhauer_08}. Here the plan is to measure the black hole
``shadow'' or ``silhouette.''

\epubtkImage{}{%
\begin{figure}[htbp]
  \centerline{\includegraphics[width=0.43\textwidth]{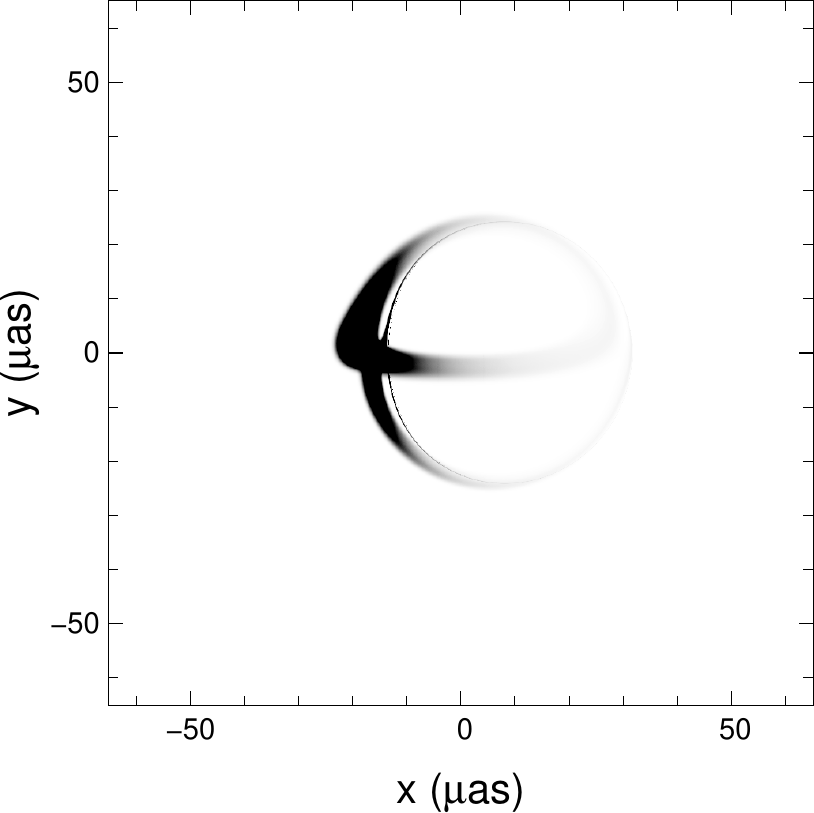}
\hfill
\includegraphics[width=0.43\textwidth]{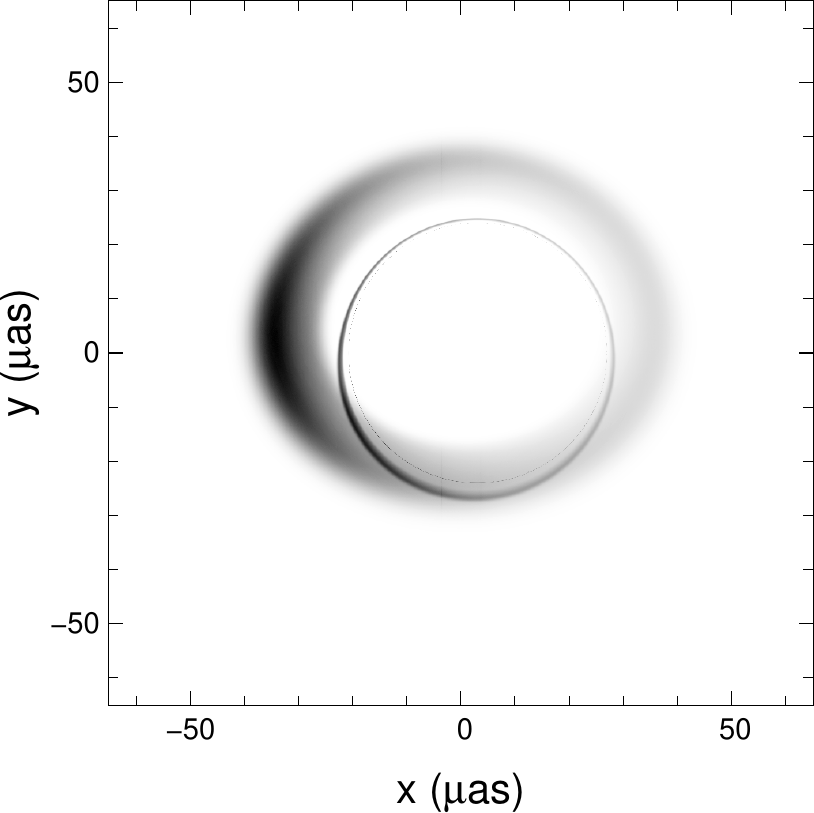}}
\centerline{\includegraphics[width=0.43\textwidth]{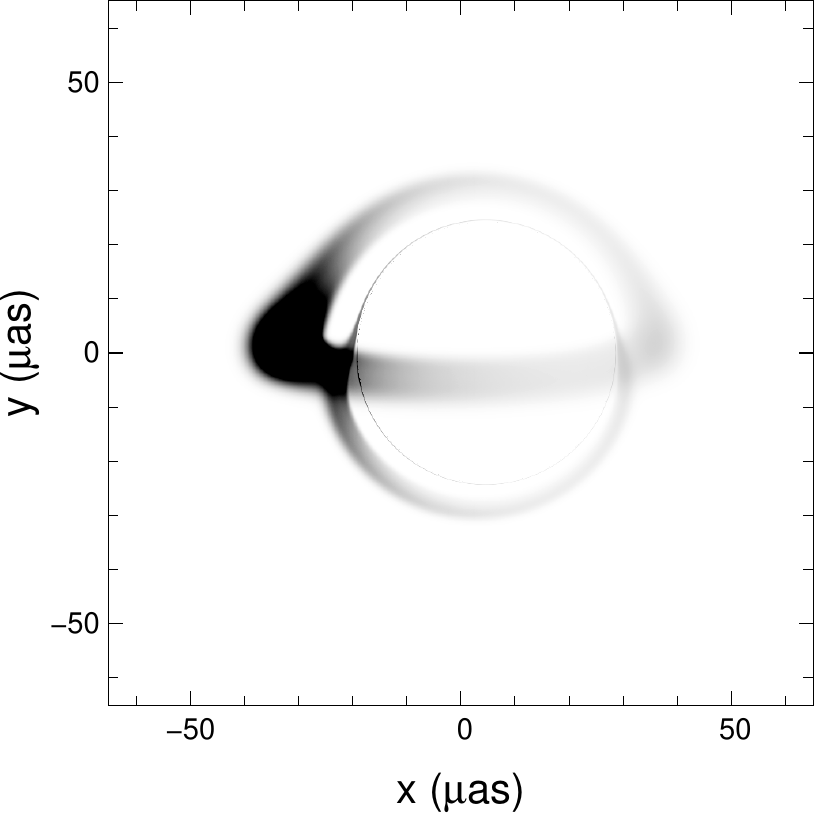}
\hfill
\includegraphics[width=0.43\textwidth]{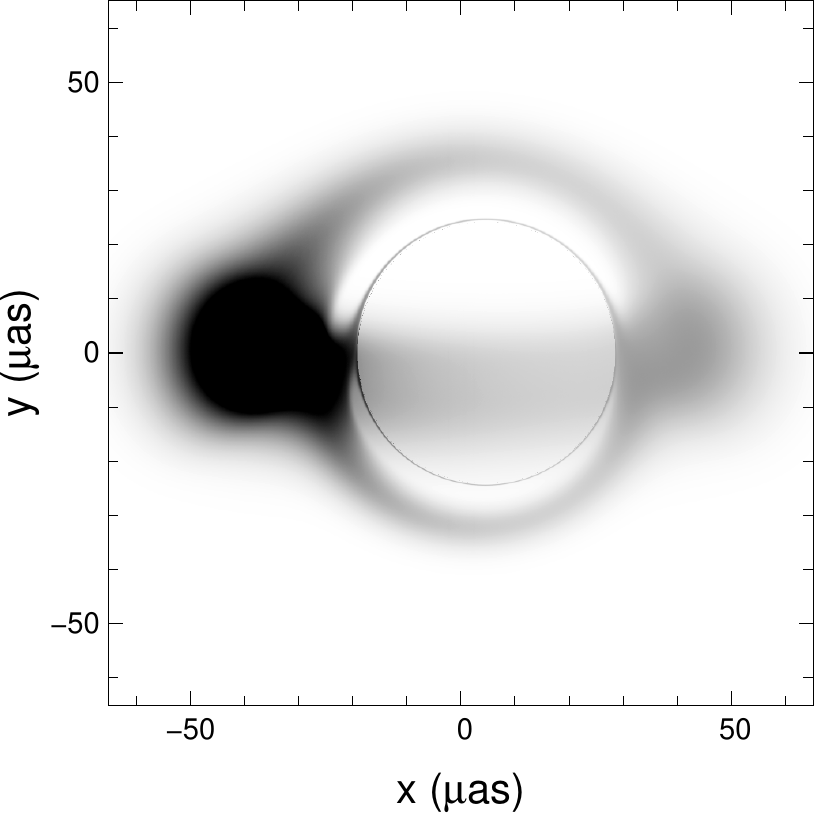}}
 \caption{Silhouettes of Sgr~A* calculated for
four optically thin accretion structures, characterized by very
different physical conditions. The display is intentionally
reversed in black-and-white and saturated in order to better show
the less luminous parts. Although ``dirty astrophysics'' makes the
most prominent differences, effects of the ``pure strong gravity''
are also seen in the form of ``the light circle'', a tiny almost
circular feature at the center. Its shape and size depends only on
the black hole mass and spin. Image reproduced by permission
from~\cite{straub_12}, copyright by ESO.}
\label{figure-SgrA-silhouette}
\end{figure}}

\textit{Evidence based on the ``no escape'' argument:} For accretion onto an object with a physical surface (such as a star), 100\% of the gravitational binding energy released by accretion must be radiated away.  This does not apply for a black hole since the event horizon allows for the energy to be advected into the hole without being radiated.  This may allow for a black hole with an event horizon to be distinguished from another, similar-mass object with a surface, such as a neutron star. This argument was first developed by Narayan and collaborators \cite{narayan_97,narayan_02b,narayan_03b}; we describe it in more detail in Section~\ref{section:horizon_argument}.

\subsection{The ergosphere}
\label{sub-section-ergosphere}

In Newtonian gravity, angular momentum $\ell$ and angular velocity $\Omega$ are related by the formula $\ell = r^2\Omega$, and therefore there is no ambiguity in defining a non-rotating frame as $\Omega = 0 = \ell$. However, in the Kerr geometry $\ell \propto (\Omega - \omega)$, where $\omega = -g_{t \phi}/g_{\phi\phi}$ is the angular velocity of the frame dragging induced by the Lense--Thirring effect. Therefore, $\Omega = 0$ does \textit{not} imply $\ell = 0$. This leads to two \textit{different} standards of ``rotational rest'': the Zero Angular Velocity Observer (ZAVO) and the Zero Angular Momentum Observer (ZAMO),
\begin{eqnarray}
\label{ZAVO} \text{ZAVO~frame}~(\Omega = 0)\text{:}&&
n^\mu = (-\eta^\nu \eta_\nu)^{-1/2}\,\eta^\mu,\\
\label{ZAMO} \text{ZAMO~frame}~(\ell = 0)\text{:}&&  N^\mu = e^\Phi(\eta^\mu +
\omega \xi^\mu) \, .
\end{eqnarray}
These two frames rotate with respect to each other with the frame-dragging angular velocity $\omega = -g_{t\phi}/g_{\phi\phi}$.

The ZAMO frame defines a \textit{local} standard of rest with respect
to the \textit{local compass of inertia}: a gyroscope stationary in
the ZAMO frame does not precess. Considering its kinematic invariants,%
\epubtkFootnote{\label{foot-kinematic-invariants} For a congruence of
  observers (or particles or photons) with four velocity $U^\mu$, the
  kinematic invariants fully describe their relative motion. Consider
  those that, in a particular moment $s_0$, occupy the surface of an
  infinitesimally small sphere.  Now, consider the deformation of that
  surface at a later moment $s_0 + ds$. The volume change $dV/ds =
  \Theta$ is called \textit{expansion}. The \textit{shear} tensor
  $\sigma_{\mu \nu}$ measures the ellipsoidal distortion of this
  sphere, and the \textit{vorticity} tensor $\omega_{\mu \nu}$
  describes its rotation (i.e., three independent rotations around
  three perpendicular axes). Expansion, shear, and vorticity are
  determined by the tensor $X_{\mu \nu} = \nabla_\mu U_\nu$ in the
  following way: $\Theta = (1/3)X^\mu_\mu$, $\sigma_{\mu \nu} =
  h^\alpha_{~\mu}\,h^\beta_{~\nu}\,(1/2)(X_{\alpha \beta} + X_{\beta
    \alpha}) - \Theta\, h_{\mu \nu}$, and $\omega_{\mu \nu} =
  h^\alpha_{~\mu}\,h^\beta_{~\nu}\,((1/2)(X_{\alpha \beta} - X_{\beta
    \alpha})$. Here $h^\alpha_{~\mu} = \delta^\alpha_{~\mu} +
  U^\alpha\, U_\mu$ is the projection tensor. The acceleration $a_\mu$
  is also considered a kinematic invariant.}
one sees that the ZAMO frame is non-inertial ($a_\mu \not = 0$), non-rigid ($\sigma_{\mu\nu}\not = 0$, $\Theta = 0$), and surface-forming ($\omega_{\mu\nu} = 0$). The ZAMO vectors ${\tilde \eta}^\mu$ and $N^\mu$ are time-like everywhere outside the horizon, i.e., outside the surface $1/g^{tt} = 0$. This means that the energy of a particle or photon with a four-momentum $p_\mu$ measured by the ZAMO is positive, $E_{\mathrm{ZAMO}} \equiv N^\mu p_\mu > 0$.

The ZAVO frame defines a \textit{global} standard of rest with respect to distant stars: a telescope that points to a fixed star does not rotate in the ZAVO frame. Considering its kinematic invariants one sees that the ZAVO frame is non-inertial ($a_\mu \not = 0$), rigid ($\sigma_{\mu\nu}= 0$, $\Theta = 0$), and not surface-forming ($\omega_{\mu\nu} \not = 0$). At infinity, i.e., for $r \rightarrow \infty$, it is $(\eta^\nu \eta_\nu)=g_{tt} \rightarrow -1$, and therefore $n^\mu \rightarrow \eta^\mu$. For this reason, $\eta^\mu$ is called the \emph{stationary observer} at infinity. The ZAVO vectors $\eta^\mu$ and $n^\mu$ are timelike outside the region surrounded by the surface $g_{tt} = 0$, called the \textit{ergosphere}. Inside the ergosphere $\eta^\mu$ and $n^\mu$ are spacelike. This means that inside the ergosphere, the \textit{conserved} energy of a particle (i.e., the energy measured ``at infinity''), as defined by (\ref{constant-motion-energy}), may be \textit{negative}.

Penrose~\cite{penrose_69} considered a process in which, inside the ergosphere, a particle with energy $E^\infty > 0$ decays into two particles with energies $E^{\infty}_+ > 0$ and $E^{\infty}_- = - |E^{\infty}_-| < 0$. The particle with positive energy escapes to infinity, and the particle with the negative energy gets absorbed by the black hole. Then, because $E^{\infty}_+ = E - E^{\infty}_- = E + |E^{\infty}_-| > E$, one gets a net gain of positive energy at infinity. The source of energy in this \textit{Penrose process} is the rotational energy of the black hole. Indeed, the angular momentum absorbed by the black hole, $J^{\infty}=p_i\xi^i$ is necessarily negative (in the sense that $J^{\infty}\omega_H <0)$, which follows from
\[
E_{\mathrm{ZAMO}} \equiv
- p_i \,e^{\Phi}(\eta^i + \omega_H \xi^i) =
e^{\Phi}(E^{\infty}_- - \omega_H J^{\infty}) > 0,
\]
and thus
\begin{equation}
\omega_H J^{\infty} < E^{\infty}_- < 0.
\label{penrose-process-particle}
\end{equation}
A more complete presentation of the Penrose process is made in~\cite{wagh_89}.  At this time it appears the most likely realization of the Penrose process would be the Blandford--Znajek mechanism~\cite{blandford_77} for launching jets from quasars and microquasars.  Observations suggest~\cite{rawlings_91, punsly_01}, and simulations confirm~\cite{tchekhovskoy_10, tchekhovskoy_11}, that through this mechanism it is possible to extract more energy from the system than is being delivered by accretion.  We discuss jets and the Blandford-Znajek mechanism more in Sections~\ref{section-jets} and \ref{section-numerical-jets}.

\subsection{ISCO: the orbit of marginal stability}
\label{sub-section-marginally-stable}


Particles (with velocity normalization $u^\mu u_\mu = -1$) and photons (with velocity normalization $u^\mu u_\mu = 0$) move freely on ``geodesic'' trajectories $x^\mu = x^\mu(s)$, with velocities $u^\mu \equiv dx^\mu/ds$, characterized by vanishing accelerations
\begin{equation}
\label{geodesics}
a^\nu \equiv u^\mu \nabla_\mu u^\nu = \frac{d^2 x^\nu}{ds^2} -
\Gamma^\nu_{~\kappa \mu}\frac{dx^\kappa}{ds}\,\frac{dx^\mu}{ds} =
0.
\end{equation}
A constant of motion ${\cal X}$, such as those defined in Eq.~(\ref{constant-motion}), is conserved along a geodesic trajectory (\ref{geodesics}) in the sense that $u^\mu \nabla_\mu {\cal X} =0$.

\textit{Circular} geodesic motion in the equatorial plane ($\theta = \pi/2$) is of fundamental importance in black hole accretion disk theory. The four velocity corresponding to circular motion is defined by,
\begin{equation}
\label{circular-geodesic}
u^\mu = A(\eta^\mu + \Omega \xi^\mu),
\end{equation}
where $\Omega = u^{\phi}/u^t = d\phi/dt$ is the angular velocity measured by the stationary observer (ZAVO, see Section~\ref{sub-section-ergosphere}), and the redshift factor, $A = u^t$, follows from $u^\mu u^\nu g_{\mu\nu} = -1$,
\begin{equation}
\label{redshift-factor} -A^{-2} = g_{tt} + 2\,\Omega\,g_{t\phi} +
\Omega^2\,g_{\phi\phi}.
\end{equation}
Other connections between these quantities that are particularly useful in our later calculations also follow from $u^\mu u^\nu g_{\mu\nu} = -1$:
\begin{equation}
\label{omega-covariant} \Omega = -\frac{\ell\,g_{tt} +
g_{t\phi}}{\ell\,g_{t\phi} + g_{\phi\phi}}, ~~~ \ell =
-\frac{\Omega\,g_{\phi\phi} + g_{t\phi}}{\Omega\,g_{t\phi} +
g_{tt}}, ~~~u_t = A(g_{tt} + \Omega g_{t\phi}).
\end{equation}
It is convenient to define the effective potential,
\begin{equation}
\label{effective-relativistic} {\cal U}_{\mathrm{eff}} = -\frac{1}{2} \ln
\vert g^{tt} - 2\ell g^{t\phi}
 + \ell^2\,g^{\phi \phi}\vert,
\end{equation}
because in terms of ${\cal U}_{\mathrm{eff}}$ and the rescaled energy ${\cal E}^* = \ln {\cal E}$, slightly non-circular motion, i.e., with $V^2 = u^r u^r g_{rr} + u^{\theta}u^{\theta}g_{\theta\theta} \ll u^{\phi}u^{\phi} g_{\phi\phi}$, is characterized by the equation,
\begin{equation}
\label{the-same-form}
\frac{1}{2}V^2 = {\cal E}^* - {\cal U}_{\mathrm{eff}},
\end{equation}
which has the same form and the same physical meaning as the corresponding Newtonian equation. Therefore, exactly as in Newtonian theory, unperturbed circular Keplerian orbits are given by the condition of an extremum (minimum or maximum) of the effective potential $(\theta=\pi/2)$,
\begin{equation}
\label{effective-Keplerian-conditions-relativistic} \left\{ \left(
\frac{\partial {\cal U}_{\mathrm{eff}}}{\partial r}\right)_\ell = 0
\right\}
\Rightarrow
\left\{ \left(\frac{\partial g^{tt}}{\partial r}\right)
- 2\ell \left(\frac{\partial g^{t\phi}}{\partial r}\right)
+ \ell^2 \left(\frac{\partial g^{\phi\phi}}{\partial r}\right) = 0
\right\}.
\end{equation}
This quadratic equation for $\ell$ has two roots $\ell = \pm \ell_K(r, a)$, corresponding to ``corotating'' and ``counterrotating'' Keplerian orbits. Their explicit algebraic form is given in Eq.~(\ref{keplerian-momentum-kerr}) in Section~\ref{section-summary-radii-frequences}.

As in Newtonian theory, slightly non-circular orbits (with $V \not = 0$ being either $\delta {\dot r}$ or $\delta {\dot \theta}$) are fully determined by the simple harmonic oscillator equations,
\begin{equation}
\label{harmonic-oscillator-relativistic} \delta{\ddot r} +
\omega^2_r \, \delta{r} = 0, ~~~ \delta{\ddot \theta} +
\omega^2_{\theta} \, \delta{\theta} = 0,
\end{equation}
where the radial $\omega_r$ and vertical $\omega_{\theta}$ epicyclic frequencies are second derivatives of the effective potential,
\begin{equation}
\label{epicyclic-frequencies-relativistic} \omega^2_r = \left(
\frac{\partial^2 {\cal U}_{\mathrm{eff}}}{\partial {r_*}^2}\right)_\ell,
~~~ \omega^2_{\theta} = \left( \frac{\partial^2 {\cal
U}_{\mathrm{eff}}}{\partial {\theta_*}^2}\right)_\ell,
\end{equation}
where $\partial {x_*}^2 = - g_{xx}\,\partial x^2$. The epicyclic frequencies~(\ref{epicyclic-frequencies-relativistic}) are measured by the comoving observer. To get the frequencies ${\omega}_x^*$ measured by the stationary ``observer at infinity'' (Section~\ref{sub-section-ergosphere}), one must rescale by the redshift factor  ${\omega}_x^* = A\,\omega_x$. Obviously, when $({\omega}_r)^2 < 0$, the epicyclic radial oscillations described by Eq.~(\ref{harmonic-oscillator-relativistic}) are \textit{unstable} -- from Eq.~(\ref{epicyclic-frequencies-relativistic}) we see that they correspond to \textit{maxima} of the effective potential. This happens for all circular orbits with radii less than $r = r_{\mathrm{ms}}(a)$, and this limiting radius is called ISCO, the innermost stable circular orbit.

\epubtkImage{}{%
\begin{figure}[htb]
  \centerline{\includegraphics[width=0.6\textwidth]{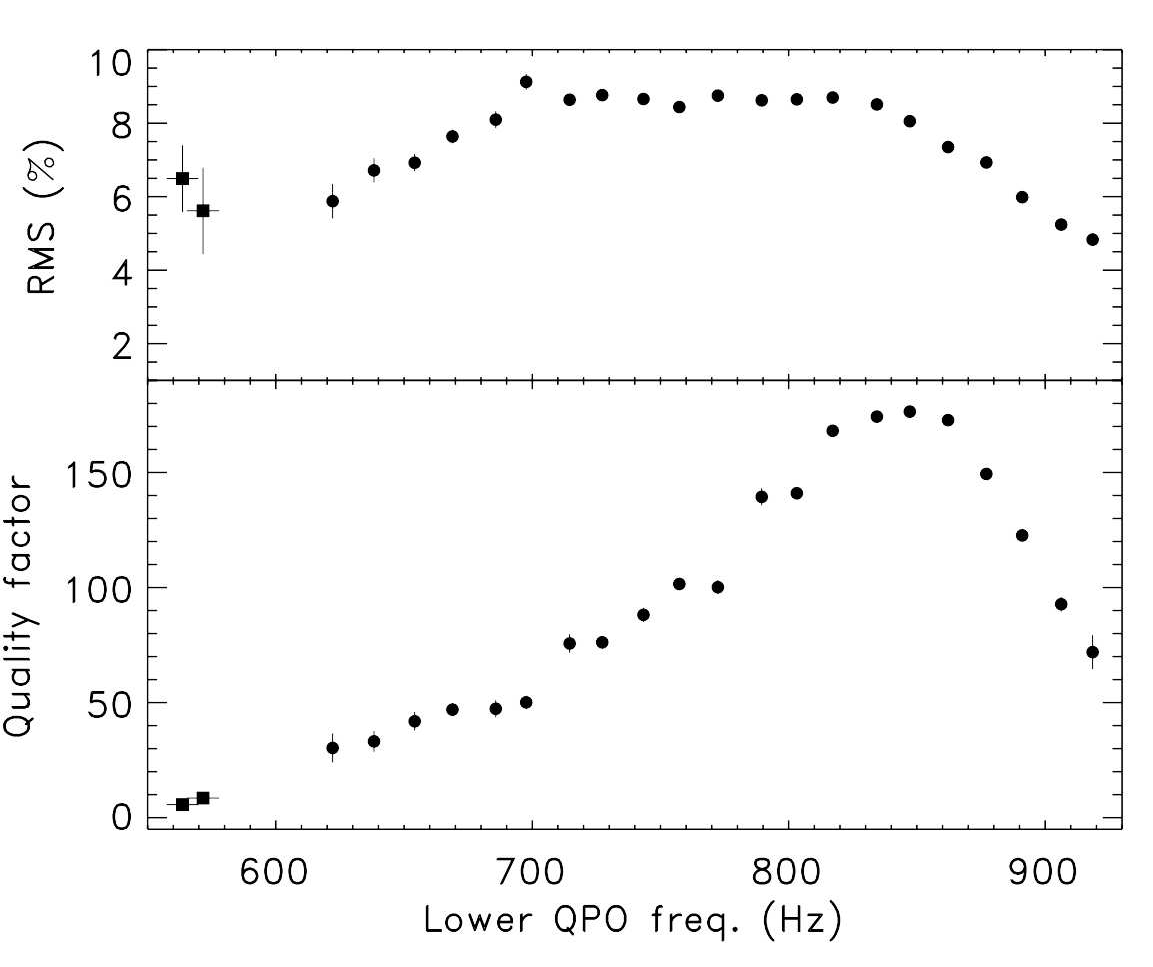}}
\caption{Evidence for the existence of the ISCO from data recorded by the Rossi X-ray Timing Explorer satellite from neutron star binary source 4U~1636--536 \cite{barret_05}. The source shows quasi-periodic oscillations (QPOs) with frequencies in the range $650\unit{Hz} < \nu <900\unit{Hz}$. The sharp drop in the quality factor ({\emph bottom panel}) seen at $\sim 870\unit{Hz}$ may be attributable to the ISCO \cite{barret_07}.}
\label{figure-didier-barret-quality}
\end{figure}}

Free circular orbits with $r > r_{\mathrm{ms}}$ are stable, while those with $r < r_{\mathrm{ms}}$ are not. Accordingly, accretion flows of almost free matter (i.e., with stresses insignificant in comparison with gravity or centrifugal effects), resemble almost circular motion for $r > r_{\mathrm{ms}}$, and almost radial free-fall for $r < r_{\mathrm{ms}}$. For thin disks, this transition in the character of the flow is expected to produce an effective inner truncation radius in the disk (see Section~\ref{section-Shakura-Sunyaev}). The exceptional stability of the inner radius of the X-ray binary LMC X-3~\cite{steiner_10}, provides considerable evidence for such a connection and, hence, for the existence of the ISCO. The transition of the flow at the ISCO may also show up in the observed variability pattern, if variability is modulated by the orbital motion. In this case, one may expect that the there will be no variability observed with frequencies $\nu > \nu_{\mathrm{ISCO}}$, i.e., higher than the Keplerian orbital frequency at ISCO, or that the quality factor for variability, $Q \sim \nu/\Delta \nu$ will significantly drop at $\nu_{\mathrm{ISCO}}$. Several variants of this idea have been discussed~\cite{barret_05, barret_07}, and some observational evidence to support them has been presented (see Figure~\ref{figure-didier-barret-quality}).

\subsection{The Paczy{\'n}ski--Wiita potential}
\label{sub-section-paczynski-wiita}

For a non-rotating black hole ($a=0$), the Kerr metric reduces to the Schwarzschild solution,
\begin{equation}
\label{schwarzschild-metric} ds^2 = -\left( 1 - \frac{2\,M}{r}
\right)\,dt^2 + \left( 1 - \frac{2\,M}{r} \right)^{-1}\,dr^2 +
r^2\, \left[ d\theta^2 + \sin^2\theta\,d\phi^2 \right].
\end{equation}
Paczy{\'n}ski and Wiita~\cite{paczynski_80} proposed a practical and accurate Newtonian model for a Schwarzschild black hole, based on the gravitational potential,
\begin{equation}
\label{paczynski-wiita} \Phi_{PW} = - \frac{GM}{r -r_S},~~~ r_S =
\frac{2GM}{c^2} \, .
\end{equation}
The Paczy{\'n}ski--Wiita potential became a very handy tool for studying black hole astrophysics. It has been used in many papers on the subject and still has applicability today.  The Schwarzschild and the Paczy{\'n}ski--Wiita expressions for the Keplerian angular momentum and locations of the marginally stable and marginally bound orbits (Section~\ref{sub-section-marginally-stable}) are identical. Similar, though less commonly adopted, pseudo-Newtonian potentials have also been found for Kerr (rotating) black holes~\cite{artemova_96,semerak_99,mukhopadhyay_02}.

\subsection{Summary: characteristic radii and frequencies}
\label{section-summary-radii-frequences}

We end this section with a few formulae for the Kerr geometry that we will use elsewhere in this review.

Keplerian circular orbits exist in the region $r > r_{\mathrm{ph}}$, with $r_{\mathrm{ph}}$ being the circular photon orbit. Bound orbits exist in the region $r > r_{\mathrm{mb}}$, with $r_{\mathrm{mb}}$ being the marginally bound orbit, and stable orbits exist for $r > r_{\mathrm{ms}}$, with $r_{\mathrm{ms}}$ being the marginally stable orbit (also called the ISCO -- Section~\ref{sub-section-marginally-stable}). The location of these radii, as well as the location of the horizon $r_H$ and ergosphere $r_0$, are given by the following formulae~\cite{bardeen_72}:
\begin{eqnarray}
\label{photon-radius-kerr}
\text{photon}~~{r_{\mathrm{ph}}} &=&
2 r_G\left\{ 1 + \cos \left[ \frac{2}{3}\cos^{-1}(a_*)\right] \right\},\\
\label{bound-radius-kerr}
\text{bound}~~{r_{\mathrm{mb}}} &=&
2 r_G \left( 1 - \frac{a_*}{2} + \sqrt{1 - a_*}\right),\\
\label{stable-radius-kerr}
\text{stable}~~{r_{\mathrm{ms}}} &=&
r_G \left\{3 + Z_2 -\left[(3 - Z_1) (3 + Z_1 + 2Z_2)\right]^{1/2}\right\}, \\
\text{horizon}~~{r_{\mathrm{H}}} &=&
r_G \left( 1 + \sqrt{1 - a_*^2} \right),\\
\text{ergosphere}~~{r_{0}} &=& r_G \left( 1 + \sqrt{1 - a_*^2
\cos^2 \theta} \right) ,
\end{eqnarray}
where $Z_1 = 1 + (1 - a_*^2)^{1/3} [(1 + a_*)^{1/3} + (1 - a_*)^{1/3}]$, $Z_2 = (3a_*^2 + Z_1^2)^{1/2}$, $a_* = a/M$, and $r_G=GM/c^2$ is the gravitational radius.

The Keplerian angular momentum $\ell_K$ and angular velocity $\Omega_K$, and the angular velocity of frame dragging $\omega$ are given by,
\begin{eqnarray}
\ell_K &=& \ell_K(r, a) =
\frac{M^{1/2}(r^2-2aM^{1/2}r^{1/2}+a^2)}{r^{3/2}-2Mr^{1/2}+aM^{1/2}},
\label{keplerian-momentum-kerr} \\
\Omega_K^2 &=& \frac{GM}{\left(r^{3/2} + aM^{1/2}\right)^2},\\
\label{keplerian-velocity-kerr} \\
\omega &=& \frac{2aMr}{(a^2 + r^2)\varrho^2 + 2a^2Mr \sin^2 \theta}.
\label{dragg-velocity-kerr}
\end{eqnarray}
The epicyclic frequencies measured ``at infinity'' are (here $x =r/M$),
\begin{eqnarray}
\label{epicyclic-radial}
(\omega_r^*)^2 &=& \Omega_K^2 \left( 1 - 6x^{-1} + 8a_*x^{-3/2} -
3a_*^2 x^{-2}\right),\\
\label{epicyclic-vertical}
(\omega^*_{\theta})^2 &=& \Omega_K^2 \left( 1 - 4a_*x^{-3/2} + 3
a_*^2 x^{-2}\right).
\end{eqnarray}
Comparing the Keplerian and epicyclic frequencies and the characteristic radii between the Schwarz\-schild metric and the Paczy{\'n}ski--Wiita potential (Section~\ref{sub-section-paczynski-wiita}), we find for the Schwarz\-schild metric,
\begin{eqnarray}
\label{schwarzschild-frequencies}
&&\Omega^2_K = \frac{GM}{r^3} = (\omega^*_{\theta})^2,
~~(\omega^*_r)^2 = \Omega_K^2 \left( 1 - 6x^{-1}\right),
\nonumber \\
&&\text{and}~~x_{\mathrm{ms}}=6, ~~x_{\mathrm{mb}}=4, ~~x_{\mathrm{ph}}=3,
\end{eqnarray}
and for the Paczy{\'n}ski--Wiita potential,
\begin{eqnarray}
\label{paczynski-frequencies}
&&\Omega^2_K = \frac{GM}{r^3(1 - 2x^{-1})^2} =
(\omega^*_{\theta})^2, ~~(\omega^*_r)^2 = \Omega_K^2 \frac{(1 - 6x^{-1})}{(1
- 2x^{-1})},
\nonumber \\
&& \text{and}~~x_{\mathrm{ms}}=6, ~~x_{\mathrm{mb}}= 4.
\end{eqnarray}

\newpage

\section{Matter Description: General Principles}
\label{section-matter}

Having provided a detailed description of the key signatures associated
with a black hole spacetime, we now move into the mirkier realm of
the accretion disk itself.  We start from the fundamental conservation
laws that govern the behavior of all matter, namely the conservation
of rest mass and conservation of energy-momentum, stated mathematically as
\begin{equation}
\label{general-conservation-equations}
\nabla_\mu (\rho\,u^\mu) = 0,
~~~\nabla_\mu T^\mu_\nu = 0.
\end{equation}
Here $\rho$ is the rest mass density, $u^\mu$ is the four velocity of matter, and
$T^\mu_\nu$ is the stress energy tensor describing properties of the matter. The
conservation equations (\ref{general-conservation-equations}) are supplemented by numerous
``material'' equations, like the equation of state, prescriptions of viscosity, opacity,
conductivity, etc. Several of them are phenomenological or simple approximations.
Nevertheless, we can give a GEN-eral form of $T^\mu_\nu$ that is relevant to accretion disk
theory as a sum of FLU-id, VIS-cous, MAX-well, and RAD-iation parts, which may be written
as,
\begin{eqnarray}
\label{stress-energy-general} (T^\mu_\nu)_{\sf GEN} &=&
(T^\mu_\nu)_{\sf FLU} + (T^\mu_\nu)_{\sf VIS} + (T^\mu_\nu)_{\sf MAX} + (T^\mu_\nu)_{\sf RAD},\\
\label{stress-energy-fluid}
(T^\mu_\nu)_{\sf FLU} &=& (\rho u^\mu)(W u_\nu) + \delta^\mu_\nu\,P, \\
\label{stress-energy-stress}
(T^\mu_\nu)_{\sf VIS} &=& \nu_*\,\sigma^\mu_\nu, \\
\label{stress-energy-Maxwell} (T^\mu_\nu)_{\sf MAX} &=& F^{\mu\alpha}F_{\alpha\nu} - \frac{1}{4} \delta^\mu_\nu F_{\alpha\beta} F^{\alpha\beta}, \\
\label{stress-energy-radiation}
(T^\mu_\nu)_{\sf RAD} &=&  \frac{4}{3} E u^\mu u_\nu + u^\mu\,F_\nu + u_\nu\,F^\mu.
\end{eqnarray}
Here $W$~=~enthalpy, $\delta^\mu_\nu$~=~Kronecker delta tensor, $P$~=~pressure, $\nu_*$~=~kinematic viscosity, $\sigma^\mu_\nu$~=~shear, $F^{\mu\nu}$~=~Faraday electromagnetic field tensor, $E$~=~radiation energy density, and $F^\mu$~=~radiation flux. In the remainder of this section we describe these components one by one, including the most relevant details. Most models of accretion disks are given by steady-state solutions of the conservation equations~(\ref{general-conservation-equations}), with particular choices of the form of the stress-energy tensor $T^\mu_\nu$, and a corresponding choice of the supplementary material equations. For example, thick accretion disk models (Section~\ref{section-thick-disks}) often assume $(T^\mu_\nu)_{\sf VIS} = (T^\mu_\nu)_{\sf MAX} = (T^\mu_\nu)_{\sf RAD} = 0$, thin disk models (Section~\ref{section-thin-disks}) assume $(T^\mu_\nu)_{\sf MAX} = 0$, and most current numerical models (Section~\ref{section:numerical_simulations}) assume $(T^\mu_\nu)_{\sf RAD} = 0$.

\subsection{The fluid part}
\label{subsection-fluid-part}

The one absolutely essential piece of the stress-energy tensor for describing accretion disks is the fluid part, $(T^\mu_\nu)_{\sf FLU} = (\rho u^\mu)(W u_\nu) + \delta^\mu_\nu\,P$. The fluid density, enthalpy, and pressure, as well as other fluid characteristics, are linked by the first law of thermodynamics, $dU = T\,dS - P\,dV$, which we write in the form,
\begin{equation}
\label{first-law-thermodynamics} d\epsilon = W\,d\rho + n T\,dS,
\end{equation}
where $U$ is the internal energy, $T$ is the temperature, $S$ is the entropy, and $\epsilon = \rho\,c^2 + \Pi$ is the total energy density, with $\Pi$ being the internal energy density, and
\begin{equation}
\label{volume-energy-enthalpy} V = \frac{1}{n}, ~~~ U =
\frac{\Pi}{n}, ~~~ W = \frac{P + \epsilon}{\rho}.
\end{equation}
The equation of state is often assumed to be that of an ideal gas,
\begin{equation}
\label{ideal-gas-radiation}
P = \frac{\cal R}{\mu}\rho T ,
\end{equation}
with ${\cal R}$ being the gas constant and $\mu$ the mean molecular weight.

Sometimes we may wish to consider a two temperature fluid, where the temperature $T_i$ and molecular weight $\mu_i$ of the ions are different from those of the electrons ($T_e$ and $\mu_e$).  For such a case
\begin{equation}
\label{two-temperature-pressure}
P = P_i + P_e = \frac{\cal R}{\mu_i}\rho T_i + \frac{\cal
R}{\mu_e}\rho T_e.
\end{equation}
Two-temperature plasmas are critical in advection-dominated flows (discussed in Section~\ref{section-ADAFs}). Two-temperature fluids are also important when one considers radiation~\cite{shapiro_76}, as it is the ions that are generally heated by dissipative processes in the disk, while it is generally the electrons that radiate. Ions and electrons normally exchange energy via Coulomb collisions.  As this process is generally not very efficient, the electrons in the inner parts of accretion flows are usually much cooler than the ions (Coulomb collisions are not able to heat the electrons as fast as they radiate or advect into the black hole).  However, there have been suggestions that more efficient processes may couple the ions and electrons~\cite{phinney_81}, such as plasma waves~\cite{begelman_88} or kinetic instabilities~\cite{sharma_07}.  At this point, this remains an open issue in plasma physics, so it is difficult to know how much heating electrons experience.

\subsubsection{Perfect fluid}
\label{section-perfect-fluid}

In the case of a \textit{perfect fluid}, the whole stress-energy tensor~(\ref{stress-energy-general}) is given by its fluid part~(\ref{stress-energy-fluid}), and all other parts vanish, i.e., $(T^\mu_\nu)_{\sf GEN} =(T^\mu_\nu)_{\sf FLU}$. In this particular case, one can use $\nabla_\mu(T^\mu_\nu \eta^\nu) = \eta^\nu \nabla_\mu(T^\mu_\nu) + T^{\mu \nu} (\nabla_\mu \eta_\nu) = 0 + 0 = 0$, and similarly derived $\nabla_\mu(T^\mu_\nu\xi^\nu) = 0$, to prove that
\begin{equation}
\label{bernoulli}
{\cal B} = - W(u\eta) = - W\,u_t, \qquad {\cal J} = W(u\xi) = W\,u_\phi,
\end{equation}
are constants of motion. We can identify ${\cal B}$ as the Bernoulli function and ${\cal J}$
as the angular momentum. Their ratio is obviously also a constant of motion,
\begin{equation}
\label{specific-momentum-fluid} \ell = \frac{\cal J}{\cal B} =
- \frac{u_\phi}{u_t},
\end{equation}
identical in form with the specific angular momentum~(\ref{constant-motion-specific}), which is a constant of geodesic motion.

\subsection{The stress part}
\label{subsection-stress-part}

In the stress part $(T^\mu_\nu)_{\sf VIS} =\nu_*\,\sigma^\mu_\nu$, the shear tensor $\sigma^\mu_\nu$ is a kinematic invariant (cf.\ Footnote~\ref{foot-kinematic-invariants}).  It is defined as
\begin{equation}
\label{shear-tensor}
\sigma_{\mu \nu} \equiv \left[ \frac{1}{2}(\nabla_{\mu}u_{\nu} + \nabla_{\nu}u_{\mu})- \Theta\,g_{\mu\nu}\right]_{\perp}
\end{equation}
where the symbol $_{\perp}$ denotes projection into the instantaneous 3-space perpendicular to $u^\mu$ in the sense that $(X^\mu)_{\perp} = X^\alpha\,(\delta^{\mu}_{~\alpha} + u^\mu\,u_\alpha)$. The other kinematic invariants are vorticity,
\begin{equation}
\label{vorticity-tensor}
\omega_{\mu\nu} \equiv \frac{1}{2} (\nabla_{\mu}u_{\nu} - \nabla_{\nu}u_{\mu})_{\perp}
\end{equation}
and expansion,
\begin{equation}
\label{expansion-scalar}
\Theta \equiv \frac{1}{3}(\nabla_\mu\,u^\mu)
\end{equation}

In the standard hydrodynamical description (e.g.~\cite{landau_59}), the viscous stress tensor, $S^\mu_\nu$, is proportional to the shear tensor,
\begin{equation}
\label{viscous-stress-landau} S^\mu_\nu =
\nu_*\,\rho\,\sigma^\mu_\nu.
\end{equation}
The rate of heat generation by viscous stress in a volume $V$ is then given by
\begin{equation}
\label{viscous-stress-energy-generation} Q^+ = \int S^\mu_\nu
\sigma^\nu_{~\mu}\,dV \,.
\end{equation}
In addition, the rates of viscous angular momentum and energy transport across a surface $S$, with a unit normal vector $N^\mu$, are
\begin{equation}
\label{viscous-transport} {\cal J}_S = \int S^\mu_\nu \xi^k\nu_* N_\mu
dS, ~~~ {\cal B}_S = \int S^\mu_\nu \eta^\nu N_\mu dS \,.
\end{equation}

For the case of purely circular motion, where $u^i = A(\eta^i + \Omega\xi^i)$, the kinematic invariants are
\begin{equation}
\label{expansion-shear-vorticity}
\Theta = 0, ~~~\sigma_{\mu\phi} = \frac{1}{2}A^3\Psi^2\partial_\mu\Omega,
~~~\omega_{\mu\phi} = \frac{1}{2}A^3\Psi^{-2}\partial_\mu\ell \,,
\end{equation}
where $\Psi^2 = g_{t\phi}^2 - g_{tt}g_{\phi\phi}$. It is a general property that $(X^\mu)_{\perp}\,u_\mu =0$, and so for purely circular motion, one has,
\begin{equation}
\label{shear-vorticity-perpendicular}
\sigma_{\mu\nu}\eta^\nu = -\Omega\, \sigma_{\mu\nu}\xi^\nu, ~~~
\omega_{\mu\nu}\eta^\nu = -\Omega\, \omega_{\mu\nu}\xi^\nu.
\end{equation}
From Eqs.~(\ref{viscous-transport}) and (\ref{shear-vorticity-perpendicular}) one deduces that for purely circular motion, the rates of energy and angular momentum transport are related as
\begin{equation}
\label{viscous-transport-circular} {\cal B}_S = -\Omega_S\,{\cal J}_S,
\end{equation}
where $\Omega_S$ is the angular velocity averaged on the surface $S$. From this, one sees that as angular momentum is transported outward, additional energy is carried inward by the fluid.

\subsubsection{The alpha viscosity prescription}
\label{subsubsection-alpha-viscosity}

As we mentioned in Section~\ref{section-introduction}, the viscosity in astrophysical accretion disks can not come from ordinary molecular viscosity, as this is orders of magnitude too weak to explain observed phenomena.  Instead, the source of stresses in the disk is likely turbulence driven by the magneto-rotational instability (MRI, described in Section~\ref{section-MRI}).  Even so, one can still parametrize the stresses within the disk as an \emph{effective viscosity} and use the normal machinery of standard hydrodynamics without the complication of magnetohydrodynamics (MHD).  This is sometimes desirable as analytic treatments of MHD can be very difficult to work with and full numerical treatments can be costly.

For these reasons, the Shakura--Sunyaev ``alpha viscosity'' prescription~\cite{shakura_73} still finds application today.  It is an \textit{ad hoc} assumption based on dimensional arguments. Shakura and Sunyaev realized that if the source of viscosity in accretion disks is turbulence, then the kinematic viscosity coefficient $\nu_*$ has the form,
\begin{equation}
\nu_* \approx l_0\,v_0 \,,
\end{equation}
where $l_0$ is the correlation length of turbulence and $v_0$ is the mean turbulent speed. Assuming that the velocity of turbulent elements cannot exceed the sound speed, $v_0 < c_S$, and that their typical size cannot be greater than the disk thickness, $l_0 < H$, one gets
\begin{equation}
\label{viscosity-ansatz} \nu_* = \alpha\, H c_S \,,
\end{equation}
where $0 <\alpha <1$ is a dimensionless coefficient, assumed by Shakura and Sunyaev to be a constant.

For thin accretion disks (see Section~\ref{section-thin-disks}) the viscous stress tensor reduces to an internal torque with the following approximate form [see Eqs.~(\ref{shear-tensor}) and (\ref{viscous-stress-landau})]
\begin{equation}
{\cal T}_{r\phi} \approx \rho\nu_* r\frac{\partial \Omega}{\partial r} \,.
\label{viscosity-ansatz-stress-thickness}
\end{equation}
However, for thin disks, $r (\partial \Omega/\partial r)\approx -\Omega$ and $c_S \approx (P/\rho)^{1/2} \approx \Omega H$, so Shakura and Sunyaev argued that the torque must have the form ${\cal T}_{r\phi} = -\alpha P$. A critical question that was left unanswered was what pressure $P$ one should consider: $P_\mathrm{gas}$, $P_\mathrm{rad}$, or $P_\mathrm{Tot} = P_\mathrm{gas} + P_\mathrm{rad}$? This question has now been answered using numerical simulations~\cite{hirose_09a}, so that we now know the appropriate pressure to be $P_\mathrm{Tot}$. Typical values of $\alpha$ estimated from magnetohydrodynamic simulations are close to 0.02~\cite{hawley_11}, while observations suggest a value closer to 0.1 (see~\cite{king_07} and references therein).

\subsection{The Maxwell part}
\label{subsection-Maxwell-part}

Magnetic fields may play many interesting roles in black hole accretion disks. Large scale magnetic fields threading a disk may exert a torque, thereby extracting angular momentum~\cite{blandford_82}. Similarly, large scale poloidal magnetic fields threading the inner disk, ergosphere, or black hole, have been shown to be able to carry energy and angular momentum away from the system, and power jets~\cite{blandford_77}. Weak magnetic fields can tap the differential rotation of the disk itself to amplify and trigger an instability that leads to turbulence, angular momentum transport, and energy dissipation (exactly the processes that are needed for accretion to happen)~\cite{balbus_91,balbus_98}.

In most black hole accretion disks, it is reasonable to assume ideal MHD, whereby the conductivity is infinite, and consequently the magnetic diffusivity is zero. Whenever this is true, magnetic field lines are effectively frozen into the fluid. A corollary to this is that parcels of fluid are restricted to moving along field lines, like ``beads'' on a wire. In ideal MHD, the Faraday tensor obeys the homogeneous Maxwell's equation
\begin{equation}
\nabla_\mu (^*F^\mu_\nu\,)=0,
\label{maxwell-equation}
\end{equation}
where $^*F^\mu _\nu$ is the dual. If we define a magnetic field 4-vector $b^\mu \equiv u^\nu\, ^*F^\mu _\nu$, then using $b^\mu u_\mu = 0$ one can show that
\begin{equation}
^*F^\mu _\nu = b^\mu u_\nu - b_\nu u^\mu.
\label{faraday-tensor-definition}
\end{equation}
Using this, it is easy to show that the spatial components of~(\ref{maxwell-equation}) give the induction equation
\begin{equation}
\partial_t (\sqrt{-g}B^i) = -\partial_j [\sqrt{-g}(B^i v^j - B^j v^i)],
\end{equation}
while the time component gives the divergence-free constraint
\begin{equation}
\partial_i (\sqrt{-g}B^i) = 0 \,,
\end{equation}
where $B^i = u^t b^i - u^i b^t$, and $g$ is the 4-metric determinant.  

\subsubsection{The magneto-rotational instability (MRI)}

We mentioned in Section~\ref{subsection-stress-part} that a hydrodynamic treatment of accretion requires an internal viscous stress tensor of the form ${\cal T}_{r\phi} < 0$.  However, we also pointed out that ordinary molecular viscosity is too weak to provide the necessary level of stress.  Another possible source is turbulence.  The mean stress from turbulence always has the property that ${\cal T}_{r\phi} < 0$, and so it can act as an effective viscosity. As we will explain in Section~\ref{section-MRI}, weak magnetic fields inside a disk are able to tap the shear energy of its differential rotation to power turbulent fluctuations. This happens through a mechanism known as the magneto-rotational (or ``Balbus--Hawley'') instability~\cite{balbus_91,hawley_91b,balbus_98}. Although the non-linear behavior of the MRI and the turbulence it generates is quite complicated, its net effect on the accretion disk can, in principle, be characterized as an effective viscosity, possibly making the treatment much simpler.  However, no such complete treatment has been developed at this time.

\subsection{The radiation part}
\label{subsection-radiation-part}

Radiation is important in accretion disks as a way to carry excess energy away from the system.  In geometrically thin, optically thick (Shakura--Sunyaev) accretion disks (Section~\ref{section-Shakura-Sunyaev}), radiation is highly efficient and nearly all of the heat generated within the disk is radiated locally.  Thus, the disk remains relatively cold.  In other cases, such as ADAFs (Section~\ref{section-ADAFs}), radiation is inefficient; such disks often remain geometrically thick and optically thin.

In the optically thin limit, the radiation emissivity $f$ has the following components: brems\-strah\-lung $f_{\mathrm{br}}$, synchrotron $f_{\mathrm{synch}}$, and their Comptonized parts $f_{\mathrm{br, C}}$ and $f_{\mathrm{synch, C}}$. In the optically thick limit, one often uses the diffusion approximation with the total optical depth $\tau = \tau_{\mathrm{abs}} + \tau_{\mathrm{es}}$ coming from the absorption and electron scattering optical depths. In the two limits, the emissivity is then
\begin{equation}
f = \begin{cases} f_{\mathrm{br}} + f_{\mathrm{synch}} +
f_{\mathrm{br, C}} + f_{\mathrm{synch, C}} & \text{~~~~optically thin case~} (\tau \ll 1),\\
\frac{8 \sigma T_e^{4}}{3 H \tau} & \text{~~~~optically thick case~} (\tau \gg 1),
\end{cases}
\end{equation}
where $\sigma$ is the Stefan--Boltzmann constant. In the intermediate case one should solve the transfer equation to get reliable results, as has been done in~\cite{shimura_95,davis_06a}. Often, though, the solution of the grey problem obtained by Hubeny~\cite{hubeny_90} can serve reliably:
\begin{equation}
f  =  \frac{4\sigma T_e^4}{H} \left[\frac{3\tau}{2} + \sqrt{3} +
       \frac{4\sigma T_e^4}{H}
       \left( f_{\mathrm{br}} + f_{\mathrm{synch}} + f_{\mathrm{br, C}}
      + f_{\mathrm{synch, C}}\right)^{-1}\right]^{-1}.
\label{cooling}
\end{equation}
%
%
%
%
In sophisticated software packages like {\tt BHSPEC}, color temperature corrections in the optically thick case (the ``hardening factor'') are often applied~\cite{davis_06a}.

In the remaining parts of this section we give explicit formulae for the bremsstrahlung and synchrotron emissivities and their Compton enhancements. These sections are taken almost directly from the work of Narayan and Yi~\cite{narayan_95b}.  Additional derivations and discussions of these equations in the black hole accretion disk context may be found in~\cite{svensson_82,stepney_83,narayan_95b,esin_96}.

\subsubsection{Bremsstrahlung}
\label{subsubsection-Bremsstrahlung}

Thermal bremsstrahlung (or free-free emission) is caused by the inelastic scattering of relativistic thermal electrons off (nonrelativistic) ions and other electrons. The emissivity (emission rate per unit volume) is $f_{\mathrm{br}} = f_{ei} + f_{ee}$. The ion-electron part is given by~\cite{narayan_95b}
\begin{equation}
f_{ei} = n_e {\bar n} \sigma_T c \alpha_f m_e c^2\times
\begin{cases}4 \left(\frac{2 \theta_e}{\pi^3}\right)^{1/2}(1+ 1.781\theta^{1.34}_e) & \theta_e < 1,\\
\frac{9\theta_e}{2 \pi} [\ln(1.123 \theta_e + 0.48) + 1.5]& \theta_e \ge 1.
\end{cases}
\end{equation}
where $n_e$ is the electron number density, ${\bar n}$ is the ion number density averaged over all species, $\sigma_T = 6.62\times 10^{-25}\mathrm{\ cm}^2$ is the Thomson cross section,
$\alpha_f=1/137$ is the fine structure constant, $\theta_e = k_BT_e/m_e c^2$ is the dimensionless electron temperature, and $k_B$ is the Boltzmann constant. The electron-electron part is given by~\cite{narayan_95b}
\begin{equation}
f_{ee} = n_e^2 c r_e^2 m_e c^2 \alpha_f \times
\begin{cases} \frac{20}{9 \pi^{1/2}}
(44 - 3 \pi^2) \theta_e^{3/2}(1 + 1.1 \theta_e + \theta_e^2 - 1.25 \theta_e^{5/2})& \theta_e < 1,\\
24 \theta_e [\ln (1.1232 \theta_e) + 1.28] & \theta_e \ge 1.
\end{cases}
\end{equation}
where $r_e = e^2/m_e c^2$ is the classical radius of the electron.

\subsubsection{Synchrotron}
\label{subsubsection-Synchrotron}

Assuming the accretion environment is threaded by magnetic fields, the hot (relativistic) electrons can also radiate via synchrotron emission.  For a relativistic Maxwellian distribution of electrons, the formula is~\cite{narayan_95b}
\begin{equation}
f^-_{\mathrm{synch}} = \frac{2\pi}{3 c^2} kT_e \frac{d}{dr} \left[
\frac{3eB\,\theta_e^2 x_M}{4 \pi m_e c} \right],
\end{equation}
where $e$ is the electric charge, $B$ is the equipartition magnetic field strength, and $x_M$ is the solution of the transcendental equation
\begin{equation}
\exp(1.8899x_M^{1/3}) = 2.49 \times 10^{-10} \left(\frac{4 \pi n_e
r}{B}\right) \frac{x_M^{-7/6} + 0.40 x_M^{-17/12} + 0.5316
x_M^{-5/3}}{\theta_e^3 K_2(1/\theta_e)},
\end{equation}
where the radius $r$ must be in physical units and $K_2$ is the modified Bessel function of the 2nd kind. This expression is valid only for $\theta_e > 1$, but that is sufficient in most applications.

\subsubsection{Comptonization}
\label{subsubsection-Comptonization}

The hot, relativistic electrons can also Compton up-scatter the photons emitted via bremsstrahlung and synchrotron radiation. The formulae for this are~\cite{narayan_95b}
\begin{eqnarray}
f_{\mathrm{br, C}} &=& f_{\mathrm{br}}\left\{ \eta_1 - \frac{\eta_1 x_c}{3\theta_e} - \frac{3\eta_1 \left[3^{-(\eta_3 + 1)} - (3\theta_e)^{-(\eta_3 + 1)}\right]}{\eta_3+1} \right\} \\
f_{\mathrm{synch, C}} &=& f_{\mathrm{synch}} \left[\eta_1 - \eta_2
\left({x_c / \theta_e}\right)^{\eta_3} \right].
\end{eqnarray}
Here $\eta = 1 + \eta_1 + \eta_2 (x/\theta_e)^{\eta_3}$ is the Compton energy enhancement factor, and
\begin{alignat}{2}
x &= \frac{h \nu}{m_e c^2},~~x_c = \frac{h \nu_c}{m_e c^2},
&\quad ~~~\eta_1 &= \frac{x_2(x_1-1)}{1 - x_1 x_2}, \nonumber  \\
x_1 &= 1 + 4\theta_e + 16 \theta^2_e,
&\quad ~~~~\eta_2  &= \frac{- \eta_1}{3^{\eta_3}}, \nonumber  \\
x_2 &= 1 - \exp{(-\tau_{es})},
&\quad ~~~~\eta_3  &=-1 - \ln x_2/\ln x_1 ,
\end{alignat}
where $h$ is Planck's constant and $\nu_c$ is the critical frequency, below which it is assumed that the emission is completely self-absorbed and above which the emission is assumed to be optically thin.

\newpage

\section{Thick Disks, Polish Doughnuts, \& Magnetized Tori}
\label{section-thick-disks}

In this Section we discuss the simplest analytic model of a black hole accretion disk -- the ``Polish doughnut.''  It is simplest in the sense that it only considers gravity (Section~\ref{section-strong-gravity}), plus a perfect fluid (Section~\ref{section-perfect-fluid}), i.e., the absolute minimal description of accretion.  We include magnetized tori in Section~\ref{section-magnetized-tori}, which allows for $(T^\mu_\nu)_{\sf MAX} \ne 0$, but otherwise $(T^\mu_\nu)_{\sf VIS} = (T^\mu_\nu)_{\sf MAX} = (T^\mu_\nu)_{\sf RAD} = 0$ throughout this section.

\subsection{Polish doughnuts}

Paczy{\'n}ski and his collaborators developed, in the late 1970s and early 1980s, a very general method of constructing perfect fluid equilibria of matter orbiting around a Kerr black hole~\cite{jaroszynski_80, paczynski_80, paczynski_81, paczynski_82}. They assumed for the stress energy tensor and four velocity,
\begin{eqnarray}
\label{assumptions-paczynski} T^{\mu}_{~\nu} = (T^{\mu}_{~\nu})_{\sf FLU} & = & \rho W
u^{\mu}u_{\nu} + P\delta^{\mu}_{~\nu},\nonumber \\
u^\mu & = & A\,(\eta^\mu + \Omega \xi^\mu),
\end{eqnarray}
and derived from $\nabla_{\mu}T^{\mu}_{~\nu}=0$ that,
\begin{equation}
\label{perfect-fluid-circular}
\nabla_\nu \ln A - \frac{\ell\nabla_\nu \Omega}{1 - \ell\Omega} = \frac{1}{\rho}\nabla_\nu\,P \, .
\end{equation}
In the case of a barytropic fluid $P = P(\epsilon)$, the right hand side of Eq.~(\ref{perfect-fluid-circular}) is the gradient of a scalar function, and thus the left hand side must also be the gradient of a scalar, which is possible \textit{if and only if}
\begin{equation}
\label{equipressure-perfect-fluid1} \ell = \ell(\Omega).
\end{equation}
This statement is one of several useful integrability conditions, collectively called \textit{von Zeipel theorems}, found by a number of authors~\cite{boyer_65,bardeen_70,abramowicz_71,komissarov_06}.

In real flows, the function $\ell = \ell(\Omega)$ is determined by dissipative processes that have timescales much longer than the dynamical timescale, and are not yet fully understood. Paczy{\'n}ski realized that instead of deriving $\ell = \ell(\Omega)$ from unsure assumptions about viscosity that involve a free function fixed \textit{ad hoc} (e.g., by assuming $\alpha(r, \theta) =$ const), one may instead \textit{assume} the result, i.e., assume $\ell = \ell(\Omega)$. Assuming $\ell = \ell(\Omega)$ is \textit{not} self-consistent, but neither is assuming $\alpha(r, \theta) =$ const.

In Boyer--Lindquist coordinates, the equation for the equipressure surfaces, $P = P(r, \theta) =\,$ const, may be written as $r_P = r_P(\theta)$, with the function $r_P(\theta)$ given by
\begin{equation}
\label{equipressure-perfect-fluid2}
- \frac {dr_P}{d\theta} = \frac{\partial_{\theta}\,P }{\partial_r\,P } = \frac {(1 - \ell\,\Omega)
\partial_{\theta} \ln\,A + \ell\,\partial_{\theta}\,\Omega} {(1 -
\ell\,\Omega ) \partial_{r} \ln\,A + \ell\,\partial_{r}\,\Omega} \,.
\end{equation}
Using the expressions for $A = A(r, \theta)$, $\Omega = \Omega(r,
\theta)$, and $\ell = \ell(r, \theta)$ from
Section~\ref{section-strong-gravity} (Eqs.~\ref{redshift-factor} and
\ref{omega-covariant}), one can integrate
Eq.~(\ref{equipressure-perfect-fluid2}) to get the equipressure
surfaces. A description of how to do this for both Schwarzschild and
Kerr black holes is given in~\cite{chakrabarti_85}.
Figure~\ref{figure:equipotential-Roche} illustrates the simplest (and
important) case of $\ell = \ell(\Omega) = \ell_0 = \mathrm{const}$.

\epubtkImage{}{%
\begin{figure}[htbp]
  \centerline{\includegraphics[scale=0.75]{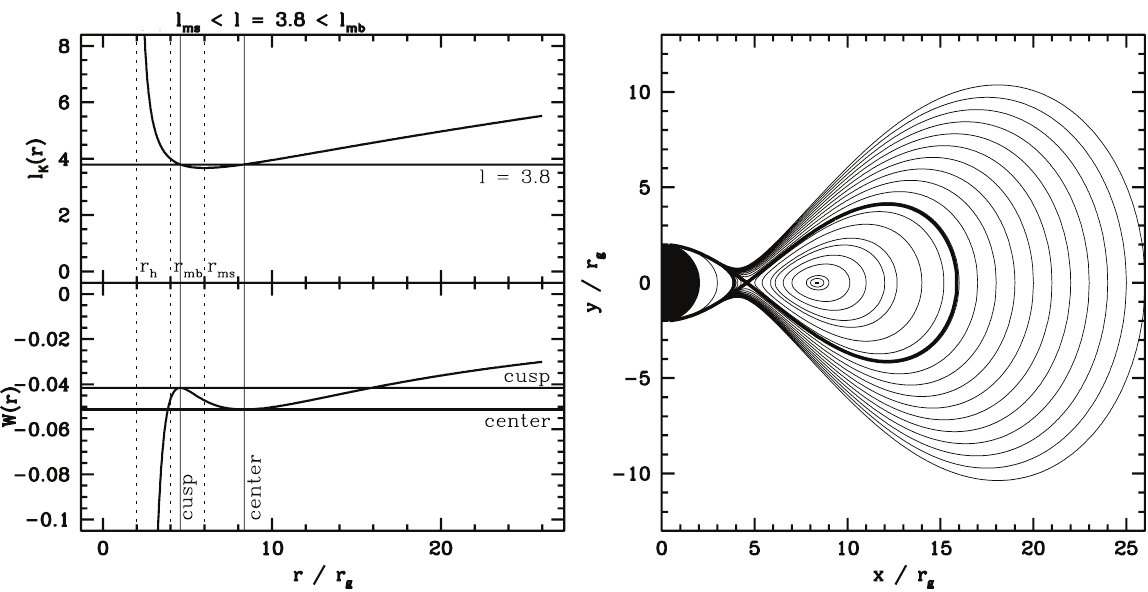}}
  \caption{In equilibrium, the equipressure surfaces should coincide with the surfaces shown by the solid lines in the right panel. Note the Roche lobe, self-crossing at the cusp. The cusp and the center,  both located at the equatorial plane $\theta = \pi/2$, are circles on which the pressure gradient vanishes. Thus, the (constant) angular momentum of matter equals the Keplerian angular momentum at these two circles, $\ell_0 = \ell_{K}(r_{\mathrm{cusp}}) = \ell_{K} (r_{\mathrm{center}})$, as shown in the upper left panel. Image adapted from~\cite{font_02}. In this figure $W$ refers to the effective potential.}
  \label{figure:equipotential-Roche}
\end{figure}}

Another useful way to think about thick disks is from the relativistic analog of the Newtonian effective potential $\Phi$,
\begin{equation}
\label{potential-gap}
\Phi - \Phi_\mathrm{in} = - \int^P_0 \frac{dP}{\rho W} \, ,
\end{equation}
where $\Phi_\mathrm{in}$ is the potential at the boundary of the thick disk. For constant angular momentum $\ell$, the form of the potential reduces to $\Phi = \ln (-u_t)$. Provided $\ell > \ell_\mathrm{ms}$, the potential $\Phi(r,\theta)$ will have a saddle point $\Phi_\mathrm{cusp}$ at $r=r_\mathrm{cusp}$, $\theta = \pi/2$. We can define the parameter $\Delta \Phi = \Phi_\mathrm{in} - \Phi_\mathrm{cusp}$ as the potential barrier (energy gap) at the inner edge of the disk. If $\Delta \Phi < 0$, the disk lies entirely within its Roche lobe, whereas if $\Delta \Phi > 0$, matter will spill into the black hole even without any loss of angular momentum.

Before leaving the topic of Polish doughnuts, we should point out
that, starting with the work of Hawley, Smarr, and
Wilson~\cite{hawley_84a}, this simple, analytic solution has been the
most commonly used starting condition for numerical studies of black
hole accretion.

\subsection{Magnetized Tori}
\label{section-magnetized-tori}

Komissarov~\cite{komissarov_06} was able to extend the Polish doughnut solution by adding a purely azimuthal magnetic field to create a magnetized torus. This is possible because a magnetic field of this form only enters the equilibrium solution as an additional pressure-like term.  For example, the extended form of Eq.~(\ref{perfect-fluid-circular}) is
\begin{equation}
\nabla_\nu\,\ln A - \frac{\ell\,\nabla_\nu\,\Omega}{1 - \ell\,\Omega} = \frac{\nabla_\nu\,P}{\rho} + \frac{\nabla_\nu (\Psi^2 P_{\mathrm{mag}})}{\Psi^2 \rho} \, ,
\end{equation}
where $\Psi^2 = g_{t\phi} g_{t\phi} - g_{tt} g_{\phi\phi}$ and Eq.~(\ref{potential-gap}) becomes
\begin{equation}
\Phi - \Phi_\mathrm{in} = - \int^P_0 \frac{dP}{\rho W} - \int^{\tilde{P}_{\mathrm{mag}}}_0 \frac{d\tilde{P}_{\mathrm{mag}}}{\Psi \rho W} \, ,
\end{equation}
where $\tilde{P}_{\mathrm{mag}} = \Psi^2 P_{\mathrm{mag}}$. Komissarov~\cite{komissarov_06} gives a procedure for solving the case of a barotropic magnetized torus with constant angular momentum ($\ell =$ const.).

\newpage

\section{Thin Disks}
\label{section-thin-disks}

Most analytic accretion disk models assume a stationary and axially symmetric state of the matter being accreted into the black hole. In such models, all physical quantities depend only on the two spatial coordinates: the ``radial'' distance from the center $r$, and the ``vertical'' distance from the equatorial symmetry plane $z$. In addition, the most often studied models assume that the disk is not vertically thick. In ``thin'' disks $z/r \ll 1$ everywhere inside the matter distribution, and in ``slim'' disks (Section~\ref{section-slim-disks}) $z/r \le 1$.

In thin and slim disk models, one often uses a vertically integrated form for many physical quantities. For example, instead of density $\rho(r, z)$ one uses the surface density defined as,
\begin{equation}
\label{surface-density}
\Sigma (r) = \int_{-H(r)}^{+H(r)} \rho(r, z) dz,
\end{equation}
where $z=\pm\,H(r)$ gives the location of the surface of the accretion disk.

\subsection{Equations in the Kerr geometry}
\label{subsection-thin-disk-quations-Kerr}

The general relativistic equations describing the physics of thin disks have been derived independently by several authors~\cite{lasota_94,abramowicz_96,abramowicz_97,gammie_98,beloborodov_98}. Here we present them in the form used in~\cite{sadowski_09}:

(i) \textit{Mass conservation (continuity)}:
\begin{equation}
 \dot M=-2\pi \Sigma\Delta^{1/2}\frac{V}{\sqrt{1-V^2}}
\label{eq_cont2}
\end{equation}
where $V$ is the gas radial velocity measured by an observer at fixed $r$ who co-rotates with the fluid, and $\Delta$ has the same meaning as in Section~\ref{section-strong-gravity}.

(ii) \textit{Radial momentum conservation}:
\begin{equation}
\frac{V}{1-V^2}\frac{dV}{dr}=\frac{\cal
A}{r}-\frac{1}{\Sigma}\frac{dP}{dr} \label{eq_rad3}
\end{equation}
where
\begin{equation}
{\cal A}= -\frac{M\tilde{A}}{r^3\Delta\Omega_K^+\Omega_K^-}\frac{(\Omega-\Omega_K^+)
(\Omega-\Omega_K^-)}{1-\tilde\Omega^2\tilde R^2} \,,
\label{eq_rad4}
\end{equation}
$\tilde{A} = (r^2+a^2)^2-a^2 \Delta \sin^2 \theta$, $\Omega=u^\phi /u^t$ is the angular velocity with respect to the stationary observer, $\tilde\Omega=\Omega-\omega$ is the angular velocity with respect to the inertial observer, $\Omega_K^\pm=\pm M^{1/2}/(r^{3/2}\pm aM^{1/2})$ are the angular frequencies of the co-rotating and counter-rotating Keplerian orbits, and $\tilde R=\tilde{A}/(r^2\Delta^{1/2})$ is the radius of gyration.

(iii) \textit{Angular momentum conservation}:
\begin{equation}
 \frac{\dot{M}}{2\pi}({\cal L}-{\cal L}_{in})=
 \frac{\tilde{A}^{1/2}\Delta^{1/2}\gamma}{r}\alpha \Pi
\label{eq_ang6}
\end{equation}
where ${\cal L}=u_\phi$ is the specific angular momentum, $\gamma$ is the Lorentz factor, $\Pi=2HP$ can be considered to be the vertically integrated pressure, $\alpha$ is the standard \textit{alpha viscosity} (Section~\ref{subsubsection-alpha-viscosity}), and ${\cal L}_{in}$ is the specific angular momentum at the horizon, which can not be known \textit{a priori}. As we explain in the next section, it provides an eigenvalue linked to the unique eigensolution of the set of thin disk differential equations, once they are properly constrained by boundary and regularity conditions.

(iv) \textit{Vertical equilibrium}:
\begin{equation}
 \frac{\Pi}{\Sigma H^2}=\frac{{\cal L}^2-a^2({\cal E}^2-1)}{2 r^4}
\label{eq_vert}
\end{equation}
with ${\cal E} = -u_t$ being the conserved energy associated with the time symmetry.

(v) \textit{Energy conservation}:
\begin{equation}
-\frac{\alpha \Pi \tilde{A}\gamma^2}{r^3}\frac{d\Omega}{dr}-\frac{32}{3}\frac{\sigma
T^4}{\kappa\Sigma}= -\frac{\dot M}{2\pi
r\rho}\frac1{\Gamma_3-1}\left(\der Pr-\Gamma_1\frac P\rho\der\rho
r\right)
\label{energy}
\end{equation}
where $T$ is the temperature in the equatorial plane, $\kappa$ is the mean (frequency-independent) opacity,
\[
\Gamma_1 = \beta^* + (4-3\beta^*)(\Gamma_3 - 1) \,,
\]
\[
\Gamma_3 = 1 + \frac{(4-3\beta^*)(\gamma_g -1)}{12(1-\beta/\beta_m)(\gamma_g-1) + \beta}\,,
\]
$\beta = P_{\mathrm{gas}}/(P_{\mathrm{gas}} + P_{\mathrm{rad}} + P_{\mathrm{mag}})$, $\beta_m = P_{\mathrm{gas}}/(P_{\mathrm{gas}} + P_{\mathrm{mag}})$, $\beta^* = \beta(4-\beta_m)/3\beta_m$, and $\gamma_g$ is the ratio of specific heats of the gas.

\subsection{The eigenvalue problem}
\label{subsection-sonic}

Through a series of algebraic manipulations one can reduce the thin disk equations to a set of two ordinary differential equations for two dependent variables, e.g., the Mach number ${\cal M}=-V/c_S=-V\Sigma/P$ and the angular momentum ${\cal L}=u_\phi$.  Their structure reveals an important point here,
\begin{equation}
\der{\ln {\cal M}}{\ln r}=\frac{{\cal N}_1(r,{\cal M},{\cal L})}{{\cal D}(r,{\cal M},{\cal L})}\\
\der{\ln {\cal L}}{\ln r}=   \frac{{\cal N}_2(r,{\cal M},{\cal L})}{{\cal
D}(r,{\cal M},{\cal L})} \, .
\label{eq_reg}
\end{equation}
In order for this to yield a non-singular physical solution, the numerators ${\cal N}_1$ and ${\cal N}_2$ must vanish at the same radius as the denominator ${\cal D}$. The denominator vanishes at the ``sonic'' radius $r_{\mathrm{sonic}}$ where the Mach number is equal to unity, and the equation ${\cal D}(r,{\cal M}, {\cal L}) = 0$ determines its location.

The extra regularity conditions at the sonic point ${\cal N}_i(r,{\cal M},{\cal L}) = 0$ are satisfied only for one particular value of the angular momentum at the horizon ${\cal L}_{in}$, which is the \textit{eigenvalue} of the problem that should be found. For a given $\alpha$ the location of the sonic point depends on the mass accretion rate. For low mass accretion rates one expects the transonic transition to occur close to the ISCO. Figure~\ref{fig:sonic-point-location} shows that this is indeed the case for accretion rates smaller than about $0.4\,{\dot M_{\mathrm{Edd}}}$, independent of $\alpha$, where we use the authors' definition of $\dot M_{\mathrm{Edd}} = 16\,L_{\mathrm{Edd}}/c^2$. At ${\dot M} = 0.4\,{\dot M_{\mathrm{Edd}}}$ a qualitative change occurs, resembling a ``phase transition'' from the Shakura--Sunyaev behavior to a very different slim-disk behavior. For higher accretion rates the location of the sonic point significantly departs from the ISCO. For low values of $\alpha$, the sonic point moves closer to the horizon, down to $\sim4\,M$ for $\alpha=0.001$. For $\alpha>0.2$ the sonic point moves outward with increasing accretion rate, reaching values as high as $8\,M$ for $\alpha=0.5$ and ${\dot M} =100\,{\dot M}_{\mathrm{Edd}}$. This effect was first noticed for small accretion rates by Muchotrzeb~\cite{muchotrzeb_83} and later investigated for a wide range of accretion rates by Abramowicz~\cite{abramowicz_89}, who explained it in terms of the disk-Bondi dichotomy.

\epubtkImage{}{%
\begin{figure}[htbp]
\centerline{\includegraphics[width=7cm]{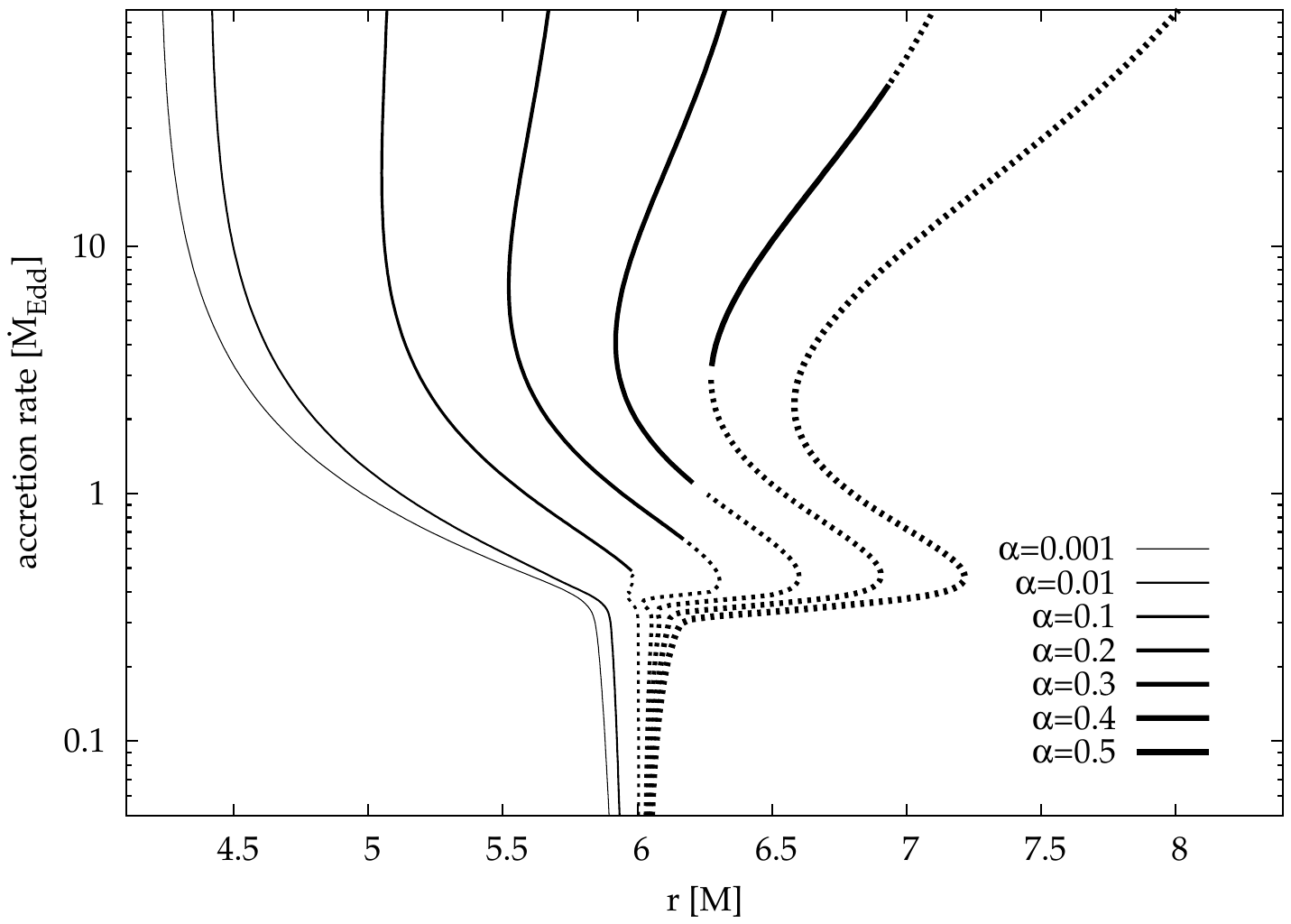}}
\caption{Location of the sonic point as a function of the accretion rate for different values of $\alpha$, for a non-rotating black hole, $a = 0$, taking $\dot M_{\mathrm{Edd}} = 16\,L_{\mathrm{Edd}}c^2$. The solid curves are for saddle type solutions while the dotted curves present nodal type regimes. Image reproduced by permission from~\cite{abramowicz_10}, copyright by ESO.}
\label{fig:sonic-point-location}
\end{figure}}

The topology of the sonic point is important, because physically acceptable solutions must be of the saddle or nodal type; the spiral type is forbidden. The topology may be classified by the eigenvalues $\lambda_1, \lambda_2, \lambda_3$ of the Jacobi matrix,
\begin{equation}
{\cal J} = \left[
\begin{array}{ccc}
 \pder{\cal D}r      &\pder{\cal D}{\cal M}      &\pder{\cal D}{\cal L}\\
 \pder{{\cal N}_1}r  &\pder{{\cal N}_1}{\cal M}  &\pder{{\cal N}_1}{\cal L}\\
 \pder{{\cal N}_2}r  &\pder{{\cal N}_2}{\cal M}  &\pder{{\cal N}_2}{\cal L}
\end{array}
\right].
\end{equation}
Because $\det({\cal J}) =0$, only two eigenvalues $\lambda_1, \lambda_2$ are non-zero, and the quadratic characteristic equation that determines them takes the form,
\begin{equation}
2\,\lambda^2 - 2\,\lambda\,\mathrm{tr}({\cal J}) - \left[\mathrm{tr}({\cal J}^2) - \mathrm{tr}^2({\cal J})\right] = 0.
\end{equation}
The nodal-type solution is given by $\lambda_1\lambda_2 > 0$ and the saddle type by $\lambda_1\lambda_2 < 0$, as marked in Figure~\ref{fig:sonic-point-location} with the dotted and the solid lines, respectively. For the lowest values of $\alpha$ only the saddle-type solutions exist. For moderate values of $\alpha$ ($0.1\le\alpha\le0.4$) the topological type of the sonic point changes at least once with increasing accretion rate. For the highest $\alpha$ solutions, only nodal-type critical points exist.

\subsection{Solutions: Shakura--Sunyaev \& Novikov--Thorne}
\label{section-Shakura-Sunyaev}

Shakura and Sunyaev~\cite{shakura_73} noticed that a few physically reasonable extra assumptions reduce the system of thin disk equations~(\ref{eq_cont2})\,--\,(\ref{energy}) to a set of algebraic equations. Indeed, the continuity and vertical equilibrium equations, (\ref{eq_cont2}) and (\ref{eq_vert}), are already algebraic. The radial momentum equation~(\ref{eq_rad4}) becomes a trivial identity $0=0$ with the extra assumptions that the radial pressure and velocity gradients vanish, and the rotation is Keplerian, $\Omega = \Omega_k^+$. The algebraic angular momentum equation~(\ref{eq_ang6}) only requires that we specify ${\cal L}_{in}$. The Shakura--Sunyaev model makes the assumption that ${\cal L}_{in} = {\cal L}_k(\mathrm{ISCO})$. This is equivalent to assuming that the torque vanishes at the ISCO.  This is a point of great interest that has been challenged repeatedly~\cite{krolik_99b,gammie_99,balbus12}.  Direct testing of this hypothesis by numerical simulations is discussed in Section~\ref{section-numerical-novikov-thorne}.

The right-hand side of the energy equation~(\ref{energy}) represents advective cooling. This is assumed to vanish in the Shakura--Sunyaev model, though we will see that it plays a critical role in slim disks (Section~\ref{section-slim-disks}) and ADAFs (Section~\ref{section-ADAFs}). Because the Shakura--Sunyaev model assumes the rotation is Keplerian, $\Omega = \Omega_k^+$, meaning $\Omega$ is a known function of $r$, the first term on the left-hand side of Eq.~(\ref{energy}), which represents viscous heating, is algebraic. The second term, which represents the radiative cooling (in the diffusive approximation) is also algebraic in the Shakura--Sunyaev model.

In addition to being algebraic, these thin-disk equations are also \textit{linear} in three distinct radial ranges: outer, middle, and inner. Therefore, as Shakura and Sunyaev realized, the model may be given in terms of explicit algebraic (polynomial) formulae. This was an achievement of remarkable consequences -- still today the understanding of accretion disk theory is in its major part based on the Shakura--Sunyaev analytic model. The Shakura--Sunyaev paper~\cite{shakura_73} is one of the most cited in astrophysics today (see Figure~\ref{figure:shakura-citations}), illustrating how fundamentally important accretion disk theory is in the field.

\epubtkImage{}{%
\begin{figure}[htbp]
  \centerline{\includegraphics[scale=0.5]{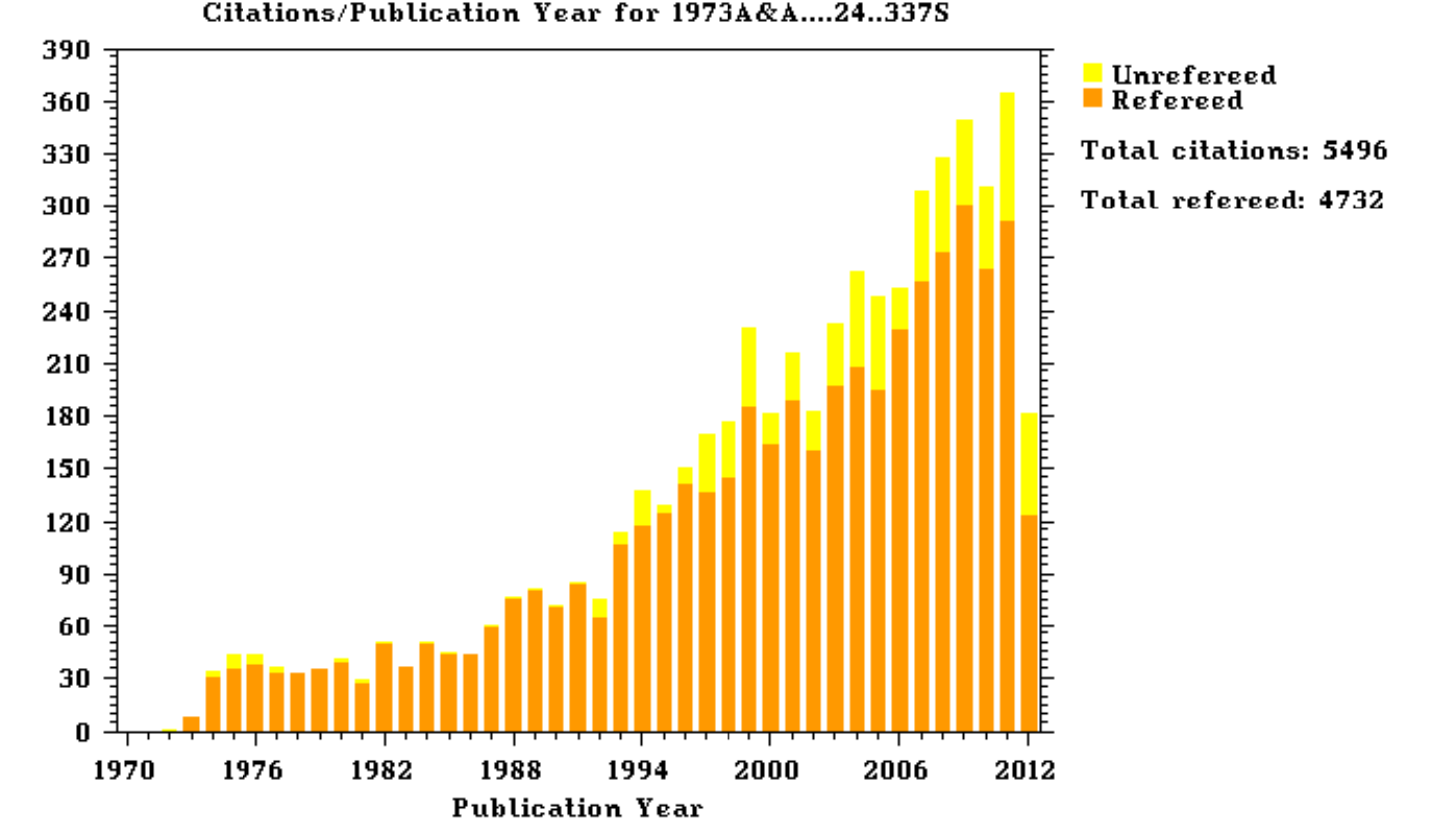}}
  \caption{The number of citations to the Shakura \& Sunyaev paper~\cite{shakura_73} is still growing exponentially, implying that the field of black hole accretion disk theory still has not reached saturation. Image reproduced from the \emph{SAO/NASA Astrophysics Data System}:
    \url{http://adsabs.harvard.edu/abs/1973A&A....24..337S}.}
  \label{figure:shakura-citations}
\end{figure}}

The general relativistic version of the Shakura--Sunyaev disk model was worked out by Novikov and Thorne~\cite{novikov_73}, with important extensions and corrections provided in subsequent papers~\cite{page_74, riffert_95, penna_12}.  Here we reproduce the solution, although with a more general scaling: $m = M/M_{\odot}$ and ${\dot m} = {\dot M} c^2/L_{\mathrm{Edd}}$.

Outer region: $P = P_{\mathrm{gas}}$, $\kappa = \kappa_{ff}$ (free-free opacity)
%
\begin{eqnarray}
F &=& [7 \times 10^{26} \unit{erg~cm^{-2}~s^{-1}}](m^{-1} {\dot m})\,r^{-3}_*\, {\cal B}^{-1} {\cal
C}^{-1/2} {\cal Q},
\nonumber \\
\Sigma &=& [4 \times 10^5 \unit{g~cm^{-2}}]
(\alpha^{-4/5}\,m^{2/10} {\dot m}^{7/10}_{0*})\,r^{-3/4}_*\,
{\cal A}^{1/10} {\cal B}^{-4/5} {\cal C}^{1/2} {\cal D}^{-17/20}
{\cal E}^{-1/20} {\cal Q}^{7/10}, \nonumber \\
H &=& [4 \times 10^2 \unit{cm}] (\alpha^{-1/10}\,m^{18/20} {\dot
m}^{3/20})\,r^{9/8}_* \,{\cal A}^{19/20} {\cal B}^{-11/10}
{\cal C}^{1/2} {\cal D}^{-23/40} {\cal E}^{-19/40} {\cal Q}^{3/20},
\nonumber \\
\rho_0 &=& [4 \times 10^2 \unit{g~cm^{-3}}]
(\alpha^{-7/10}\,m^{-7/10} {\dot m}^{11/20})\,r^{-15/8}_*
\,{\cal A}^{-17/20} {\cal B}^{3/10} {\cal D}^{-11/40} {\cal
E}^{17/40} {\cal Q}^{11/20},
\nonumber \\
T &=& [2 \times 10^8 \unit{K}] (\alpha^{-1/5}\,m^{-1/5} {\dot
m}^{3/10})\,r^{-3/4}_*\, {\cal A}^{-1/10} {\cal B}^{-1/5}
{\cal D}^{-3/20} {\cal E}^{1/20} {\cal Q}^{3/10},
\nonumber \\
\beta/(1-\beta) &=& [3](\alpha^{-1/10}\,m^{-1/10} {\dot
m}^{-7/20})\,r^{3/8}_*\,{\cal A}^{-11/20} {\cal B}^{9/10}
{\cal D}^{7/40} {\cal E}^{11/40} {\cal Q}^{-7/20},
\nonumber \\
\tau_{ff}/{\tau}_{es} &=& [2 \times 10^{-3}]({\dot m}^{-1/2})\,r^{3/4}_*\,{\cal A}^{-1/2} {\cal B}^{2/5} {\cal
D}^{1/4} {\cal E}^{1/4} {\cal Q}^{-1/2},
 \label{novikov-thorne-outer-region}
\end{eqnarray}
where $r_* = r c^2/GM$.

Middle region: $P = P_{\mathrm{gas}}$, $\kappa = \kappa_{es}$ (electron-scattering opacity)
%
\begin{eqnarray}
F &=& [7 \times 10^{26} \unit{erg~cm^{-2}~s^{-1}}](m^{-1} {\dot m})\,r^{-3}_*\, {\cal B}^{-1} {\cal
C}^{-1/2} {\cal Q},
\nonumber \\
\Sigma &=& [9 \times 10^4 \unit{g~cm^{-2}}](\alpha^{-4/5}
m^{1/5} {\dot m}^{3/5})\,r^{-3/5}_*\, {\cal B}^{-4/5}
{\cal C}^{1/2} {\cal D}^{-4/5} {\cal Q}^{3/5}, \nonumber \\
H &=& [1 \times 10^3 \unit{cm}](\alpha^{-1/10} m^{9/10} {\dot
m}^{1/5})\,r^{21/20}_*\,{\cal A} {\cal B}^{-6/5}
{\cal C}^{1/2} {\cal D}^{-3/5} {\cal E}^{-1/2} {\cal Q}^{1/5}, \nonumber \\
\rho_0 &=& [4 \times 10^1 \unit{g~cm^{-3}}](\alpha^{-7/10}
m^{-7/10} {\dot m}^{2/5})\,r^{-33/20}_*\,{\cal A}^{-1}
{\cal B}^{3/5}
{\cal D}^{-1/5} {\cal E}^{1/2} {\cal Q}^{2/5}, \nonumber \\
T &=& [7 \times 10^8 \unit{K}](\alpha^{-1/5} m^{-1/5} {\dot
m}^{2/5})\,r^{-9/10}_*\,{\cal B}^{-2/5}
{\cal D}^{-1/5}{\cal Q}^{2/5}, \nonumber \\
\beta/(1-\beta) &=& [7 \times 10^{-3}](\alpha^{-1/10} m^{-1/10} {\dot
m}^{-4/5})\,r^{21/20}_*\,{\cal A}^{-1} {\cal B}^{9/5} {\cal
D}^{2/5} {\cal E}^{1/2} {\cal Q}^{-4/5}, \nonumber \\
\tau_{ff}/{\tau}_{es} &=& [2 \times 10^{-6}]({\dot
m}^{-1})\,r^{3/2}_*\,{\cal A}^{-1} {\cal B}^{2} {\cal
D}^{1/2}{\cal E}^{1/2} {\cal Q}^{-1},
\label{novikov-thorne-middle-region}
\end{eqnarray}
%

Inner region: $P = P_{\mathrm{rad}}$, $\kappa = \kappa_{es}$
%
\begin{eqnarray}
F &=& [7 \times 10^{26} \unit{erg~cm^{-2}~s^{-1}}](m^{-1} {\dot m})\,r^{-3}_*\, {\cal B}^{-1} {\cal
C}^{-1/2} {\cal Q},
\nonumber \\
\Sigma &=& [5 \unit{g~cm^{-2}}](\alpha^{-1} {\dot
m}^{-1})\,r^{3/2}_*\, {\cal A}^{-2}{\cal B}^{3}
{\cal C}^{1/2}{\cal E} {\cal Q}^{-1}, \nonumber \\
H &=& [1 \times 10^5 \unit{cm}]({\dot m})\,{\cal A}^2 {\cal B}^{-3}
{\cal C}^{1/2} {\cal D}^{-1} {\cal E}^{-1} {\cal Q}^{}, \nonumber \\
\rho_0 &=& [2 \times 10^{-5} \unit{g~cm^{-3}}](\alpha^{-1}
m^{-1} {\dot m}^{-2})\,r^{3/2}_*\,{\cal A}^{-4} {\cal B}^{6}
{\cal D}^{} {\cal E}^{2} {\cal Q}^{-2}, \nonumber \\
T &=& [5 \times 10^7 \unit{K}](\alpha^{-1/4}
m^{-1/4})\,r^{-3/8}_*\,{\cal A}^{-1/2}{\cal B}^{1/2}
{\cal E}^{1/4}, \nonumber \\
\beta/(1-\beta) &=& [4 \times 10^{-6}](\alpha^{-1/4} m^{-1/4}
{\dot m}^{-2})\,r^{21/8}_*\,{\cal A}^{-5/2} {\cal B}^{9/2}
{\cal
D}^{} {\cal E}^{5/4} {\cal Q}^{-2}, \nonumber \\
(\tau_{ff}{\tau}_{es})^{1/2} &=& [1 \times 10^{-4}](
\alpha^{-17/16} m^{-1/16} {\dot
m}^{-2})\,r^{93/32}_*\,{\cal A}^{-25/8} {\cal B}^{41/8}{\cal
C}^{1/2} {\cal D}^{1/2}{\cal E}^{25/16} {\cal Q}^{-2}.
\label{novikov-thorne-inner-region}
\end{eqnarray}
The radial functions ${\cal A}, ..., {\cal Q}$ that appear in Eqs.~(\ref{novikov-thorne-outer-region}),
(\ref{novikov-thorne-middle-region}), (\ref{novikov-thorne-inner-region}), are defined in terms of $y = (r/M)^{1/2}$ and $a_* = a/M$ as ~\cite{page_74}:
\begin{alignat}{2}
{\cal A} &= 1 + a_*^2 y^{-4} + 2a_*^2 y^{-6},
&\quad ~~~{\cal B} &= 1 + a_*y^{-3}, \nonumber  \\
%
{\cal C} &= 1 - 3y^{-2} + 2a_*y^{-3},
&\quad ~~~{\cal D} &= 1 - 2y^{-2} + a_*^2 y^{-4}, \nonumber  \\
{\cal E} &= 1 + 4a_*^2y^{-4} - 4a_*^2y^{-6} + 3a_*^4y^{-8}
&\quad ~~~{\cal Q}_0 &= \frac{1 + a_*y^{-3}}{y(1 - 3y^{-2} +
2a_*y^{-3})^{1/2}}, \nonumber
\end{alignat}
\begin{eqnarray}
{\cal Q} = &{\cal Q}_0&\left[ y - y_0 - \frac{3}{2}a_*\ln\left(
\frac{y}{y_0}\right) - \frac{3(y_1 - a_*)^2}{y_1(y_1 - y_2)(y_1 -
y_3)} \ln\left( \frac{y - y_1}{y_0 - y_1}\right)\right] \nonumber
\\
- &{\cal Q}_0&\left[ \frac{3(y_2 - a_*)^2}{y_2(y_2 - y_1)(y_2 -
y_3)}\ln \left( \frac{y - y_2}{y_0 - y_2}\right) -\frac{3(y_3 -
a_*)^2}{y_3(y_3 - y_1)(y_3 - y_2)}\ln \left( \frac{y - y_3}{y_0 -
y_3} \right) \right] \nonumber
\end{eqnarray}
Here $y_0 = (r_\mathrm{ms}/M)^{1/2}$, and $y_1$, $y_2$, and $y_3$ are the three roots of $y^3 - 3y + 2a_* = 0$; that is
\begin{eqnarray}
y_1 & = & 2 \cos [(\cos^{-1} a_* - \pi)/3], \nonumber \\
y_2 & = & 2 \cos [(\cos^{-1} a_* + \pi)/3], \nonumber \\
y_3 & = & -2 \cos [(\cos^{-1} a_*)/3]. \nonumber
\end{eqnarray}
For the numerical solutions reproduced in
Figure~\ref{figure:inner-problem}, the opacities were assumed to be
$\kappa_{es} = 0.34 \unit{cm^2 \ g^{-1}}$ and $\kappa_{ff} = 6.4
\times 10^{22} \rho_\mathrm{cgs} T_\mathrm{K}^{-7/2} \unit{cm^2
  \ g^{-1}}$, where $\rho_\mathrm{cgs}$ is the density in
$\mathrm{g\ cm}^{-3}$ and $T_\mathrm{K}$ is the temperature in Kelvin.

The Shakura--Sunyaev and Novikov--Thorne solutions are only \textit{local} solutions; this is because they do not take into account the full eigenvalue problem described in Section~\ref{subsection-sonic}. Instead, they make an assumption that the viscous torque goes to zero at the ISCO, which makes the model singular there. For very low accretion rates, this singularity of the model does not influence the electromagnetic spectrum~\cite{straub_11}, nor several other important astrophysical predictions of the model. However, in those astrophysical applications in which the inner boundary condition is important (e.g., global modes of disk oscillations), the Novikov--Thorne model is inadequate. Figure~\ref{figure:inner-problem} illustrates a few ways in which the model fails to capture the true physics near the ISCO.

\epubtkImage{}{%
\begin{figure}[htbp]
  \centerline{\includegraphics[scale=0.5]{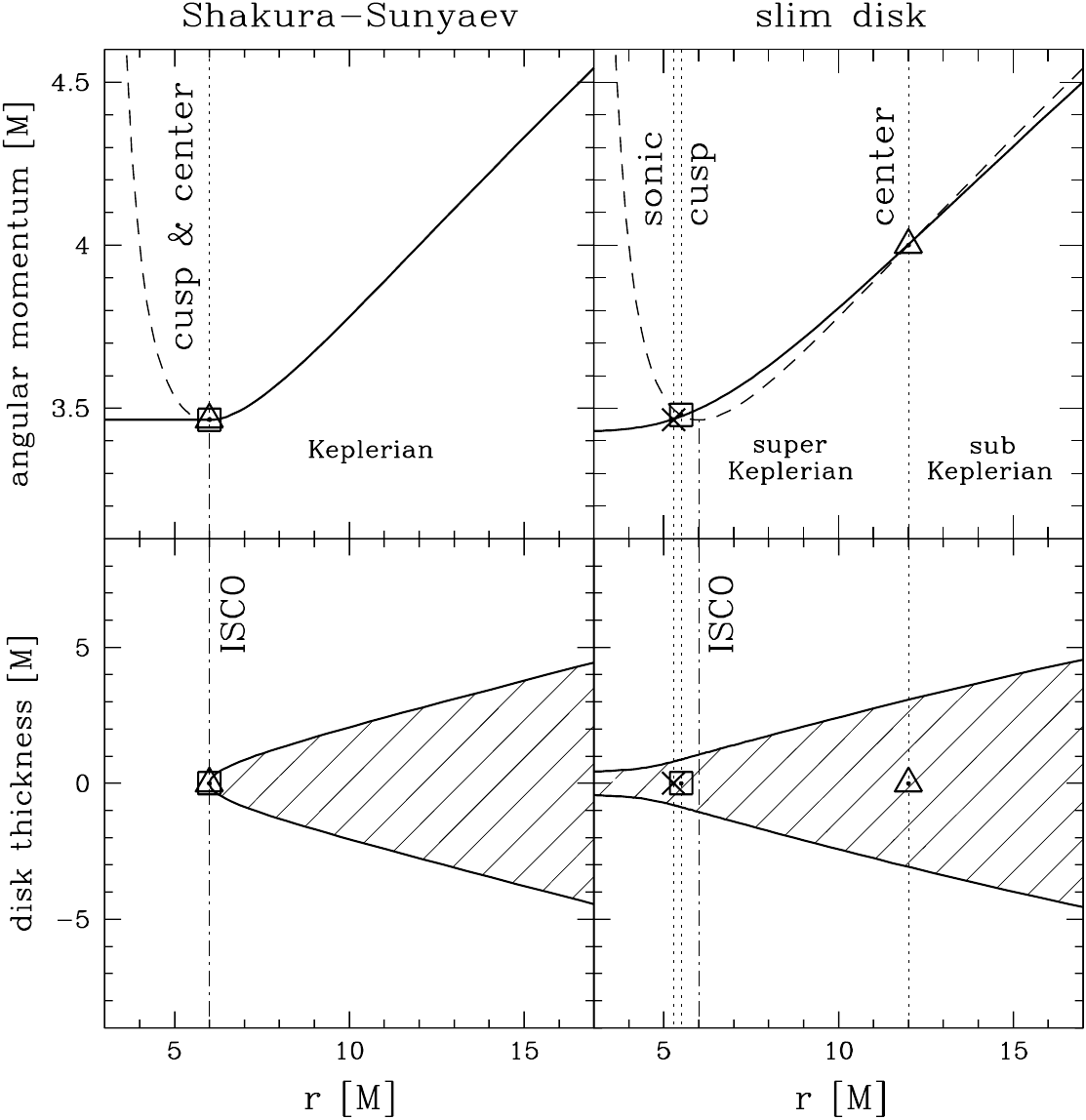}}
  \caption{The innermost part of the disk. In the Shakura--Sunyaev and
    Novikov--Thorne models, the locations of the maximum pressure
    (a.k.a.\ the center) $r_{\mathrm{center}}$ and the cusp
    $r_{\mathrm{cusp}}$, as well as the sonic radius
    $r_{\mathrm{sound}}$, are assumed to coincide with the
    ISCO. Furthermore, the angular momentum is assumed to be strictly
    Keplerian outside the ISCO and constant inside it. In real flows,
    $r_{\mathrm{center}} \not = r_{\mathrm{cusp}} \not =
    r_{\mathrm{sound}} \not = \mathrm{ISCO}$, and angular momentum is
    super-Keplerian between $r_{\mathrm{cusp}}$ and
    $r_{\mathrm{center}}$.  Image reproduced by permission
    from~\cite{abramowicz_10}, copyright by ESO.}
  \label{figure:inner-problem}
\end{figure}}


\newpage

\section{Slim Disks}
\label{section-slim-disks}

The Shakura--Sunyaev and Novikov--Thorne models of thin disks assume that accretion is radiatively efficient. This assumption means that all the heat generated by viscosity at a given radius is immediately radiated away. In other words, the viscous heating is balanced by the radiative cooling locally and no other cooling mechanism is needed. This assumption can be satisfied as long as the accretion rate is small. At some luminosity ($L\approx 0.3\,L_{\mathrm{Edd}}$), however, the radial velocity is large, and the disk is thick enough, to trigger another mechanism of cooling: advection. It results from the fact that the viscosity-generated heat does not have sufficient time to transform into photons and leave the disk before being carried inwards by the motion of the gas. The higher the luminosity, the more significant advective cooling is. At the highest luminosities, it becomes comparable to the radiative cooling (see Figure~\ref{f-fadv}), and the standard, thin disk approach can no longer be applied.

\epubtkImage{}{%
\begin{figure}[htbp]
\centerline{\includegraphics[width=.6\textwidth,angle=0]{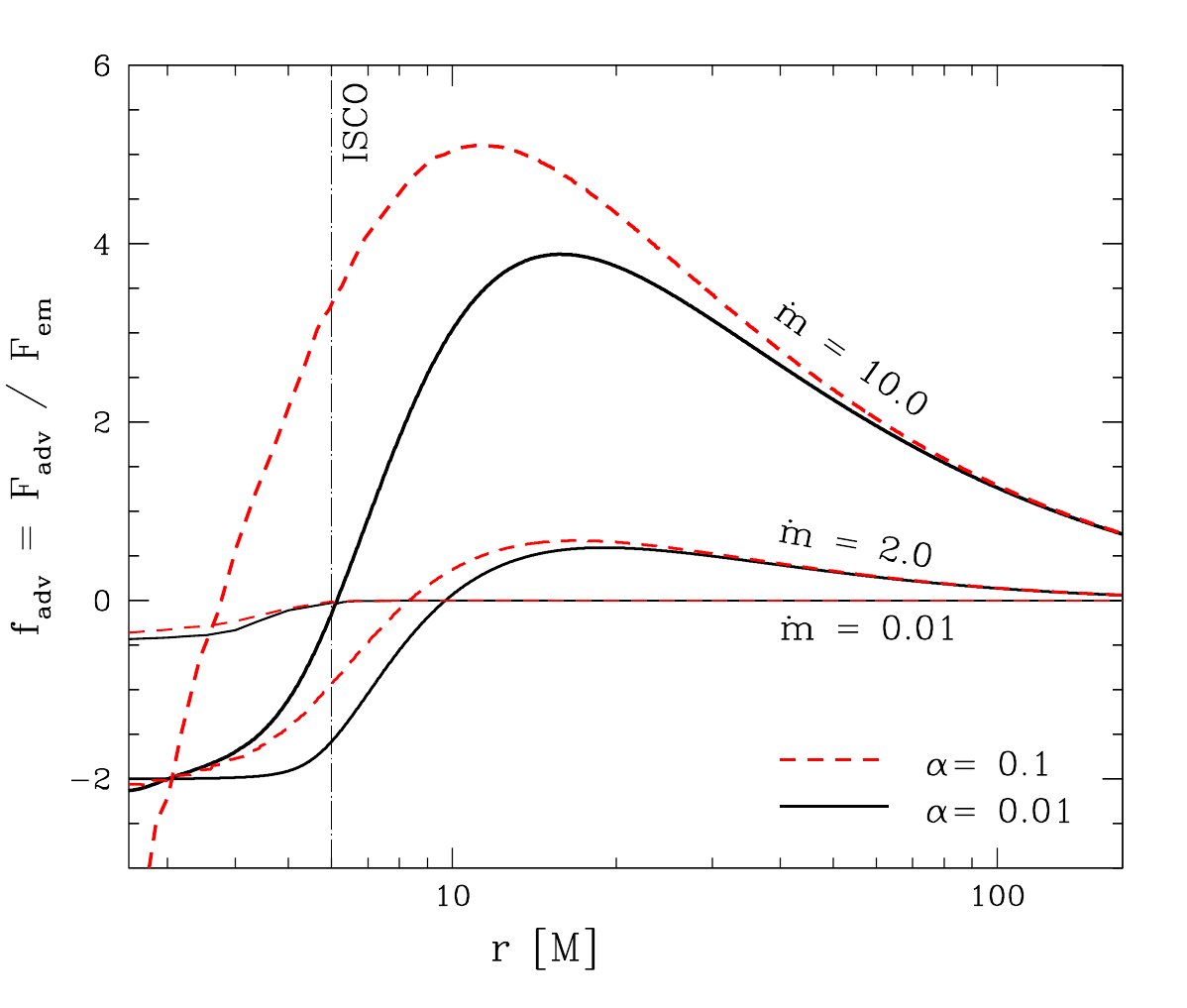}}
\caption{The advection factor (ratio of advective to radiative
  cooling) profiles for $\dot m=0.01$, $1.0$ and $10.0$ (here, $\dot m
  = \dot M c^2/16\,L_{\mathrm{Edd}}$). Profiles for $\alpha=0.01$ and
  $0.1$ are presented with solid black and dashed red lines,
  respectively. The fraction $f_{\mathrm{adv}}/(1+f_{\mathrm{adv}}$)
  of heat generated by viscosity is carried along with the flow. In
  regions with $f_{\mathrm{adv}}<0$ the advected heat is
  released. Image reproduced by permission
  from~\cite{sadowski-thesis}.}
  \label{f-fadv}
\end{figure}}

The problem of accretion with an additional cooling mechanism has to be treated in a different way than radiatively efficient flows. Without the assumptions of radiative efficiency and Keplerian angular momentum, it is no longer possible to find an analytic solution to the system of equations presented in Section~\ref{subsection-thin-disk-quations-Kerr}. Instead, one has to solve a two-dimensional system of ordinary differential equations (\ref{eq_reg}) with a critical point -- the radius at which the gas velocity exceeds the local speed of sound (the sonic radius). This was first done in the pseudo-Newtonian limit by Abramowicz~\cite{abramowicz_88}, who forged the term ``slim disks''.  It has since been done using a fully relativistic treatment by Beloborodov~\cite{beloborodov_98}. Recently, S{\c a}dowski~\cite{sadowski_09} constructed slim disk solutions for a wide range of parameters applicable to X-ray binaries.

These slim disks are in some sense more physical than thin disks and offer a more general set of solutions, while in the limit of low accretion rates they converge to the standard thin disk solutions. Slim disks are more physical in that they extend down to the black hole horizon, as opposed to thin disks that formally terminate at the ISCO. Slim disks are more general in that they may rotate with an angular momentum profile significantly different than the Keplerian one -- the higher the accretion rate, the more significant the departure (see Figure~\ref{f-l-a0}). The disk thickness also increases with the accretion rate. For rates close the Eddington limit, the maximal $H/R$ ratio reaches 0.3. Finally, the flux emerging from the slim disk surface is modified by the advection. At high luminosities, a large fraction of the viscosity-generated heat is advected inward and released closer to the black hole or not released at all. As a result, the slope of the radial flux profile changes, and radiation is even emitted from within the ISCO (see Figure~\ref{f-flux}). Due to the increasing rate of advection, the efficiency of transforming gravitational energy into radiative flux decreases with increasing accretion rate. Despite highly super-Eddington accretion rates, the disk luminosity remains only moderately super-Eddington (see Figure~\ref{f-eff}).  The Eddington luminosity may be exceeded because the geometry of the flow is not spherical and the classical definition of this quantity does not apply -- most of the accretion takes place in the equatorial plane while the radiation escapes vertically. Thus, the radiation is not capable of stopping the inflow, though it may cause outflows from the surface.

\epubtkImage{}{%
\begin{figure}[htbp]
  \centerline{
    \includegraphics[height=.43\textwidth]{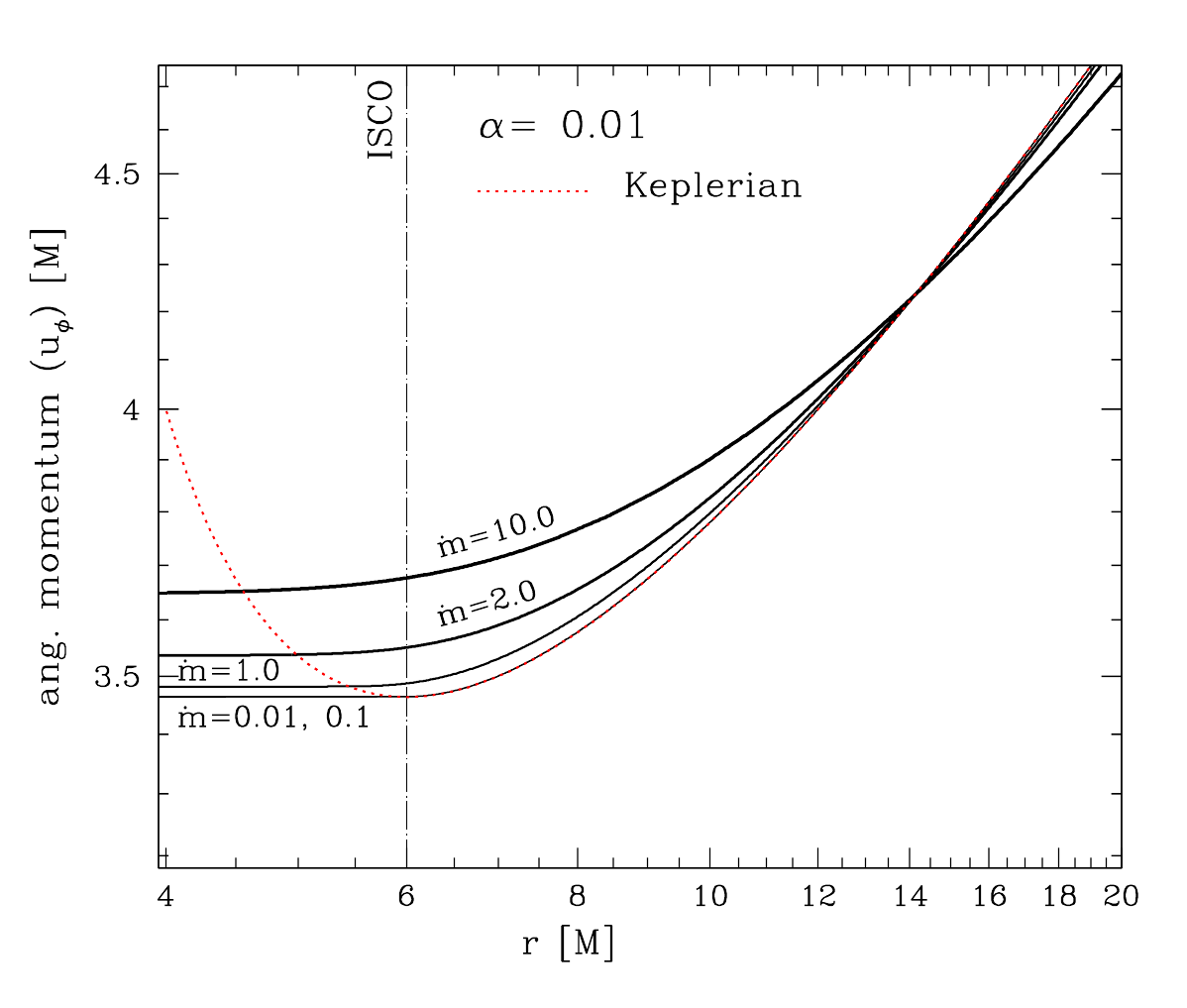}
    \includegraphics[height=.43\textwidth]{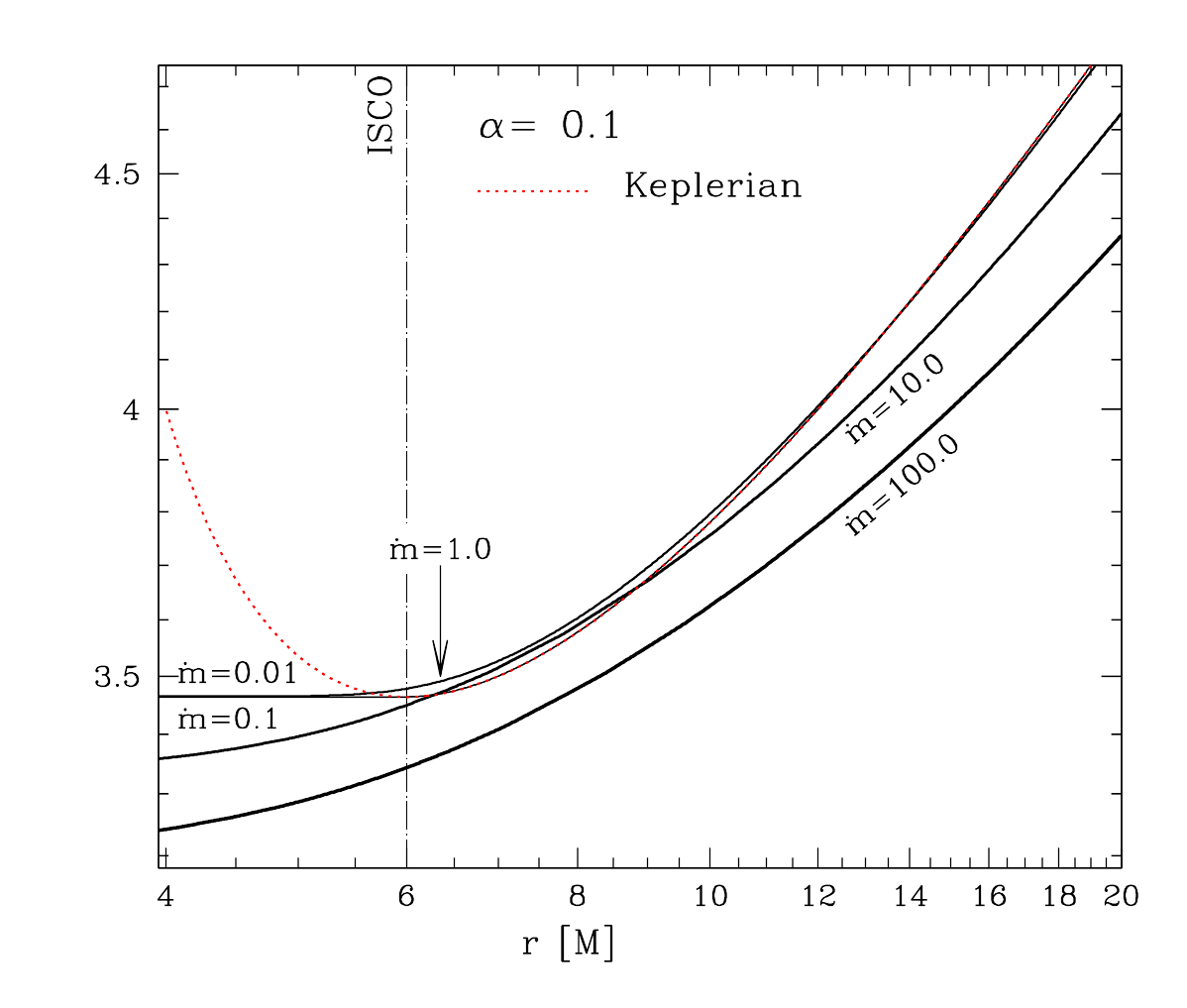}
  }
  \caption{Profiles of the disk angular momentum ($u_\phi$) for
    $\alpha=0.01$ (left) and $\alpha=0.1$ (right panel) for different
    accretion rates (as a reminder, $\dot m = \dot M
    c^2/16\,L_{\mathrm{Edd}}$), showing the departures from the
    Keplerian profile. These plots are for a non-rotating black
    hole. Image reproduced by permission from~\cite{sadowski_11},
    copyright by ESO.}
\label{f-l-a0}
\end{figure}}

\epubtkImage{}{%
\begin{figure}[htbp]
  \centerline{\includegraphics[width=.6\textwidth,angle=0]{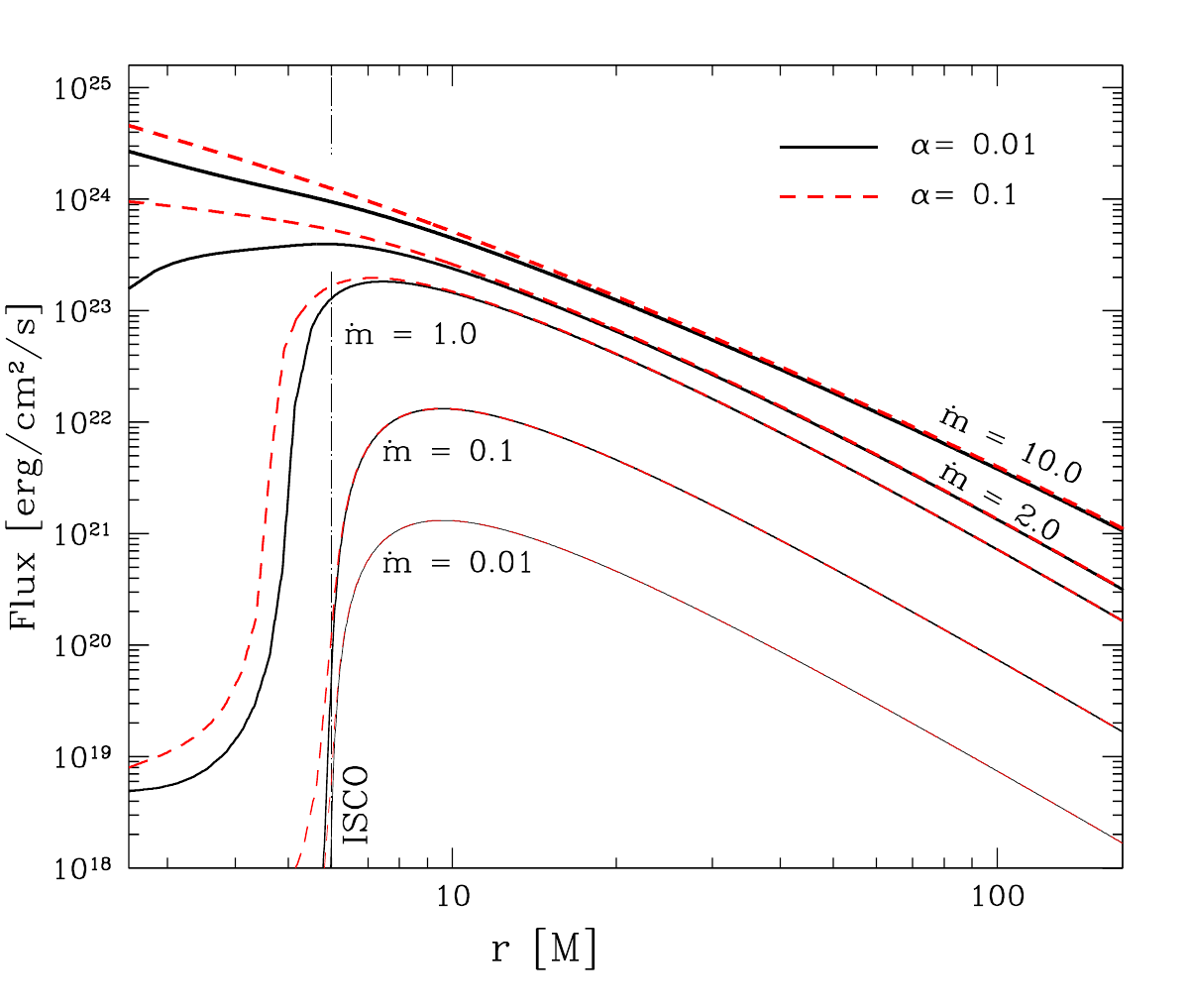}}
  \caption{Flux profiles for different mass accretion rates in the
    case of a non-rotating black hole and two values of $\alpha$: 0.01
    (black solid), 0.1 (red dashed lines). For each value of $\alpha$
    there are five lines corresponding to the following mass accretion
    rates: $\dot m = 0.01$, 0.1, 1.0, 2.0 and 10.0 (as a reminder,
    $\dot m = \dot M c^2/16\,L_{\mathrm{Edd}}$). The black hole mass
    is $10\,M_{\odot}$. Image reproduced by permission
    from~\cite{sadowski_11}, copyright by ESO.}
  \label{f-flux}
\end{figure}}

\epubtkImage{}{%
\begin{figure}[htbp]
  \centerline{\includegraphics[width=.6\textwidth,angle=0]{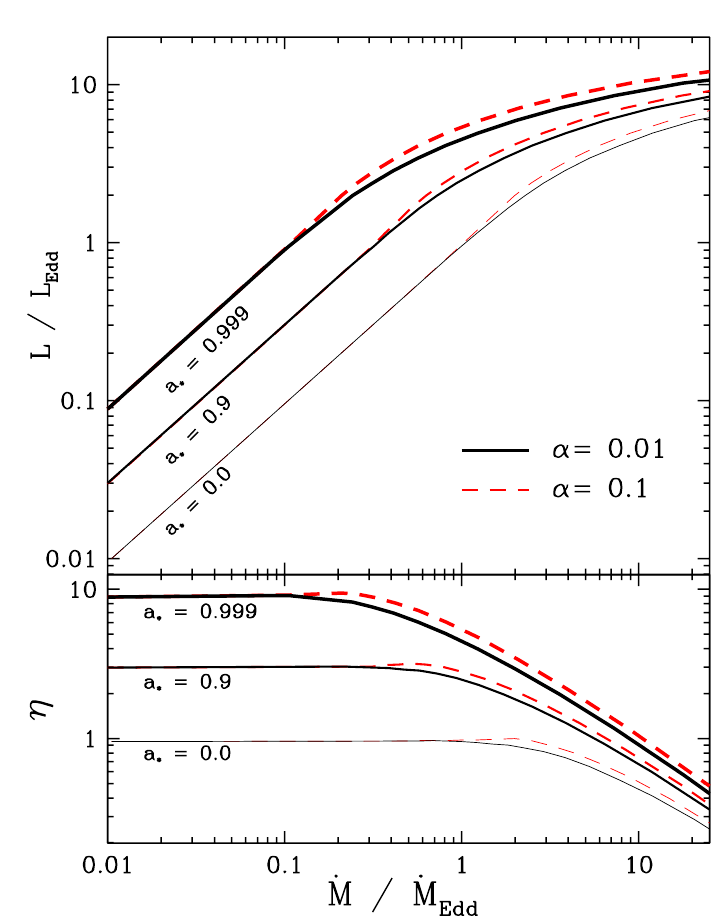}}
  \caption{\emph{Top panel:} Luminosity vs accretion rate for three
    values of black hole spin ($a_* = a/M=0.0$, $0.9$, $0.999$) and
    two values of $\alpha=0.01$ (black) and 0.1 (red
    line). \emph{Bottom panel:} efficiency of accretion
    $\eta=(L/L_{\mathrm{Edd}}) / (\dot M / \dot M_{\mathrm{Edd}})$
    (here $\dot M_{\mathrm{Edd}} = 16\,L_{\mathrm{Edd}}/c^2$). Image
    reproduced by permission from~\cite{sadowski-thesis}.}
  \label{f-eff}
\end{figure}}

\clearpage

\section{Advection-Dominated Accretion Flows (ADAFs)}
\label{section-ADAFs}

The ADAF, or advection-dominated accretion flow, solution also involves advective cooling. In fact, it carries it to an extreme -- nearly all of the viscously dissipated energy is advected into the black hole rather than radiated.  Unlike the slim disk solution, which is usually invoked at high luminosities, the ADAF applies when the luminosity (and generally the mass accretion rate) are low.

Because of their low efficiency, ADAFs are much less luminous than the Shakura--Sunyaev thin disks.  The solutions tend to be hot (close to the virial temperature), optically thin, and quasi-spherical (see Figure~\ref{fig:yuan}).  Their spectra are non-thermal, appearing as a power-law, often with a strong Compton component.  This makes them a good candidate for the Hard state observed in X-ray binaries (discussed in Section~\ref{section:SpectralStates}).

\epubtkImage{}{%
\begin{figure}[htb]
  \centerline{\includegraphics[width=.75\textwidth,angle=0]{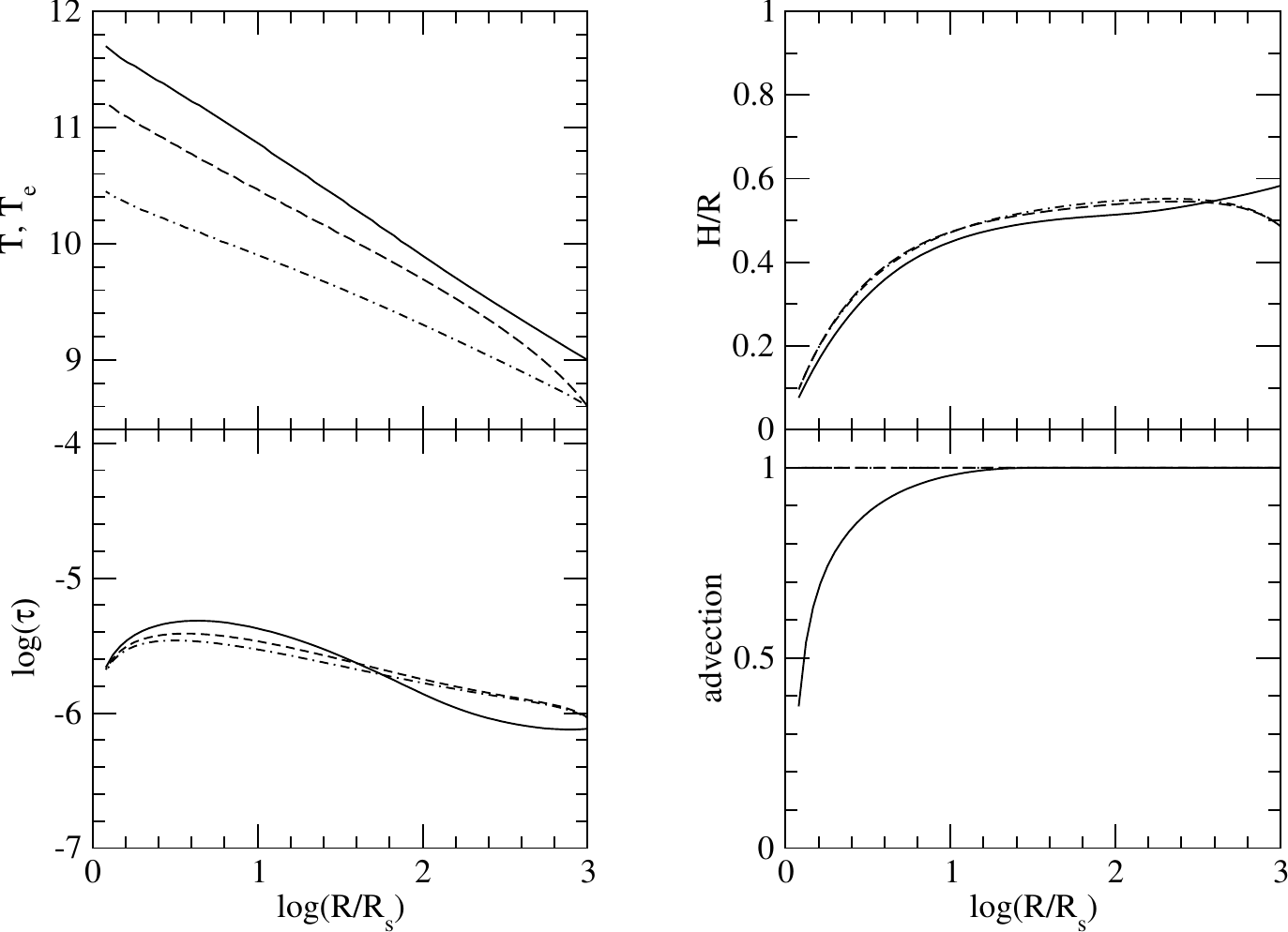}}
  \caption{Profiles of temperature, optical depth, ratio of scale
    height to radius, and advection factor (the ratio of advective
    cooling to turbulent heating) of a hot, one-$T$ ADAF (solid
    lines). The parameters are $M=10\,M_{\odot}$, $\dot{M} =
    10^{-5}\, L_{\mathrm{Edd}}/c^2$, $\alpha = 0.3$, and $\beta =
    P_{\mathrm{gas}}/(P_{\mathrm{gas}} + P_{\mathrm{mag}}) = 0.9$. The
    outer boundary conditions are $R_{\mathrm{out}} = 10^3 R_S$,
    $T=10^9\unit{K}$, and $v/c_s = 0.5$.  Two-$T$ solutions with
    the same parameters and $\delta = 0.5$ (dashed lines) and 0.01
    (dot-dashed lines) are also shown for comparison, where $\delta$
    is the fraction of the turbulent viscous energy that directly
    heats the electrons. Image reproduced by permission
    from~\cite{yuan_06}, copyright by AAS.}
  \label{fig:yuan}
\end{figure}}

ADAFs were formally introduced in the Newtonian limit through a series of papers by Narayan and Yi~\cite{narayan_94,narayan_95a,narayan_95b}, followed closely by Abramowicz~\cite{abramowicz_95, abramowicz_96} and others~\cite{gammie_98}, although the existence of this solution had been hinted at much earlier~\cite{ichimaru_77,rees_82}. In the same spirit as we gave the equations for the Novikov--Thorne solution in Section~\ref{section-Shakura-Sunyaev} for thin disks, we report the self-similar ADAF solution found by Narayan and Yi~\cite{narayan_95a}. Again we present the solution with the following scaling: $r_* = r c^2/GM$, $m = M/M_{\odot}$ and ${\dot m} = {\dot M} c^2/L_{\mathrm{Edd}}$.

\begin{eqnarray}
v &=& [-3.00 \times 10^{10} \unit{cm~s^{-1}}]\alpha c_1 r_*^{-1/2}, \nonumber \\
\Omega &=& [2.03 \times 10^5 \unit{s^{-1}}] c_2 m^{-1} r_*^{-3/2}, \nonumber \\
c_S^2 &=& [9.00 \times 10^{20} \unit{cm~s^{-2}}] c_3 r_*^{-1}, \nonumber \\
\rho &=& [1.07 \times 10^{-5} \unit{g~cm^{-3}}] \alpha^{-1} c_1^{-1} c_3^{-1/2} m^{-1} \dot{m} r_*^{-3/2}, \nonumber \\
P &=& [9.67 \times 10^{15} \unit{g~cm^{-1}~s^{-2}}] \alpha^{-1} c_1^{-1} c_3^{1/2} m^{-1} \dot{m} r_*^{-5/2}, \nonumber \\
B &=& [4.93 \times 10^8 \unit{G}] \alpha^{-1/2} (1-\beta_m)^{1/2} c_1^{-1/2} c_3^{1/4} m^{-1/2} \dot{m}^{1/2} r_*^{-5/4}, \nonumber \\
q^+ &=& [2.94 \times 10^{21} \unit{erg~cm^{-3}~s^{-1}}] \epsilon^\prime c_3^{1/2} m^{-2} \dot{m} r_*^{-4}, \nonumber \\
\tau_\mathrm{es} &=& [1.75] \alpha^{-1} c_1^{-1} \dot{m} r_*^{-1/2},
 \label{adaf}
\end{eqnarray}
where $v$ is the radial infall velocity and $q^+$ is the viscous dissipation of energy per unit volume. The constants $c_1$, $c_2$, and $c_3$ are given by
\begin{eqnarray}
c_1 & = & \frac{(5+2\epsilon^\prime)}{3\alpha^2}g(\alpha, \epsilon^\prime) \nonumber \\
c_2 & = & \left[ \frac{2\epsilon^\prime (5 + 2\epsilon^\prime)}{9\alpha^2} g(\alpha, \epsilon^\prime) \right]^{1/2} \nonumber \\
c_3 & = & \frac{2(5+2\epsilon^\prime)}{9\alpha^2}g(\alpha, \epsilon^\prime), \nonumber
\end{eqnarray}
where
\begin{eqnarray}
\epsilon^\prime & = & \frac{1}{f_\mathrm{adv}}\left(\frac{5/3 - \gamma_g}{\gamma_g-1}\right) \nonumber \\
g(\alpha, \epsilon^\prime) & \equiv & \left[1+\frac{18\alpha^2}{(5+2\epsilon^\prime)^2}\right]^{1/2} - 1,\nonumber
\end{eqnarray}
and the parameter $f_\mathrm{adv}$ represents the fraction of viscously dissipated energy which is advected.  The remaining amount, $1-f_\mathrm{adv}$, is radiated locally.

The rapid advection in ADAFs generally has two effects: 1) dissipated orbital energy can not be radiated locally before it is carried inward and 2) the rotation profile is generally no longer Keplerian, although Abramowicz~\cite{abramowicz_95} found solutions where the dominant cooling mechanism was advection, even when the angular momentum profile was Keplerian.  Fully relativistic solutions of ADAFs have also been found numerically~\cite{abramowicz_97,beloborodov_97}. Further discussion of ADAFs is given in the review article by Narayan and McClintock~\cite{narayan_08}.

\newpage

\section{Stability}
\label{section-stability}

Having reviewed some of the main analytic models of accretion disks, it is important now to discuss the issue of stability. Since all analytic models presume steady-state solutions, such models are only useful if the resulting solutions are stable.  One reason to suspect accretion disks may not be stable is that the systematic differential rotation that is a signature feature of accretion is a potential source of energy, and therefore, of instability. Another is that some level of instability may be essential in accretion disks as it can provide a pathway to the kind of sustained turbulence anticipated by Shakura and Sunyaev (see Section~\ref{subsubsection-alpha-viscosity}).


\subsection{Hydrodynamic stability}

Within ideal hydrodynamics, local linear stability of an axisymmetric rotating flow is guaranteed if the H{\o}iland criterion is satisfied~\cite{tassoul_00}:
\begin{subeqnarray}
\frac{1}{R^3} \frac{\partial \ell^2}{\partial R} - \frac{1}{C_p \rho} \nabla p \cdot \nabla S & > & 0, \\
\frac{\partial p}{\partial z} \left( \frac{\partial \ell^2}{\partial R} \frac{\partial S}{\partial z} - \frac{\partial \ell^2}{\partial z} \frac{\partial S}{\partial R} \right) & < & 0,
\end{subeqnarray}
where $C_p$ is the specific heat at constant pressure and $R$ is the cylindrical radius (see \cite{seguin_75} for the criterion for relativistic stars). This criterion can be easily understood in two limits: For non-rotating equilibria (e.g., a non-rotating star), the criterion reduces to the Schwarzschild criterion ($\partial S/\partial r > 0$) that the entropy must not increase toward the interior (for stability against convection).  Provided this is true, local fluid elements will simply oscillate under stable buoyancy forces.  To see the effects of rotation, we can consider an equilibrium that has constant entropy everywhere. Then the H{\o}iland criterion reduces to the Rayleigh criterion ($d \ell^2/dR > 0$): the specific angular momentum must not decrease outward. Physically, if one perturbs a fluid element radially outward, it conserves its own specific angular momentum. If the ambient specific angular momentum decreases outward, then the fluid element will be rotating too fast to stay in its new position, and centrifugal forces will push it further outward. Stability would be a fluid element that oscillates at the local epicyclic frequency.

As it turns out, the H{\o}iland criterion is a huge disappointment for understanding why turbulence might exist in accretion disks. This is because it indicates that accretion disks with rotation profiles that do not differ too much from Keplerian should be strongly stable!

\subsubsection{Papaloizou--Pringle Instability (PPI)}
\label{section-PPI}

The H{\o}iland criterion is only a local stability criterion. Flows can be locally stable, yet have global instabilities. An example of this occurs in the Polish doughnut solution (Section~\ref{section-thick-disks}).  Papaloizou and Pringle~\cite{papaloizou_84} showed that this solution is marginally stable with respect to local axisymmetric perturbations yet unstable to low-order nonaxisymmetric modes. As with all global instabilities, the existence of the Papaloizou--Pringle instability (PPI) is sensitive to the assumed boundary conditions~\cite{blaes_87}.  In cases where the disk overflows its potential barrier (Roche lobe) and accretes through pressure-gradient forces across the cusp, the PPI is generally suppressed~\cite{hawley_91a}.


\subsubsection{Runaway instability}
\label{section-runaway-instability}

Another instability associated with the Polish doughnut is the runaway instability \cite{abramowicz_83}. If matter is overflowing its Roche lobe and accreting onto the black hole, then one of two evolutionary tracks are possible: (i) As the disk loses material it contracts inside its Roche lobe, slowing the mass transfer and resulting in a stable situation, or (ii) as the black hole mass grows, the cusp moves deeper inside the disk, causing the mass transfer to speed up, leading to the runaway instability. Recent numerical simulations show that, while this instability grows very fast, on timescales of a few orbital periods, over a wide range of disk-to-black hole mass ratios when $\ell = \text{const.}$, i.e., a constant specific angular momentum profile~\cite{font_02}, it is strongly suppressed whenever the specific angular momentum of the disk increases with the radial distance as a power law, $\ell \propto r^p$~\cite{daigne_04}. Even values of $p$ much smaller than the Keplerian limit ($p=1/2$) suffice to suppress this particular instability.  [This is equivalent to angular velocity profiles, $\Omega \propto r^{-q}$, with $q > 3/2$.]

\subsection{Magneto-rotational instability (MRI)}
\label{section-MRI}

Although it had long been suspected that some sort of MHD instability might provide the necessary turbulent stresses to make accretion work, the nature of this instability remained a mystery until the rediscovery of the magneto-rotational instability by Balbus and Hawley~\cite{balbus_91,hawley_91b,balbus_98}.  Originally discovered by Velikhov~\cite{velikhov_59}, and generalized by Chandrasekhar~\cite{chandrasekhar_60}, in the context of vertically magnetized Couette flow between differentially rotating cylinders, the application of this instability to accretion disks was originally missed.

The instability itself can be understood through a simple mechanical model. Consider two particles of gas connected by a magnetic field line. Arrange the particles such that they are initially located at the same cylindrical distance from the black hole but with some vertical separation. Give one of the particles (say the upper one) a small amount of extra angular momentum, while simultaneously taking away a small amount of angular momentum from the lower one.  The upper particle now has too much angular momentum to stay where it is and moves outward to a new radius. The lower particle experiences the opposite behavior and moves to a smaller radius. In the usual case where the angular velocity of the flow drops off with radius, the upper particle will now be orbiting slower than the lower one. Since these two particles are connected by a magnetic field line, the differing orbital speeds mean the field line will get stretched. The additional tension coming from the stretching of the field line provides a torque, which transfers angular momentum from the lower particle to the upper one. This just reinforces the initial perturbation, so the separation grows and angular momentum transfer is enhanced. This is the fundamental nature of the instability.

In more concrete terms, consider a disk threaded with a vertical magnetic field $B_z$ and having an  Alf\'ven speed $v^2_A = B^2_z/(4\pi \rho)$. The dispersion relation for perturbations of a fluid quantity $\delta X \sim \exp[i(kz - \omega t)]$ is~\cite{balbus_98}
\begin{equation}
\omega^4 -(2kv_A + \omega^2_r)\omega^2 +kv_A(kv_A + rd\Omega^2/dr)=0 \,.
\end{equation}
This equation has an unstable solution ($\omega^2 < 0$), if and only if, $kv_A + rd\Omega^2/dr < 0$. Since
\begin{equation}
\frac{\partial \Omega}{\partial r} < 0,
\label{MIR-criterion}
\end{equation}
in accretion disks, the instability criteria can generally be met for weakly magnetized disks. More specifically, the MRI exists for intermediate magnetic field strengths. In terms of the natural length scale of the instability ($\sim v_A/\Omega$),  the field strength is constrained at the upper limit by the requirement that the unstable MRI wavelength fit inside the vertical thickness of the disk ($v_A/\Omega \lesssim H$). This corresponds to field energy densities that are less than the thermal pressure, i.e., $b^2 < P_\mathrm{gas}$. At the lower end, dissipative processes set a floor on the relevant length scales, and hence, field strengths.

If the conditions for the instability are met, the fastest-growing mode, which dominates the early evolution, has the form of a ``channel flow'' involving alternating layers of inward- and outward-moving fluid.  The amplitude of this solution grows exponentially until it becomes unstable to three-dimensional ``parasitic modes'' that feed off the gradients of velocity and magnetic field provided by the channel flow. The flow rapidly reaches a state of magnetohydrodynamic turbulence~\cite{hawley_91b,hawley_92}. This instability can be self-sustaining through a nonlinear dynamo process~\cite{brandenburg_95} -- nonlinear because the motion that sustains or amplifies the magnetic field is driven by the field itself through the MRI. A more complete description of the linear and non-linear evolution of the MRI is provided in the review article by Balbus and Hawley~\cite{balbus_98}. A general relativistic linear analysis is presented in~\cite{araya-gochez_02}.



\subsection{Thermal and viscous instability}

It was realized by Shakura and Sunyaev themselves \cite{shakura_76}, as well as other authors \cite{lightman_74, shibazaki_75}, that the Shakura--Sunyaev solution (Section~\ref{section-Shakura-Sunyaev}) should be thermally and viscously unstable for disks in which radiation pressure dominates (when the opacity is governed by electron scattering). The most general and elegant arguments are presented in the classic paper by Piran \cite{piran_78}. This discovery started a long debate, which continues unresolved to this day. A recent update is provided in \cite{ciesielski_12}.

To understand the thermal instability better, we consider a disk cooling through radiative diffusion. The local emergent flux at radius $r$ is given by
\begin{equation}
F^- = \frac{acT^4}{\tau} \,,
\end{equation}
where $a$ is the radiation density constant, $T$ is a measure of the disk interior temperature, and $\tau$ is half the total vertical optical depth. If the opacity is dominated by electron scattering, then $\tau \sim \kappa_T \Sigma/2$, where $\Sigma$ is the surface density of the disk, which is constant on the time scales of the instabilities (very much less than the radial flow time scale). The Thomson opacity $\kappa_T$ is also constant, being independent of temperature, provided there is already sufficient ionization.  Therefore, the optical depth is independent of temperature and the cooling rate per unit area is $F^- \propto T^4$.

The dissipation rate per unit area is
\begin{equation}
F^+ \sim rH {\cal T}_{r\phi} \frac{d\Omega}{dr} \,.
\label{eqn:dissipation}
\end{equation}
Vertical hydrostatic equilibrium implies that the disk half thickness $H \sim 2P/(\Omega_K^2 \Sigma)$, so that Eq.~(\ref{eqn:dissipation}) becomes
\begin{equation}
F^+ \sim \frac{2rP{\cal T}_{r\phi}}{\Omega_K^2 \Sigma} \frac{d\Omega}{dr} \,.
\label{eqn:dissipation2}
\end{equation}
In the radiation pressure dominated inner region, $P \simeq aT^4/3$, so that Eq.~(\ref{eqn:dissipation2}) plus the Shakura--Sunyaev assumption ${\cal T}_{r\phi} = -\alpha P$ imply that $F^+ \propto T^8$! Hence a perturbative increase in temperature increases both the local cooling and heating rates, but the heating rate increases much faster, leading to a thermal runaway.

Note, though, that this argument only applies when the viscous stress is proportional to the total pressure $P_\mathrm{Tot}$ ($\alpha$ being the proportionality constant). For some time it seemed that a plausible way to avoid this instability was to argue that the stress is proportional instead to the gas pressure $P_\mathrm{gas}$.  Recent numerical simulations, though, of the magneto-rotational instability in radiation-pressure dominated disks have shown that the stress is, in fact, proportional to the total pressure~\cite{hirose_09a}.  Interestingly, these simulations exhibit no sign of the predicted thermal instability.

Most observations also argue against the existence of this instability.  In the case of accretion onto black holes, the instability is supposed to set in for luminosities in excess of $L > 0.01\,L_{\mathrm{Edd}}$. However, during outbursts, many stellar-mass black hole sources cross this limit both during their rise to peak luminosity and on their decline to quiescence, showing no dramatic symptoms (although they do undergo state changes, as described in Section~\ref{section:SpectralStates}).  On the contrary, observations suggest that disks in black hole X-ray binaries are stable up to at least $L \approx 0.5\,L_{\mathrm{Edd}}$~\cite{done_04}.  Certainly there is no evidence for the sensational behavior anticipated by some models~\cite{lasota_91a,taam_84}.

\newpage

\section{Oscillations}
\label{section-oscillations}

Even when analytic disk solutions are stable against finite perturbations, it is often the case that these perturbations will, nevertheless, excite oscillatory behavior. Oscillations are a common dynamical response in many fluid (and solid) bodies.  Here we briefly explore the nature of oscillations in accretion disks.  This topic is particularly relevant to understanding the physical mechanisms that may be behind quasi-periodic oscillations (or QPOs, which are discussed in Section~\ref{section-QPOs}).

There are a number of local restoring forces available in accretion disks to drive oscillations.  Local pressure gradients can drive oscillations via sound waves.  Buoyancy forces can act through gravity waves.  The Coriolis force can operate through inertial waves.  Surface waves can also exist, with the restoring force given by the local effective gravity.

Of particular interest are families of low order modes that may exist in various accretion geometries. Such modes will tend to have the largest amplitudes and produce more easily observed changes than their higher-order counterparts. Here we briefly review a couple relatively simple examples for the purpose of illustration.  More details can be found in the references given.

\subsection{Dynamical oscillations of thick disks}
\label{sub-section-dynamical-stability}

A complete analysis of the spectrum of modes in thick disks has not yet been done.  Some progress has been made by considering the limiting case of a slender torus, where slender here means that the thickness of the torus is small compared to its radial separation from the central mass (i.e. the torus has a small cross-sectional area). In this limit, the complete set of modes have been determined for the case of constant specific angular momentum in a Newtonian gravitational potential~\cite{blaes_85}. A more general analysis of slender torus modes is given in~\cite{blaes_06}.

Any finite, hydrodynamic flow orbiting a black hole, such as the Polish doughnuts described in Section~\ref{section-thick-disks}, is susceptible to axisymmetric, incompressible modes corresponding to global oscillations at the radial ($\sigma=\omega_r$) and vertical ($\sigma=\omega_{\theta}$) epicyclic frequencies. Other accessible modes are found by solving the relativistic Papaloizou--Pringle equation~\cite{abramowicz_06}
\begin{eqnarray}
\frac{1}{(-g)^{1/2}}\left\{\partial_i \left[(-g)^{1/2}g^{ij}f^n\partial_j W\right]\right\}
- \left(m^2g^{\phi\phi} - 2m\omega g^{t\phi} + \omega^2
g^{tt}\right)f^n W \nonumber \\
 = -\frac{(u^t)^2(\omega-m\Omega)^2}{c_{s0}^2}f^{n-1}W \,,
 \label{papa-pringle-relativistic}
 \end{eqnarray}
where the function $f$ is defined by
\begin{equation}
\frac{P}{\rho} = \frac{P_0}{\rho_0} f(r,\theta), ~~~~ f(r,\theta) = 1 - \frac{1}{nc_{s0}^2} \left[ \frac{{\cal E}_0^2}{2}({\cal U}_{\mathrm{eff}} - {\cal U}_{\mathrm{eff},0}) + \Phi \right],
\end{equation}
$m$ is the azimuthal wave number, $n$ is the polytropic index (assuming an equation of state of the form $P = K \rho^{1+1/n}$), $c_{s0}$ is the sound speed at the pressure maximum $r_0$, and $g$ is the determinant of the metric. The necessary boundary condition comes from requiring that the Lagrangian pressure perturbation vanish at the unperturbed surface ($f=0$):
\begin{equation}
\Delta P = (\delta P + \xi^{\alpha}\nabla_{\alpha}P) = 0 \, ,
\label{papa-pringle-surface}
\end{equation}
where $\xi^\alpha$ is the Lagrangian displacement vector. Eqs.~(\ref{papa-pringle-relativistic}) and (\ref{papa-pringle-surface}) describe a global eigenvalue problem for the modes of Polish doughnuts, with three characteristic frequencies: the radial epicyclic frequency $\omega_r$, the vertical epicyclic frequency $\omega_\theta$, and the characteristic frequency of inertial modes $\kappa$.

In the slender-torus limit, one can write down analytic expressions for a few of the lowest order modes~\cite{blaes_06} besides just the radial and vertical epicyclic ones.  In terms of local coordinates measured from the equilibrium point,
\begin{equation}
x \equiv g_{rr0}^{1/2} \left(\frac{r-r_0}{r_0} \right) \quad \mathrm{and} \quad y\equiv g_{\theta\theta 0}^{1/2} \left(\frac{\pi/2 - \theta}{r_0} \right) \,,
\end{equation}
an eigenfunction of the form $W = A xy$, for some constant $A$, yields two modes with eigenfrequencies~\cite{blaes_06}
\begin{equation}
\bar{\sigma}_0^2 = \frac{1}{2}\left\{\omega_r^2+\omega_{\theta}^2 \pm
\left[(\omega_r^2+\omega_{\theta}^2)^2+4\kappa_0^2\omega_{\theta}^2 \right]^{1/2} \right\} \,,
\end{equation}
where $\bar{\sigma}_0=\sigma_0/\Omega_0$. The positive square root gives a surface gravity mode that has the appearance of a cross($\times$)-mode. The negative square root gives a purely incompressible inertial (c-) mode, whose poloidal velocity field represents a circulation around the pressure maximum.

\epubtkImage{}{%
\begin{figure}[htbp]
  \centerline{\includegraphics[scale=0.5,angle=0]{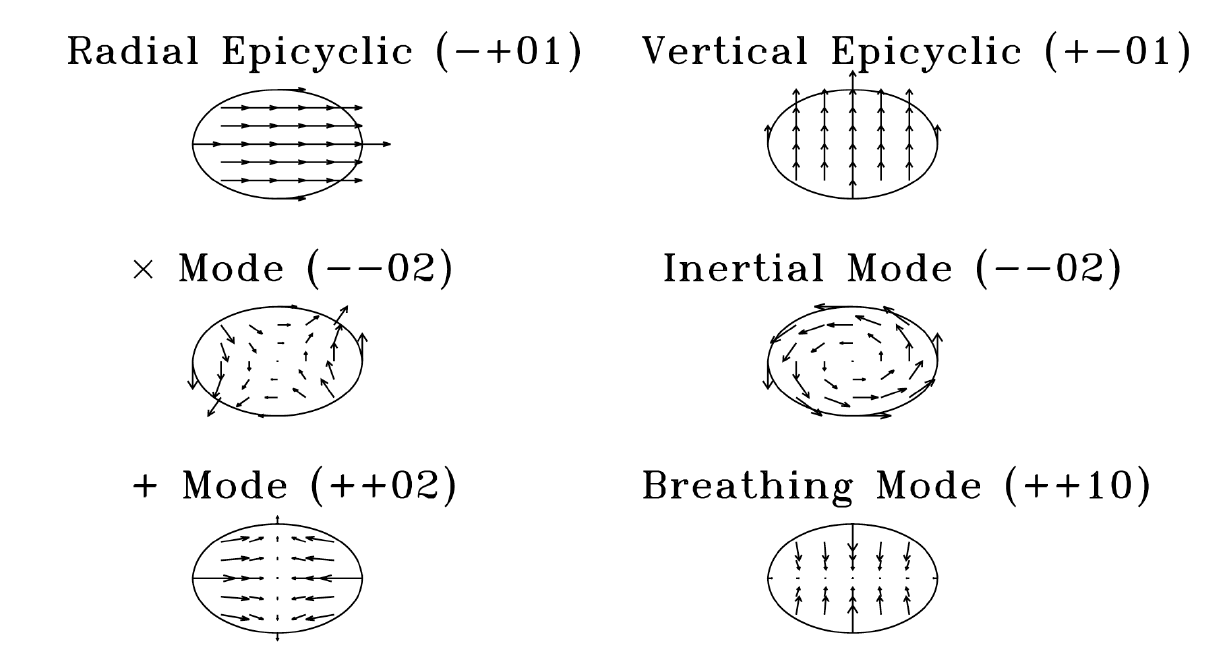}}
  \caption{Poloidal velocity fields ($\delta u_x$, $\delta u_y$) of
    the lowest order, non-trivial thick disk modes. Image reproduced
    by permission from~\cite{blaes_06}.}
  \label{figure:modes}
\end{figure}}

An eigenfunction of the form $W = A + Bx^2 + Cy^2$, with arbitrary constants $A,B,C$, also has two modes with eigenfrequencies~\cite{blaes_06}
\begin{eqnarray}
\bar{\sigma}_0^2 & = &
\frac{1}{2n}\left\{(2n+1)(\omega_r^2+\omega_{\theta}^2)-(n+1)\kappa_0^2 \right. \\ \nonumber
& & \left. \pm [ ((2n+1)(\omega_{\theta}^2-\omega_r^2)^2 + (n+1)\kappa_0^2)^2
+ 4(\omega_r^2 - \kappa_0^2) \omega_{\theta}^2 ]^{1/2} \right\} \,.
\end{eqnarray}
The upper sign results in an acoustic mode with the velocity field of a breathing mode. This mode is comparable to the acoustic mode in the incompressible Newtonian limit for $\ell = \mathrm{const.}$, while in the Keplerian limit, the mode frequency becomes that of a vertical acoustic wave. The lower sign corresponds to a gravity wave, with a velocity field reminiscent of a plus($+$)-mode. The  poloidal velocity fields of all these lowest order modes are illustrated in Figure~\ref{figure:modes}.

\subsection{Diskoseismology: oscillations of thin disks}
\label{section-diskoseismology}

To analyze oscillations of thin disks, one can express the Eulerian perturbations of all physical quantities through a single function $\delta W \propto \delta P/\rho$. Since the accretion disk is considered to be stationary and axisymmetric, the angular and time dependence are factored out
as $\delta W = W(r,z)e^{i(m\phi - \sigma t)}$, where the eigenfrequency $\sigma(r,z) = \omega - m\Omega$. Here it is useful to assume that the variation of the modes in the radial direction is much stronger than in the vertical direction, $z = r\cos\theta$. The resulting two (separated) differential equations for the functional amplitude $W(r,z) = W_r(r) W_y(r,y)$ are given by~\cite{perez_97,wagoner_99,wagoner_08}
\begin{eqnarray}
&&\frac{d^2W_r}{dr^2}-\frac{1}{(\omega^2-\omega_r^2)}
\left[\frac{d}{dr}\left(\omega^2-\omega_r^2\right)\right]\frac{dW_r}{dr}
+\frac{(u^t)^2 g_{rr}}{c_s} \left(\omega^2-\omega_r^2\right)\left(1-\frac{\Psi}{\tilde{\omega}^2}\right)W_r
= 0, \\
&&(1-y^2)\frac{d^2W_y}{dy^2}-2gy\frac{dW_y}{dy}+2g
\left[\tilde{\omega}^2 y^2 + \Psi(1-y^2)\right]W_y
= 0.
\label{seismology-master-equations}
\end{eqnarray}
The radial eigenfunction, $W_r$, varies rapidly with $r$, while the vertical eigenfunction, $W_y$, varies slowly. Here $y = (z/H)[\gamma_g/(\gamma_g - 1)]^{1/2}$ is the re-scaled vertical coordinate, $\gamma_g$ is the adiabatic index, $\tilde{\omega}(r) = \omega_r/\omega_{\theta}$ is the ratio of the epicyclic frequencies from Section~\ref{sub-section-marginally-stable}, $\Psi$ is the eigenvalue of the (WKB) separation function, and $g = (P/P_c)/(\rho/\rho_c)$, where $P_c$ and $\rho_c$ are the midplane values. The radial boundary conditions depend on the type of mode and its capture zone (see below). Oscillations in thin accretion disks, then, are described in terms of $\Psi(r,\sigma)$, along with the angular, vertical,  and radial mode numbers (number of nodes in the corresponding eigenfunction) $m$, $j$, and $n$, respectively. Modes oscillate in the radial range where
\begin{equation}
(\omega^2 - \omega_r^2)\left(1 - \frac{\Psi}{\tilde{\omega}^2}\right)
> 0 \,
\label{seismology-dispersion}
\end{equation}
outside the inner radius, $r > r_{\mathrm{i}}$.

\textbf{p-modes} are inertial acoustic modes defined by $\Psi < \tilde{\omega}^2$ and are trapped where $\omega^2 > \omega_r^2$, which occurs in two zones. The inner p-modes are trapped between the inner disk edge and the inner ``Lindblad'' radius, i.e., $r_{\mathrm{i}} < r < r_-$, where gas is accreted rapidly. The outer p-modes occur between the outer Lindblad radius and the outer edge of the disk, i.e., $r_+ < r < r_{\mathrm{o}}$. The Lindblad radii, $r_-$ and $r_+$, occur where $\omega = \omega_r$. The outer p-modes are thought to be more consequential as they produce stronger luminosity modulations~\cite{ortega_02}. In the corotating frame these modes appear at frequencies slightly higher than the radial epicyclic frequency. Pressure is the main restoring force of p-modes.

\textbf{g-modes} are inertial gravity modes defined by $\Psi > \tilde{\omega}^2$. They are trapped where $\omega^2 < \omega_r^2$ in the zone $r_- < r < r_+$ given by the radial dependence of
$\omega_r$, i.e., g-modes are gravitationally captured in the cavity of the radial epicyclic frequency and are thus the most robust among the thin-disk modes. Since this is the region where the temperature of the disk peaks, g-modes are also expected to be most important observationally~\cite{perez_97}. In the corotating frame these modes appear at low frequencies. Gravity is their main restoring force.

\textbf{c-modes} are corrugation modes defined by $\Psi = \tilde{\omega}^2$. They are non-radial ($m=1$) and vertically incompressible modes that appear near the inner disk edge and precess slowly around the rotational axis. These modes are controlled by the radial dependence of the vertical epicyclic frequency. In the corotating frame they appear at the highest frequencies.

All modes have frequencies $\propto 1/M$. Upon the introduction of a small viscosity ($\nu \propto \alpha, \,\, \alpha \ll 1$), most of the modes grow on a dynamical timescale $t_{\mathrm{dyn}}$, such that the disk should become unstable. However, evidence for these modes has so far mostly been lacking in MRI turbulent simulations (see Section~\ref{section-numerical-oscillations}). This leaves their relevance in some doubt.

\newpage

\section{Relativistic Jets}
\label{section-jets}

Although the main focus of this review is on black hole accretion disk theory, we note that there has long been a strong observational connection between accreting black holes and relativistic jets across all scales of black hole mass. For supermassive black holes this includes quasars and active galactic nuclei; for stellar-mass black holes this includes microquasars. However, the theoretical understanding of disks and jets has largely proceeded separately and the physical link between the two still remains uncertain. Therefore, we present only a few brief comments on the subject in this review.  More complete discussions of the theory of relativistic jets may be found in~\cite{meier_03}.  A review of their observational connection to black holes is given in~\cite{mirabel_99}.

In Section~\ref{sub-section-ergosphere}, we described the ``Penrose process,'' whereby rotational energy may be extracted from a black hole and carried to an observer at infinity.  To briefly recap, Penrose~\cite{penrose_69, penrose_71} imagined a freely falling particle with energy $E^\infty$ disintegrating into two particles with energies $E^\infty_- < 0$ and $E^\infty_+ > 0$.  Then, the particle with negative energy $E^\infty_-$ falls into the black hole, and the other one escapes to infinity. Clearly, $E^\infty_+ > E^\infty$, so that there is a net gain of energy.

It was first suggested by Wheeler at a 1970 Vatican conference and soon after by others~\cite{mashhoon_73, fishbone_73} that such a Penrose process may explain the energetics of superluminal jets commonly seen emerging from quasars and other black hole sources. However, a number of authors~\cite{bardeen_72, wald_74, kovetz_75} showed that for $E^\infty_+$ to be greater than $E^\infty$, the disintegration process must convert most of the rest mass energy of the infalling particle to kinetic energy, in the sense that, in the center-of-mass frame, the $E^\infty_-$ particle must have velocity $v > c/2$. The argument of Wald~\cite{wald_74} is powerful, short and elegant, so we give it here \textit{in extenso}.

Let $p^\nu = mu^\nu$ be the four momentum of a particle with the mass $m$. We assume that in the ZAMO frame (Section~\ref{sub-section-ergosphere}) the particle has a four velocity of the form, $u^\nu = \Gamma (N^\nu + V\Lambda^\nu)$, where $\Lambda^\nu$ is a timelike-unit vector (for simplicity we assume $\Lambda^\nu\eta_\nu = 0$), $V$ is the particle 3-velocity in the ZAMO frame, and $\Gamma^{-2} = 1 - V^2$. If the disintegration fragments move in the directions $\pm\Lambda^\nu$ (which one may prove is energetically most favorable), then the four velocities of these fragments in the center-of-mass frame are,
\begin{eqnarray}
u_{\pm}^\nu &=& \gamma (u^\nu \pm v \tau^\nu),
\label{velocities-disintegration}
\\
\tau^\nu &=& \Gamma (V N^\nu + \Lambda^\nu).
\label{velocities-disintegration-tau}
\end{eqnarray}
The form of (\ref{velocities-disintegration-tau}) follows from the Lorentz transformation $\{N^\nu, \Lambda^\nu \} \rightarrow \{u^\nu, \tau^\nu \}$. Multiplying (\ref{velocities-disintegration}) by $\eta_\nu$ gives
\begin{equation}
\frac{E_\pm^\infty}{m_\pm} = \gamma \left( \frac{E^\infty}{m}\right)  \pm \gamma v \left[
\left(\frac{E^\infty}{m}\right)^2 + \frac{1}{g^{tt}}\right]^{1/2}.
\label{penrose-process-energy-fragments}
\end{equation}
Since $1/g^{tt} < 1$ and realistic particles have $E/m > 1/\sqrt{3}$, the condition $E_-^\infty < 0$ necessarily requires $v > 1/2$. Such highly relativistic disintegration events are not generally seen in nature. To make matters worse, from the upper limit of (\ref{penrose-process-energy-fragments}), Wald deduced that the presence of the black hole limits the energy increase to a maximal factor of $1 + \sqrt{2}$. Thus, he concluded~\cite{wald_74}: ``\textit{The Penrose mechanism cannot serve as a useful energy source for astrophysical processes. In no case can one obtain energies which are greater by a significant factor than those which already could be obtained by a similar breakup process without the presence of the black hole.}''

Replacing particle disintegration with particle collision does not help, even though the center-of-mass energy of such a collision happening arbitrarily close to the horizon of the maximally rotating Kerr black hole may be arbitrarily large~\cite{piran_77, banados_09}.  This is because the Wald limit of $1 + \sqrt{2}$ still holds \cite{bejger_12}.  It would seem that even under idealized conditions, the maximal energy of a particle escaping via the Penrose process is only a modest factor above the total initial energy \cite{bejger_12}.

Therefore, we consider a general matter distribution, described by an unspecified stress-energy tensor $T^\mu_\nu$. In this case, the energy flux in the ZAMO frame is $E^i = - T^i_{~k}N^k$, and the energy absorbed by the black hole is
\begin{equation}
\label{komissarov-total-energy}
E = - \int T^i_{~k}n^k\,dN_i > 0,
\end{equation}
where $\int\,dN_i$ is the surface integral over the horizon. The inequality sign follows
from the fact that the locally measured energy must be positive. The above integral may by
transformed into
\begin{equation}
\label{komissarov-condition}
0 < E = - \int e^\Phi\,T^i_{~k}(\eta^k + \omega \xi^k)\,dN_i = e^\Phi_H(E^{\infty}_- -
\omega_H J^{\infty}).
\end{equation}
As in the classic Penrose process, the \textit{necessary} condition for the energy gain is:
\begin{equation}
\omega_H J^\infty < E^\infty_- < 0.
\label{penrose-process-fluid}
\end{equation}
Thus, in a way fully analogous to the Penrose process for particles (\ref{penrose-process-particle}), one may say that if the energy at infinity increases because the black hole absorbed negative-at-infinity energy, then the black hole rotation must also slow down by absorbing matter with negative angular momentum.

Blandford and Znajek \cite{blandford_77} made the brilliant discovery that an electromagnetic form of the Penrose process may work. In their model, the energy for the jet is extracted from the spin energy of the black hole via a torque provided by magnetic field lines that thread the event horizon or ergosphere. The estimated luminosity of the jet is given by \cite{macdonald_82} (although see \cite{tchekhovskoy_10} for higher order expressions that apply when $a/M \sim 1$)
\begin{equation}
L_\mathrm{BZ} = \frac{1}{32} \omega_F^2 B_\perp^2 r_H^2 (a/M)^2 \, ,
\end{equation}
where $\omega_F^2 \equiv \Omega_F(\Omega_H-\Omega_F)/\Omega_H^2$ is a measure of the effect of the angular velocity of the field $\Omega_F$ relative to that of the hole $\Omega_H \equiv a/(r_H^2+a^2)$, $B_\perp$ is the magnetic field normal to the horizon, and $r_H$ is the radius of the event horizon (\ref{horizon-radius-Kerr}). In this model, the only purpose of the disk is to act as the current sheet which continually provides magnetic field to the black hole. This last point led to one of the main objections to the Blandford--Znajek model: Ghosh and Abramowicz~\cite{ghosh_97} argued on astrophysical grounds that accretion disks simply cannot feed the required fields into the black hole. However, recent work by Rothstein and Lovelace~\cite{rothstein_08} has countered this claim and suggested that indeed the disk can serve this role. There are also more fundamental reservations with the Blandford--Znajek model, some of which are presented in~\cite{punsly_01, komissarov_05}. Such claims and counter-claims were for many years characteristic of the uncertainty in the theory of relativistic jets (see~\cite{komissarov_09} for a discussion). However, direct numerical simulations may be helping to clarify the picture, as we discuss in Section~\ref{section-numerical-jets}.  Plus, there is now observational evidence suggesting a possible connection between black hole spin and jet power, exactly as predicted by the Blandford--Znajek model~\cite{narayan_12a}, although again there are countering claims~\cite{fender_10}

\newpage

\section{Numerical Simulations}
\label{section:numerical_simulations}

In simulating accretion disks around black holes, there are a number of challenging issues. First, there is quite a lot of physics involved: relativistic gravity, hydrodynamics, magnetic fields, and radiation being the most fundamental. Then there is the issue that accretion disks are inherently multi-dimensional objects. The computational expense of including extra dimensions in a numerical simulation is not a trivial matter. Simply going from one to two dimensions (still assuming axisymmetry for a disk) increases the computational expense by a very large factor ($10^2$ or more). Going to three-dimensions and relaxing all symmetry requirements increases the computational expense yet again by a similar factor. Simulations of this size have only become feasible within the last decade and still only with a subset of the physics one is interested in and usually with a very limited time duration.

Another hindrance in simulating accretion disks is the very large range of scales that can be present. In terms of a grid based code, a disk with a scale height of $H/R$ requiring $N_z$ zones to resolve in the vertical direction at some radius $R_\mathrm{in}$, would require something of the order $N_z/(H/R)$ zones to cover \emph{each} factor of $R_\mathrm{in}$ that is treated in the radial direction. The azimuthal direction in a full three-dimensional simulation would require a comparable number of zones to what is used in the radial direction. Given that a \emph{very} large calculation by today's standards is $10^9\mbox{\--\,}10^{10}$ zones, we can see that treating very thin disks ($H/R \lesssim 0.01$) in a full three-dimensional simulation, even with a very modest number of zones in the vertical direction ($N_z \ge 50$) would take all the resources one could muster. However, two-dimensional (axisymmetric) simulations of thin disks and three-dimensional simulations of thick, slim, or moderately thin disks have now become quite common.


The approach of numericists in many ways parallels that of theorists, so we structure this section much the same as the first part of this Living Review.  That is, after a brief introduction to numerical techniques, we discuss how the various components of the matter description (\`a la Section~\ref{section-matter}) are implemented in numerical simulations. We then review a few special cases illustrating how analytic models (Sections~\ref{section-thick-disks}\,--\,\ref{section-ADAFs}) and numerical simulations can complement one another.  We finish this numerical section with a few topics of special interest.

\subsection{Numerical techniques}

\subsubsection{Computational fluid dynamics codes}

There are many numerical codes available today that include relativistic hydrodynamics or MHD that are, or can be, used to simulate accretion disks.  A partial list includes: Cosmos++~\cite{anninos_05}, ECHO~\cite{delzanna_07}, HARM~\cite{gammie_03a}, and RAISHIN~\cite{mizuno_06} plus the codes developed by Koide et al.~\cite{koide_98}, De~Villiers and Hawley~\cite{devilliers_03a}, Komissarov~\cite{komissarov_04}, Ant\'on et al.~\cite{anton_06}, and Anderson et al.~\cite{anderson_06}. This represents tremendous growth in the field as prior to 1998 there were only a couple such codes, largely derived from the early work of Wilson~\cite{wilson_77} and later developments by Hawley~\cite{hawley_84b,hawley_84a}. Those early codes were only available to a handful of researchers. This was simply a reflection of the fact that the computational resources available to most researchers at that time were insufficient to make use of such codes, so there was little incentive to develop or acquire them. This has clearly changed.

Today there are primarily two types of numerical schemes being used to simulate relativistic accretion disks, differentiated by how they treat discontinuities (shocks) that might arise in the flow: the artificial viscosity scheme, still largely based upon formulations developed by Wilson~\cite{wilson_77}; and Godunov-type approaches, using exact or approximate Riemann solvers. Both are based on finite difference representations of the equations of general relativistic hydrodynamics, although the latter tend to also incorporate finite-volume representations. The artificial viscosity schemes have the advantages that they are more straightforward to implement and easier to extend to include additional physics. More importantly, they are computationally less expensive to run than Godunov schemes. Furthermore, recent work by Anninos and collaborators~\cite{anninos_03,anninos_05} has shown that variants of this scheme can be made as accurate as Godunov schemes even for ultrarelativistic flows, which historically was one of the principal weaknesses of the artificial viscosity approach. Godunov schemes, on the other hand, appreciate the advantage that they are fully conservative, and therefore, potentially more accurate. They also require less tuning since there are no artificial viscosity parameters that need to be set for each problem. More thorough reviews of these two methods, with clear emphasis on the High-Resolution Shock-Capturing variant of the Godunov approach, are provided in the Living Reviews by Mart\'i \& M\"uller~\cite{marti_03} and Font~\cite{font_08}. Other numerical schemes, such as smooth-particle hydrodynamics (SPH) and (pseudo-)spectral methods are less well developed for work on relativistic accretion disks.


\subsubsection{Global vs.\ shearing-box simulations}

Along with settling on a numerical scheme, a decision must also be made whether or not to try to treat the disk as a whole or to try to understand it in parts. The latter choice includes ``shearing-box'' simulations, the name coming from the type of boundary conditions one imposes on the domain to mimic the shear that would be present in a real disk~\cite{hawley_95}. The obvious advantage of treating the disk in parts is that you circumvent the previously noted problem of the large range in scales in the disk by simply ignoring the large scales. Instead one treats a rectangular volume generally no larger than a few vertical scale heights on a side and in some cases much smaller. In this way, for a moderate number of computational zones, one can get as good, or often much better, resolution over the region being simulated than can be achieved in global simulations. The obvious disadvantage is that all information about what is happening on larger scales is lost. By construction, no structures larger than the box can be captured and the periodic boundaries impose some artificial conditions on the simulations, such as no background gradients in pressure or density and no flux of material into or out of the box, meaning the surface density remains fixed. Perhaps most important in the context of a review on relativity is that the box is treated as a local patch within the disk, with a scale smaller than that of the curvature of the metric. Thus most general relativistic effects are not, and by construction need not be, included in shearing box simulations. Since this is a \emph{Living Review in Relativity}, we will not dwell much on this class of simulations. Still, there are some significant highlights that should be mentioned.

By far the most extensive work using shearing box simulations has been aimed at better understanding the magneto-rotational instability (MRI).  Starting from the earliest ``proof-of-concept'' simulations~\cite{hawley_95,stone_96}, shearing box simulations have been used to demonstrate various properties of the saturated state of MRI turbulence~\cite{sano_01,sano_04,lesur_07,shi_10} and much about the vertical structure of MRI turbulent disks~\cite{blaes_11}.  Shearing box simulations have also proven valuable in studying radiation-dominated disks. For example, Turner~\cite{turner_04} studied the vertical structure of radiation dominated accretion disks and Hirose et~al.~\cite{hirose_06} studied the gas-pressure dominated case, both including radiation using flux-limited diffusion. Hirose and collaborators subsequently used shearing box simulations to demonstrate that radiation-dominated disks are thermally stable~\cite{hirose_09a}, though they had long been held not to be~\cite{lightman_74,shakura_76,piran_78}. As we said, though, there are limits imposed by the finite size of the shearing box.  For instance, the viscous, Lightman--Eardley instability~\cite{lightman_74} can only be studied through global simulations.

\subsection{Matter description in simulations}

The \emph{minimum} physics required for a global simulation of an accretion disk are gravity and hydrodynamics (assuming the disk is dense enough for the continuum approximation to hold). Since many disks have masses that are small compared to the mass of the central compact object, the self-gravity of the disk can often be ignored. Therefore, in the next three sections, gravity will simply mean that of the central black hole.

\subsubsection{Hydrodynamics + gravity}
\label{numerics-hydro}

The first researcher to develop and use numerical algorithms for simulating relativistic accretion flows was Wilson~\cite{wilson_72}, who considered the spherical infall of material with a non-zero specific angular momentum toward a Kerr black hole using the full metric, although restricted to two spatial dimensions. Wilson was able to confirm the additional centrifugal support that the infalling material experienced due to the rotating black hole. For sufficiently high values of angular momentum, the material could not immediately accrete onto the black hole, instead forming a fat disk
of the type described in Section~\ref{section-thick-disks}.

In many ways, Wilson's pioneering work was at least a decade ahead of its time and there was consequently a lull in activity until Hawley and Smarr collaborated with Wilson to revive his work~\cite{hawley_84b, hawley_84a}. As an example of how analytic and numerical work can complement each other, it is worth noting that one of the test cases they used for their new two-dimensional relativistic hydrodynamics code was based on the analytic theory for relativistic thick disks, which had been worked out in the time since Wilson's original simulations. Using the analytic theory, they constructed a series of disks with different (constant) specific angular momenta, from $\ell < \ell_\mathrm{ms}$ to $\ell > \ell_\mathrm{mb}$. The $\ell > \ell_\mathrm{mb}$ case yielded a static solution as expected and confirmed the ability of their code to accurately evolve such a solution in multi-dimensions over a dynamical time. The $\ell < \ell_\mathrm{mb}$ cases showed greater time variability and illustrated the power of direct numerical simulations to extend our understanding of black hole accretion.

\subsubsection{Magnetohydrodynamics + gravity}

Magnetic fields can play many important roles in relativistic accretion disks, from providing local viscous stresses through turbulence that results from the magnetorotational instability (Section~\ref{section-MRI}), to providing a mechanism for launching and confining jets (Section~\ref{section-jets}). Thus, the inclusion of magnetic fields in numerical simulations of relativistic accretion disks is important. Although the techniques for doing this were first described in~\cite{wilson_75}, relatively little work was done in this area until fairly recently. Perhaps the two most important contributions so far from relativistic MHD simulations of accretion disks have been: 1) elucidating the behavior of MRI turbulent disks in the vicinity of black holes, and 2) exploring the many interrelations between magnetic fields and relativistic jets.

In at least one case, the inclusion of MHD + gravity is sufficient to adequately capture the dynamics of a real black hole accretion disk. Sgr~A*, the black hole at the center of the Milky Way galaxy has such an anemically low luminosity~\cite{melia_01} that numerical simulations that ignore radiation can safely be applied~\cite{dibi_12}, as has been done by many authors~\cite{goldston_05, noble_07, moscibrodzka_09, dexter_10}. In most other cases, radiative processes must somehow be accounted for.

\subsubsection{Radiation-Magnetohydrodynamics + Gravity}
\label{sec:rad-MHD}

Probably the most glaring shortcoming of almost all numerical simulations of accretion disks and many other phenomena in astrophysics to date is the unrealistic treatment of radiation, which is most often simply ignored. This is not due to a lack of appreciation of its importance on the part of numericists, but simply a reflection of the fact that there are very few efficient ways to treat radiation computationally in multiple dimensions.

The optically-thin limit is an exception.  In this case, the radiative cooling only enters the energy and momentum conservation equations as a source term
\begin{equation}
\nabla_\mu T^{\mu\nu} = f u^\nu,
\end{equation}
where $f$ is the cooling function of the gas (Section~\ref{subsection-radiation-part}). The subsequent propagation of the radiation through the matter can be ignored or post-processed.  Even so, including radiative cooling can be critically important to capturing the true dynamics of the disk~\cite{fragile_09a,dibi_12}, as shown in Figure~\ref{figure:fragile09}.

\epubtkImage{}{%
\begin{figure}[htbp]
  \centerline{\includegraphics[scale=0.5]{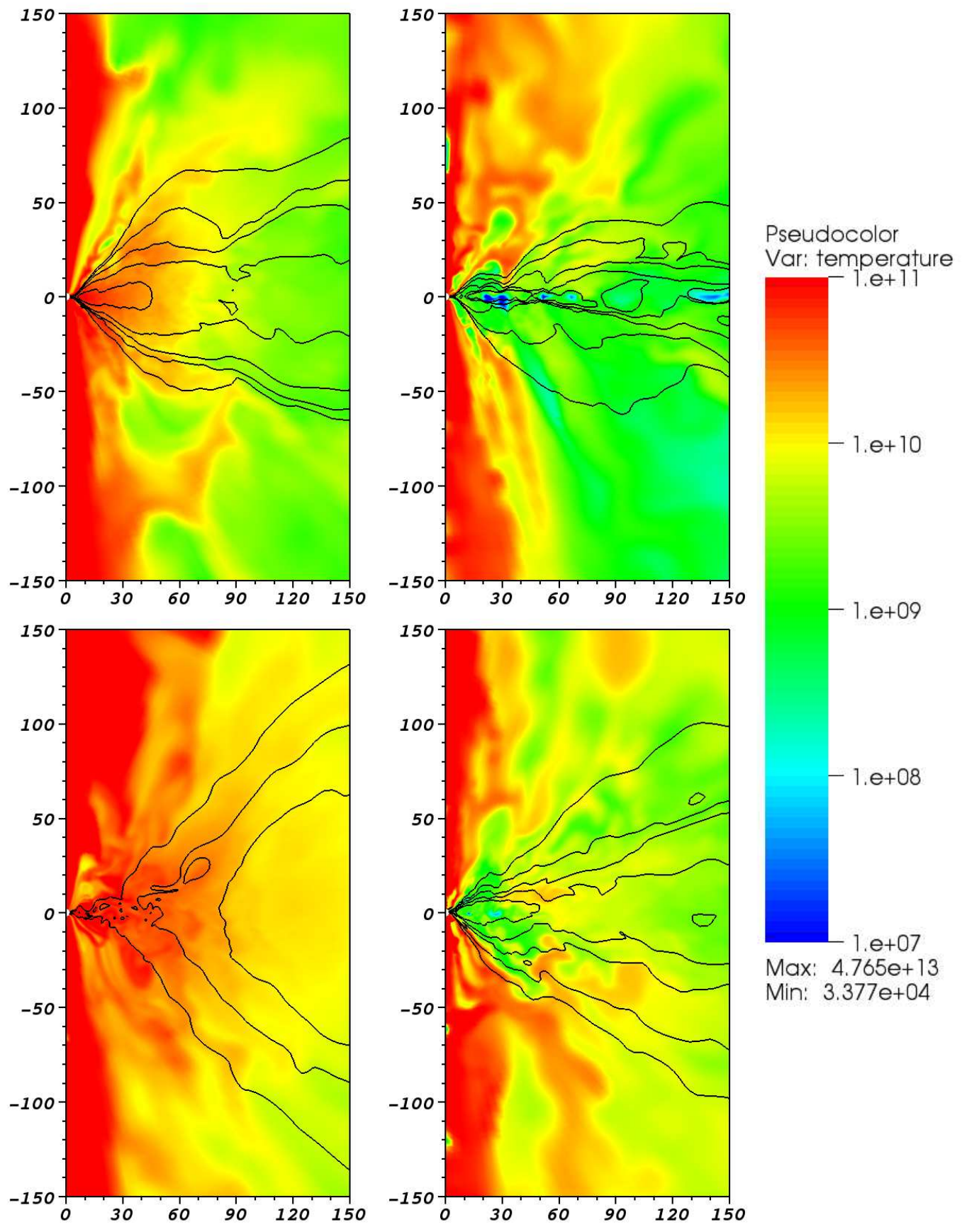}}
  \caption{Pseudo-color plots of $\log(T)$ with contours of
    $\log{\rho}$ from 4 different general relativistic MHD
    simulations. The simulations all begin with the same initial
    conditions, but have different energy conservation and cooling
    treatments: The upper-left panel conserves internal energy and
    ignores cooling; the upper-right panel conserves internal energy
    and includes cooling; the lower-left panel conserves total energy
    and ignores cooling; the lower-right panel conserves total energy
    and includes cooling.  The very different end states illustrate
    the importance of properly capturing thermodynamic
    processes. Image reproduced by permission from~\cite{fragile_09a},
    copyright by AAS.}
 \label{figure:fragile09}
\end{figure}}

In the optically-thick limit, some progress can be made by employing the flux-limited diffusion approximation~\cite{levermore_81}. Global simulations of disks including a flux-limited diffusion treatment of radiation (but using pseudo-Newtonian gravity) are now being done by Ohsuga and collaborators~\cite{ohsuga_09,ohsuga_11}. A few steps toward the goal of \emph{relativistic} radiation MHD simulations of black hole accretion disks have also been taken in recent years.  A method for treating optically thick accretion using a conservative, Godunov scheme was developed by Farris and collaborators~\cite{farris_08}.  The same basic method has now been used to examine both Bondi~\cite{fragile_12a} and Bondi--Hoyle~\cite{zanotti_11,roedig_12} accretion.  Simulations of accretion disks, though, must await the generalization of this method to treat radiation both in the optically thick and thin limits.

\subsubsection{\emph{Evolving} GRMHD}

In most simulations of accretion disks around black holes, the self-gravity of the disk is ignored. In many cases this is justified as the mass of the disk is often much smaller than the mass of the black hole. This is also much simpler as it allows one to treat gravity as a background condition, either through a Newtonian potential or a relativistic metric (the so-called Cowling approximation in relativity). However, there are plausible astrophysical scenarios in which this approximation is not valid. Two of the more interesting are: 1) a tidally disrupted neutron star accreting onto a stellar-mass black hole; and 2) an overlying stellar envelope accreting onto a nascent black hole during the final dying moments of a massive star. Interestingly these scenarios are currently the most popular models of gamma-ray bursts~\cite{eichler_89, woosley_93}, which are the most powerful explosions in the Universe since the Big Bang and are observed at a rate of about one per day.

Treating a self-gravitating disk in general relativity requires a code that can simultaneously evolve the spacetime metric (by solving the Einstein field equations) and the matter, magnetic, and radiation fields. Codes capable of doing this have very recently become available, divided into the following ``flavors'': self-gravity + hydro~\cite{baiotti_05}; self-gravity + MHD~\cite{duez_05, shibata_05, giacomazzo_07, etienne_10}; and self-gravity + radiation MHD ~\cite{farris_08}. More detailed discussion of these methods and their application to problems in astrophysics is provided in the Living Review by Font~\cite{font_08}.

The importance of evolving the gravitational field in the case of black hole accretion is well illustrated in the case of the runaway instability, which we described in Section~\ref{section-runaway-instability}. That instability has now been simulated directly, including evolving the spacetime metric~\cite{montero_08,montero_10,korobkin_11}. These studies have confirmed earlier results that the runaway instability does not develop.

Another application of these types of codes that is relevant to our review is the reexamination of the Papaloizou--Pringle instability (discussed in Section~\ref{section-PPI}).  Through this instability, a massive, self-gravitating Polish doughnut orbiting around a black hole could become a strong emitter of large amplitude, quasiperiodic gravitational waves~\cite{kiuchi_11}.

\subsection{Polish doughnuts (thick) disks in simulations}
\label{section-numerical-polish-doughnut}

As we already mentioned, it is computationally less expensive to simulate thick disks than thin, so it comes as no surprise that most of the numerical work for nearly four decades, starting with the work of Wilson~\cite{wilson_72}, has been on thick disks.  In this section we touch on a few highlights from this area of research

Some of this work has focused on confirming predictions from analytic theory.  Igumenshchev and Beloborodov~\cite{igumenshchev_97} used two-dimensional relativistic hydrodynamic simulations to confirm that: (i) The structure of the innermost disk region strongly depends on the black hole spin, and (ii) the mass accretion rate is proportional to the size of the energy gap between the inner edge of the disk and the cusp (see Figure~\ref{figure:dotMdeltaW}), as predicted by Koz{\l}owsk and collaborators~\cite{kozlowski_78}.

\epubtkImage{}{%
\begin{figure}[htbp]
  \centerline{\includegraphics[width=0.6\textwidth]{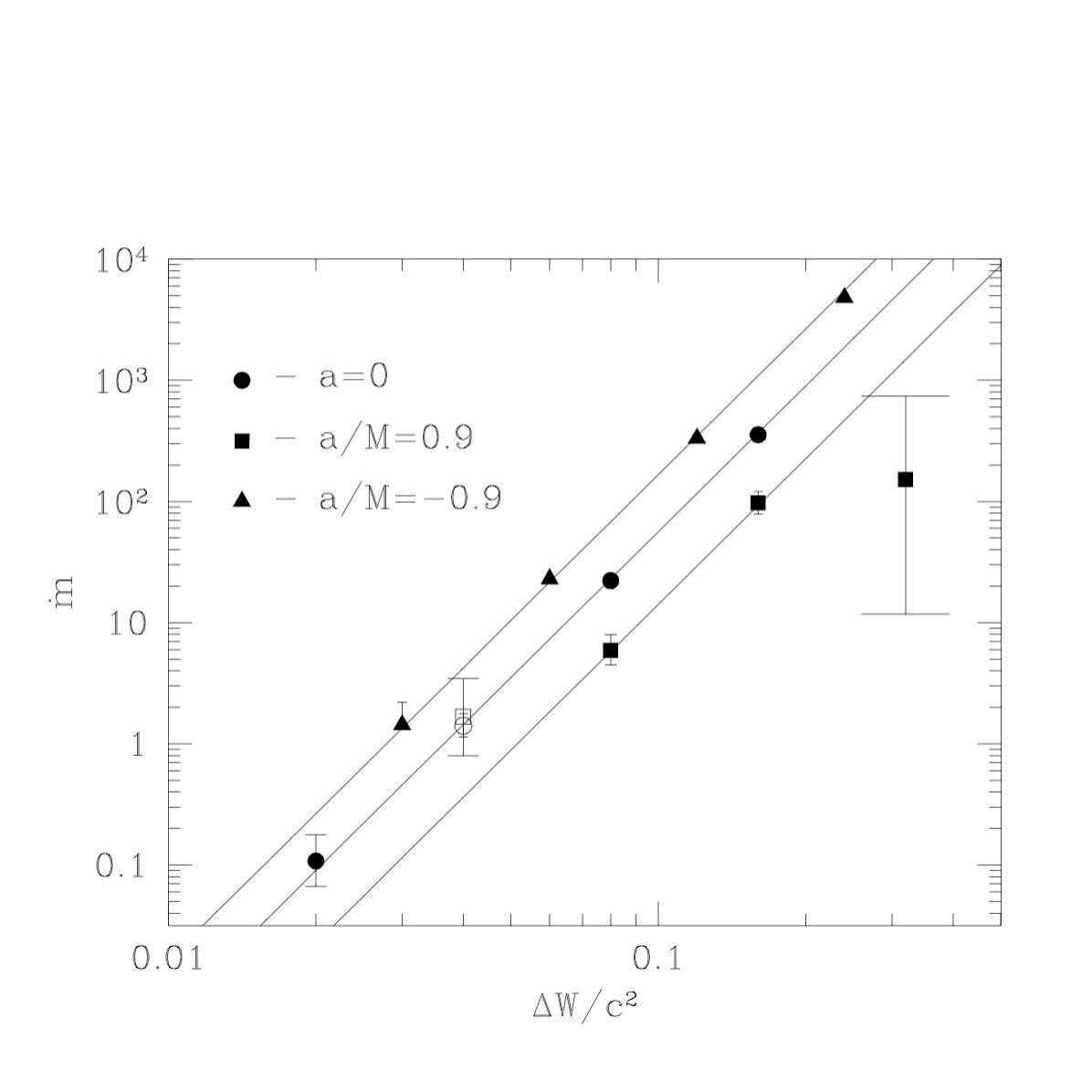}}
  \caption{Time-average mass accretion rate $\dot{m}=\dot{M}
    c^2/L_{\mathrm{Edd}}$ as a function of the energy gap $\Delta
    \Phi$ for models with $a=0$ (circles), $a/M=0.9$ (squares), and
    $a/M=-0.9$ (triangles). The bars show the variability of
    $\dot{m}$.  The lines represent the predicted dependencies
    $\dot{m}\propto(\Delta \Phi)^{\gamma_g/(\gamma_g-1)}$ where
    $\gamma_g=4/3$ is the adiabatic index. Image reproduced by
    permission from~\cite{igumenshchev_97}, copyright by RAS.}
  \label{figure:dotMdeltaW}
\end{figure}}

Hawley~\cite{hawley_90, hawley_91a} studied the nonlinear evolution of the non-axisymmetric Papaloizou--Pringle instability (Section~\ref{section-PPI}). He discovered that for radially slender tori, the principal mode of the PPI saturates with the formation of ellipsoidal overdensity regions he termed ``planets'' (see Figure~\ref{figure:PPIplanets} for an example). In cases where several wavenumbers were unstable, multiple planets would begin to form but then nonlinear mode-mode coupling would cause them to merge. For wide tori, particularly when the tori extended beyond their Roche limits, overflowed their cusps, and accreted into the hole, the PPI saturated at low amplitude or was prevented from growing altogether. The initial work only considered Schwarzschild black holes, but all of the results were later qualitatively confirmed for Kerr black holes~\cite{devilliers_02}.

\epubtkImage{}{%
\begin{figure}[htbp]
  \centerline{\includegraphics[scale=0.75]{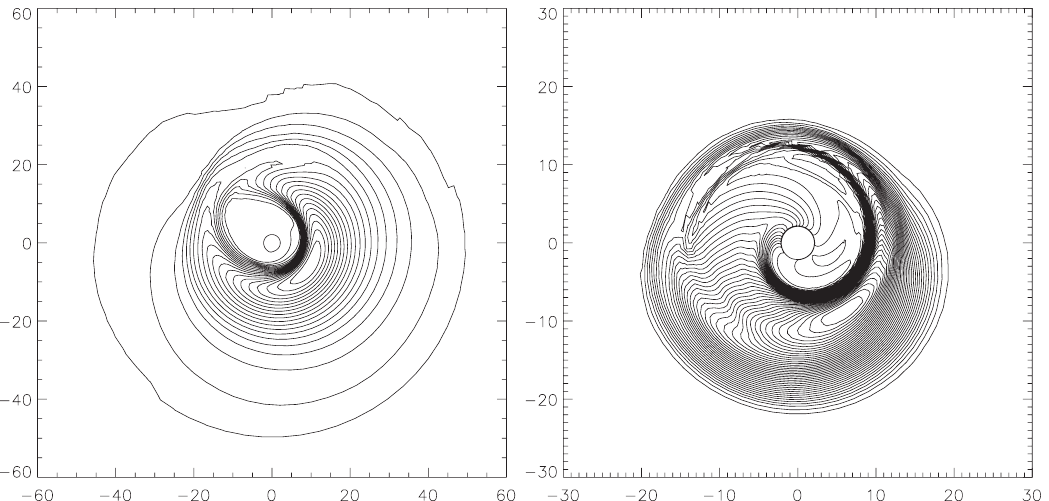}}
  \caption{Equatorial slice through hydrodynamic tori at
  saturation of the Papaloizou--Pringle instability showing formation of significant non-axisymmetric ($m=1$) overdensity clumps. The density contours are linearly spaced between $\rho_{\max}$
  and 0.0. This figure represents models A3p (left) and B3r (right) of~\cite{devilliers_02}.}
  \label{figure:PPIplanets}
\end{figure}}

Another interesting note is that most of the global numerical simulations of MRI turbulent disks have used a Polish doughnut as the starting condition.  This was first done by De~Villiers and Hawley~\cite{devilliers_03b} (animations can be viewed at \cite{devilliers_movies,hawley_movies}) and later by a number of other groups~\cite{gammie_03a,anninos_05}. As expected from linear analysis and earlier Newtonian simulations, the Maxwell stresses produced by the MRI-driven turbulence naturally force the final distribution of the angular momentum to be nearly Keplerian (see Figure~\ref{figure:devilliers03_9}), independent of the initial matter distribution.

\epubtkImage{}{%
\begin{figure}[htbp]
  \centerline{\includegraphics[width=0.75\textwidth]{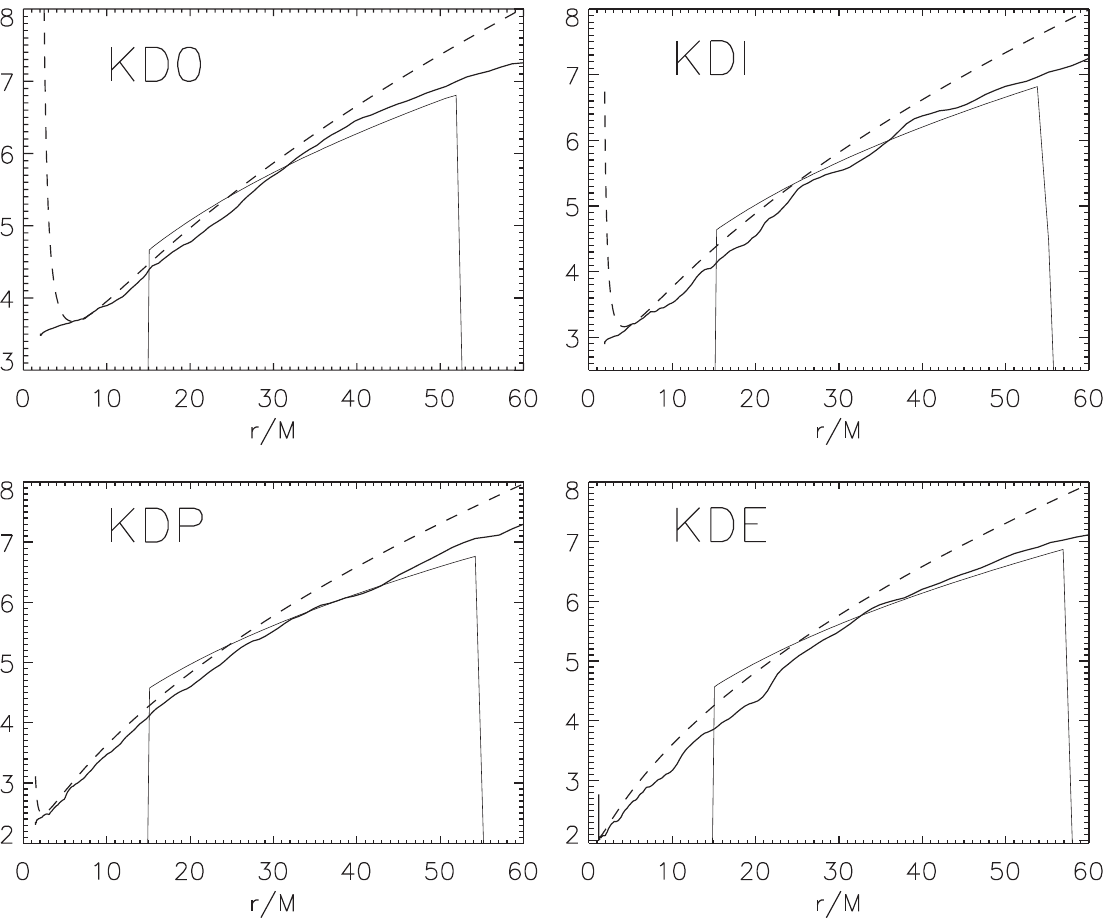}}
  \caption{Specific angular momentum $\ell$ as a function of radius
  at $t=0$ (thin line) and at $t=10.0$ orbits (thick line). The
  individual plots are labeled by model. In each case the Keplerian
  distribution for a test particle, $\ell_\mathrm{Kep}$, is shown as a
  dashed line. Image reproduced by permission
  from~\cite{devilliers_03c}, copyright by AAS.}
  \label{figure:devilliers03_9}
\end{figure}}

Although the magnetic field plays a crucial role through the action of the MRI, its amplitude saturates at a relatively weak level, always remaining subthermal in the disk (see Figure~\ref{figure:devilliers03_8}). On the other hand, simulations show that magnetic fields do tend to dominate regions outside the disk, including in the hot, magnetized corona that sandwiches the disk and in the evacuated, highly magnetized funnel region. The funnel region and the boundary between the funnel and the corona are where jets and outflows are generically observed in these simulations (Section~\ref{section-numerical-jets}).

\epubtkImage{}{%
\begin{figure}[htbp]
  \centerline{\includegraphics[width=0.75\textwidth]{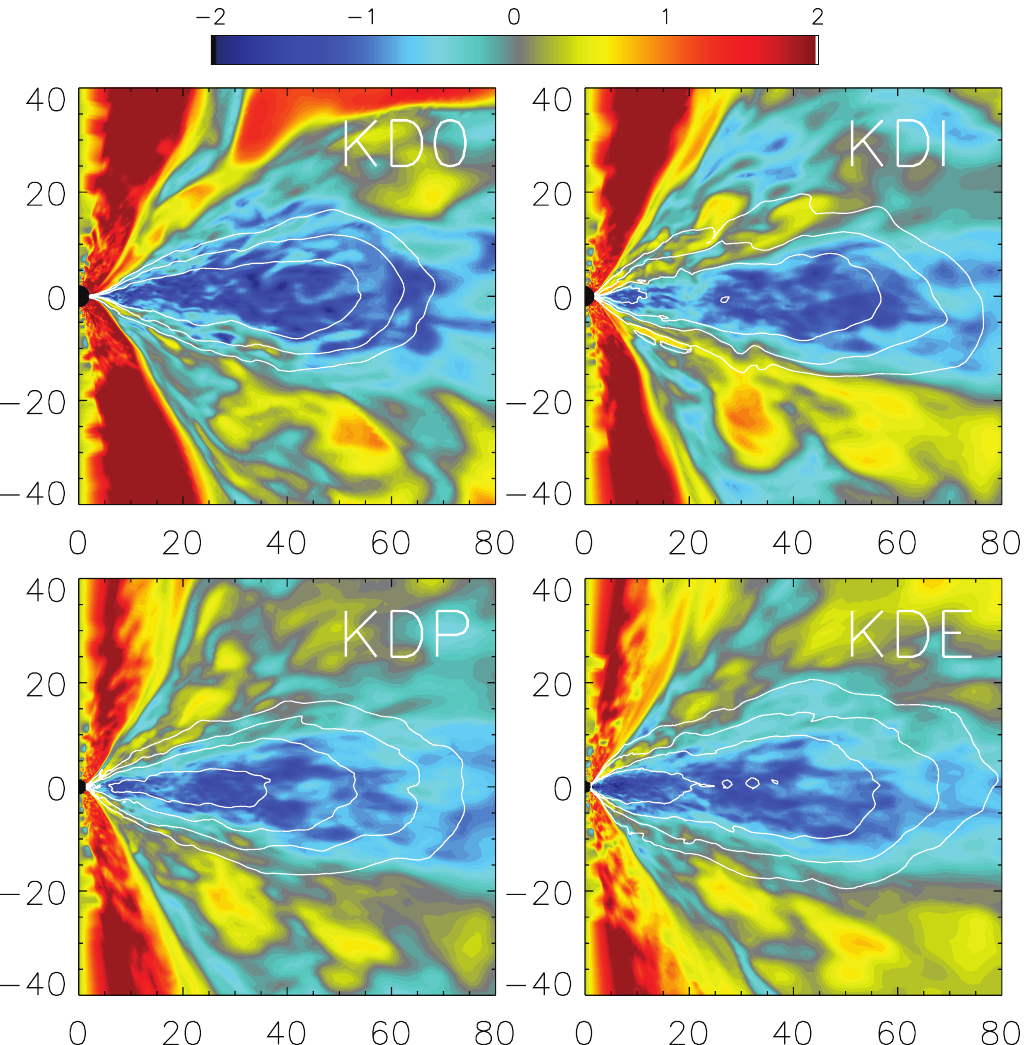}}
  \caption{Color contours of the ratio of azimuthally averaged
    magnetic to gas pressure, $P_\mathrm{mag}/P_\mathrm{gas}$. The
    scale is logarithmic and is the same for all panels; the color
    maps saturate in the axial funnel. The body of the accretion disk
    is identified with overlaid density contours at $10^{-2}$,
    $10^{-1.5}$, $10^{-1}$, and $10^{-0.5}$ of $\rho_{\max}(t=0)$. The
    individual plots are labeled by model. In all cases, the magnetic
    pressure is low ($P_\mathrm{mag}/P_\mathrm{gas} \ll 1$) in the
    disk, comparable to gas pressure ($P_\mathrm{mag}/P_\mathrm{gas}
    \sim 1$) in the corona above and below the disk, and high
    ($P_\mathrm{mag}/P_\mathrm{gas} \gg 1$) in the funnel region.
    Image reproduced by permission from~\cite{devilliers_03c},
    copyright by AAS.}
  \label{figure:devilliers03_8}
\end{figure}}

Despite the increased complexity of the disk when MRI-driven turbulence is at play, many features and properties remain strikingly similar to what is revealed in hydrodynamic-only studies. For instance, a commonly seen feature in global, non-radiative MHD simulations of black hole accretion disks is an ``inner torus'' \cite{hawley_02}. The inner torus is predominantly a hydrodynamic structure, as it is largely gas pressure supported. Generically, this inner torus consists of a sequence of equidensity and equipressure surfaces with a pressure maximum at $\sim 10\,r_G$, a cusp at $\sim 3\,r_G$, and a roughly parabolic, evacuated, though magnetized, funnel along the rotation axis. This structure is remarkably similar to the Polish doughnut model, as shown in Figure~\ref{figure:PolishDoughnut}. Thus, many hydrodynamic torus results may retain relevance even in MRI turbulent disks~\cite{qian_09}.

\epubtkImage{}{%
\begin{figure}[htbp]
  \centerline{\includegraphics[width=0.7\textwidth]{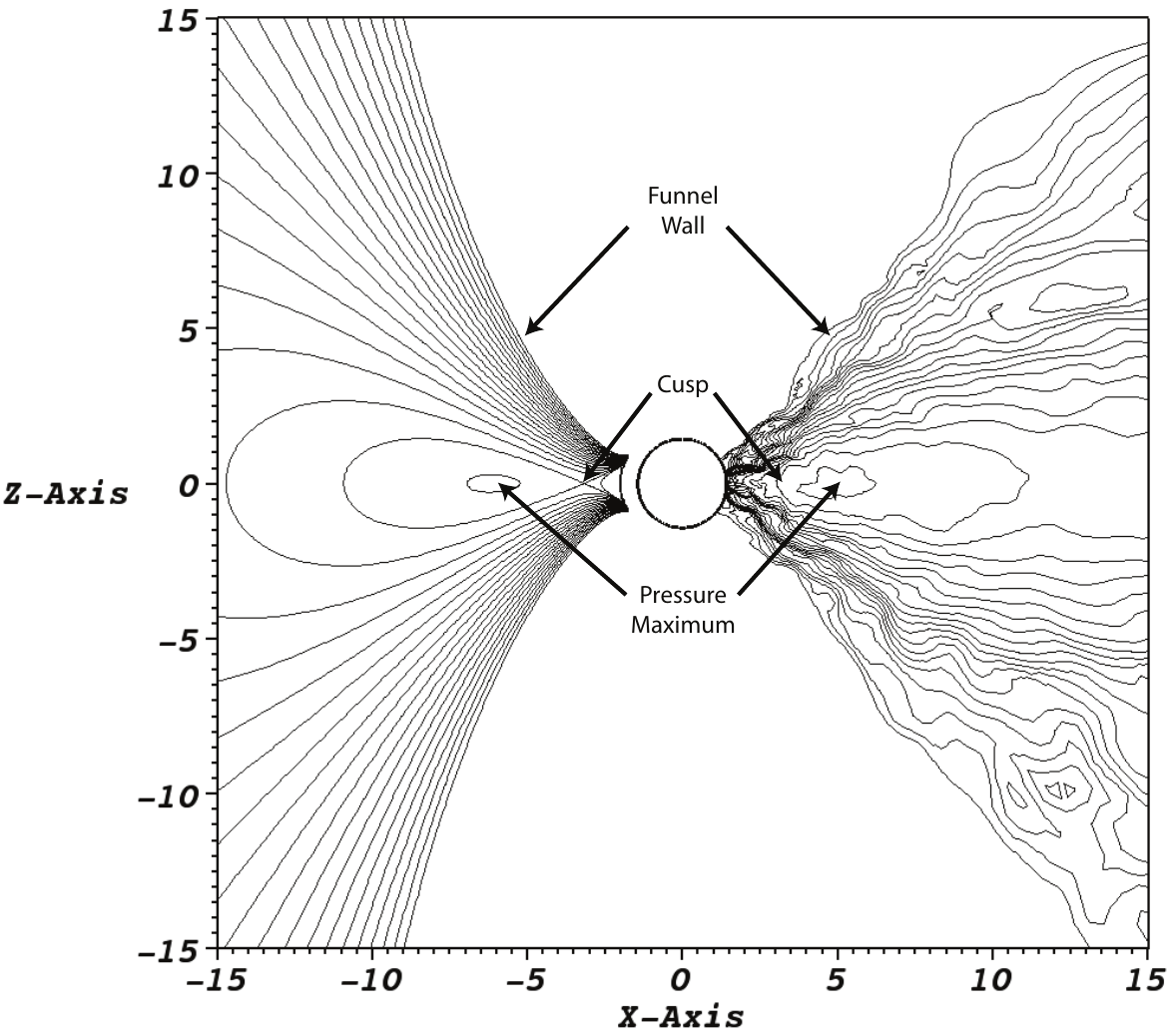}}
  \caption{On the left, equidensity contours calculated from an
  analytic Polish doughnut. On the right, equidensity contours from a
  numerical MHD simulation (model 90h from~\cite{fragile_07}). Note,
  though, that the contours on the left are linearly spaced, while those
  on the right are logarithmically spaced. Thus, the gradients
  represented on the left are shallower than those on the right. Image
  reproduced by permission from~\cite{qian_09}, copyright by ESO.}
  \label{figure:PolishDoughnut}
\end{figure}}

\subsection{Novikov--Thorne (thin) disks in simulations}
\label{section-numerical-novikov-thorne}

Early numerical simulations of black hole accretion disks nearly all focused on thick disks.  This was mostly for computational convenience as thicker disks require fewer resources than thin ones.  In recent years, though, the resources have become available to start testing thinner disks.  This has enabled researchers to begin rigorously testing the Novikov--Thorne model (Section~\ref{section-Shakura-Sunyaev}), especially the assumption that the internal torque of the disk vanishes at the ISCO. This is important as some researchers have argued that magnetic fields might nullify this hypothesis by maintaining stresses inside the ISCO~\cite{krolik_99b,gammie_99,balbus12}.

Initial global MHD simulations did show some variations from the Novikov--Thorne model at the ISCO at the level of $\sim$~10\%~\cite{hawley_02b,noble_09}, while others~\cite{penna_10} showed much smaller deviations $\lesssim$~3\%. Penna and collaborators~\cite{penna_10} discussed possible reasons for the discrepancies, and suggested that the results are, in fact, in better agreement than they may have initially appeared.  Their conclusion is that, in fact, the Novikov--Thorne model matches numerical results quite well (see Figure~\ref{figure:penna10}, plus animations can be viewed at~\cite{penna_movies}). Any remaining differences are small enough to have negligible effect on common applications of the Novikov--Thorne model, such as measuring black hole spin via spectral fitting~\cite{kulkarni_11,zhu_12} (Section~\ref{section:spin}). 

\epubtkImage{}{%
\begin{figure}[htbp]
  \centerline{
    \includegraphics[scale=0.05]{Penna_a0}\quad
    \includegraphics[scale=0.4]{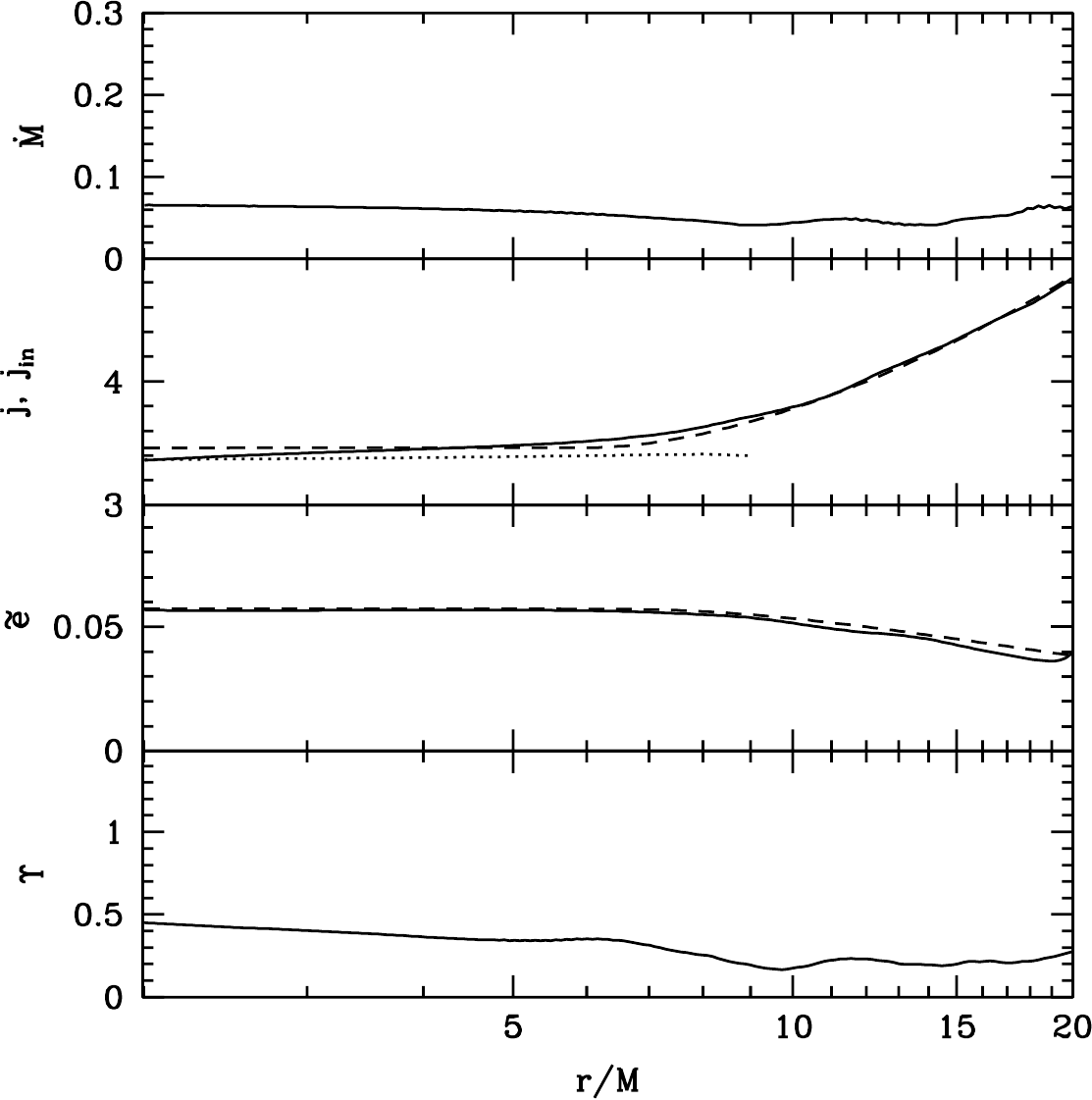}
  }
  \caption{\emph{Left:} Time-averaged rest mass density in the
    $r$\,--\,$z$ plane for four GRMHD simulations with $a = 0$ and
    various disk thicknesses. The dashed vertical line marks the
    ISCO. The disk opening angle, $h=H/r$, and effective
    Shakura--Sunyaev viscosity, $\alpha$, are reported in each
    panel. The top three panels have $h \ll \alpha$ and the inner edge
    of the disk is located outside the ISCO. The bottom panel has $h
    \gg \alpha$ and the density increases monotonically down to the
    event horizon.  Figure from~\cite{penna_12}. \emph{Right:} Various
    fluxes as functions of radius for a numerical Novikov--Thorne disk
    simulation. \emph{Top:} Mass accretion rate. \emph{Second panel:}
    Accreted specific angular momentum. \emph{Solid} line is
    simulation data; \emph{dashed} line gives Novikov--Thorne
    solution; \emph{dotted} line is ISCO value. Note that the specific
    angular momentum does not drop significantly inside the ISCO,
    unlike for thick disks, such as in
    Figure~\ref{figure:devilliers03_9}. \emph{Third panel:} The
    ``nominal'' efficiency, which is the total loss of specific energy
    from the fluid. \emph{Bottom panel:} Specific magnetic flux. The
    near constancy of this quantity inside the ISCO is an indication
    that magnetic stresses are not significant in this region. Image
    reproduced by permission from~\cite{penna_10}.}
 \label{figure:penna10}
\end{figure}}

We mention that since present day relativistic numerical simulations do not treat the radiation in geometrically thin, optically thick disks, the simulations of Novikov--Thorne disks have so far employed an \emph{ad hoc} cooling prescription~\cite{shafee_08,noble_09}. This prescription conveniently makes the same assumption that the Novikov--Thorne model does: that all energy dissipated as heat in the disk is radiated away locally on roughly the orbital timescale. This is probably reasonable for an appropriate range of mass accretion rates, though it will be good to test this assumption with future global radiation-MHD simulations.

\subsection{ADAFs in simulations}
\label{section-numerical-adaf}

A lot of recent simulation work has been focused on exploring more ADAF-like flows under the action of realistic MRI turbulence~\cite{narayan_12b, yuan_12a, yuan_12b}. To some extent, this is an extension of the Polish-doughnut simulations of the last decade, yet goes beyond it in at least two important respects: 1) the simulations cover a significantly larger spatial range (a few hundreds of $GM/c^2$ versus a few tens); and 2) the simulations explore much longer temporal evolution (hundreds of thousands of $GM/c^3$ versus tens of thousands). The result of this is that the simulations are able to explore steady-state accretion out to much larger radii ($\sim 100\,GM/c^2$ as opposed to $\sim 10\,GM/c^2$). More can be expected from this work in the coming years.


\subsection{Oscillations in simulations}
\label{section-numerical-oscillations}

Hydrodynamic simulations have also been used extensively to study the natural oscillation modes of relativistic disks, particularly as they might relate to QPOs (Section~\ref{section-QPOs}). Rezzolla and collaborators~\cite{rezzolla_03a, rezzolla_03b} identified a mode they referred to as a $p$-mode that occurred in a near 3:2 ratio with the radial epicyclic mode, which was subsequently confirmed through numerical simulations~\cite{zanotti_03,zanotti_05} (animations can be viewed at~\cite{rezzolla_movies}). The 3:2 ratio of these modes is important as the highest-frequency QPOs in black hole low-mass X-ray binaries are observed to occur in this ratio.  Later, Blaes and collaborators~\cite{blaes_06} identified a different pair of modes, the vertical epicyclic and axisymmetric breathing mode, that have a near 3:2 ratio for a broader range of parameters. Similar to~\cite{zanotti_03, zanotti_05}, they also demonstrated that these modes could be identified in numerical simulations.

A number of MHD simulations have also been performed which claim to observe QPOs. Kato~\cite{kato_04b} performed a global MRI simulation in a pseudo-Newtonian potential and claimed to see QPOs in the power spectrum in a measure of the luminosity associated with radial infall. Machida and Matsumoto~\cite{machida_06} have reported the formation of a one-armed spiral within the inner torus of a global MRI simulation, and suggested that this could be responsible for the low frequency QPO in the hard state. Schnittman, Krolik, and Hawley~\cite{schnittman_06} have found tentative evidence for high frequency QPO features in light curves generated by coupling a general relativistic ray tracing code to a global GRMHD simulation with an assumed emission model. Henisey and collaborators~\cite{henisey_09} found similar tentative evidence in a sample of tilted disk simulations, possibly confirming earlier suggestions that disk tilt may be an important mechanism for driving high frequency QPOs in black hole accretion disks \cite{kato_04c,ferreira_08}.  Most recently Dolence and collaborators~\cite{dolence_12} report evidence for near-infrared and X-ray QPOs in numerical simulations of Sgr~A*.

Aside from these few interesting examples, though, no robust QPO signal has generically emerged in MHD simulations~\cite{arras_06, reynolds_09}.  It is unclear what the implications of this are.  It may be that some missing physics, such as radiation transport, plays a fundamental role in exciting QPOs.  This is an important open problem in numerical simulations of black hole accretion disks.

\subsection{Jets in simulations}
\label{section-numerical-jets}

In recent years several numerical simulations have demonstrated the
generation of jets self-consistently from simulations of disks. A few
researchers have claimed to produce jetted outflows from purely
hydrodynamic interactions. Nobuta and Hanawa~\cite{nobuta_99}, for
instance, were able to achieve jetted outflows driven by shock waves
when infalling gas with high specific angular momentum collided with
the centrifugal barrier. However, such simulations usually require
very special starting conditions and the jets tend to be transient
features of the flow; therefore, it is unlikely these are related to
the well-collimated expansive jets observed in many black hole
systems.%
\epubtkFootnote{For the benefit of readers who are unfamiliar with the
  phenomenology of relativistic jets we mention that many jets
  demonstrate incredibly consistent collimation, even along hundreds
  or thousands of kiloparsecs (e.g.~\cite{harris_06}). From this we
  can infer that such jets have maintained their orientation and
  outflow for millions of years.}

Among MHD simulations, some distinction should be drawn between those that impose large scale magnetic fields that extend beyond the domain of the simulation and those that, at least initially, impose a self-contained magnetic field. In the first case, the disk is often treated merely as a boundary condition for the evolution of large-scale magnetic field. Shibata and Uchida \cite{shibata_85,shibata_86} used this technique to show that jets could be driven both by a pinching of the fields due to radial inflow through the disk and by the $j \times B$ force caused when twisted fields unwind.

Since 1999, many MHD disk simulations, usually starting with locally confined, weak magnetic fields, have shown some sort of magnetically driven bipolar outflow~\cite{koide_99, kato_04a, devilliers_05, hawley_06, mckinney_04, mckinney_06}. Since these jets are propagating at the Alfv\'en speed in a magnetically dominated region ($\beta_m \ll 1$), they proceed supersonically relative to the gas. However, some authors claim the outflow is only a transient state~\cite{kato_04a} and most do not achieve the very high Lorentz factors observed in astrophysical jets \cite{devilliers_05}. It is also unclear whether or not the jets in these simulations are connected with the Blandford--Znajek process \cite{devilliers_05} (Section~\ref{section-jets}), although see~\cite{mckinney_05}. McKinney~\cite{mckinney_06} and Komissarov~\cite{komissarov_05} suggest that part of the problem is that the jets arising from MHD simulations of disks are often contaminated by the application of numerical floors on density and energy and through numerical diffusion, both of which can lead to anomalous mass loading of the jet. In their own carefully constructed simulations, they independently found that the magnetic structures exhibited by their simulated jets are, in fact, consistent with the expectations of the Blandford--Znajek model, as illustrated in Figure~\ref{figure:blandford_znajek} (animations can be viewed at~\cite{mckinney_movies}).

\epubtkImage{}{%
\begin{figure}[htbp]
  \centerline{\includegraphics[width=0.9\textwidth]{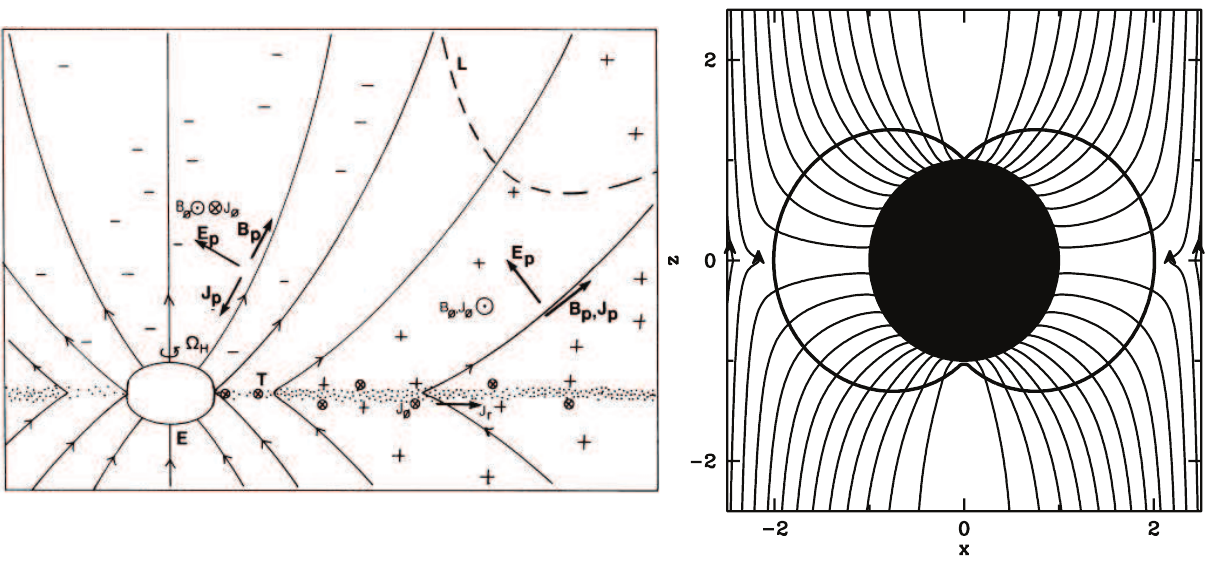}}
  \caption{On the left is a schematic diagram of the
  Blandford--Znajek mechanism~\cite{blandford_77} for an assumed parabolic field
  distribution. On the right is the result of a numerical simulation
  from~\cite{komissarov_07} showing a very similar structure. Images
  reproduced by permission; copyright RAS.}
  \label{figure:blandford_znajek}
\end{figure}}

Tchekhovskoy and collaborators~\cite{tchekhovskoy_10} have used GRMHD
simulations of black hole accretion disks plus jets to investigate
possible explanations for the observed radio loud/quiet dichotomy in
AGN.%
\epubtkFootnote{Radio loud AGN are $\sim 10^3\mbox{\,--\,}10^4$ times brighter
  in radio than radio quiet AGN of comparable optical
  luminosities~\cite{sikora_07}.}
For a black hole surrounded by a thin disk, the Blandford--Znajek mechanism predicts the luminosity of the jet should scale as $L_\mathrm{BZ} \propto \Omega_H^2$.  This limits the range of power expected for realistic AGN spins to a factor of a few tens at most -- too small to explain the observed differences.  Thicker disks, on the other hand, produce jets whose luminosity can scale as $L_\mathrm{BZ} \propto \Omega_H^4$ or even $\Omega_H^6$~\cite{tchekhovskoy_10}, providing sufficient range to perhaps explain the dichotomy.

There has also been some work recently trying to understand a different jet dichotomy -- one that is observed in black hole low-mass X-ray binaries (LMXBs).  These exhibit radio emission (associated with jets) whose properties change with the observed X-ray spectral and, to a less well determined extent, temporal properties of the accretion disk~\cite{fender_04,fender_09}. Briefly, compact, steady jets are observed in the Low/Hard state, whereas jets appear absent in the High/Soft state.  Fragile and collaborators~\cite{fragile_12b} used numerical simulations to rule out disk scale height as the controlling factor, suggesting instead that perhaps the jets are intimately connected with the corona or failed MHD wind. Alternatively, it could be that magnetic field topology is the key factor~\cite{igumenshchev_09, mckinney_09}.

\subsection{Highly magnetized accretion in simulations}

Recently, work has begun to focus on highly magnetized disk configurations, for which some modes of the MRI may be suppressed. One motivation is that these might provide an alternate explanation for the Low/Hard state~\cite{barrio_03} (Section~\ref{section:SpectralStates}).

One way a highly magnetized state could come about is as a result of a thermal instability in an initially hot, thick, weakly-magnetized disk~\cite{machida_06,fragile_09a}. Machida and collaborators~\cite{machida_06} simulated this process for an optically thin disk, assuming bremsstrahlung cooling and pseudo-Newtonian gravity. They found that, indeed, the densest inner regions of the disk collapse down to a cool, thin, magnetically-supported structure. Fragile and Meier~\cite{fragile_09a} extended these results by including more cooling processes and using a general relativistic MHD code.

Another way to achieve a highly magnetized state is to have the disk ``drag'' the magnetic field in from some distant region~\cite{bisnovatyi_74,spruit_05}. If the field has a consistent net flux, then doing so will necessarily increase the strength of the field near the hole due simply to the smaller area through which the flux must thread.  Recent numerical simulations have demonstrated this effect, first in the pseudo-Newtonian limit~\cite{igumenshchev_08} (Figure~\ref{figure:igumenshchev08}), and more recently in relativistic simulations~\cite{tchekhovskoy_11,mckinney_12}.  These examples are interesting because they have led directly to a phenomenon known as a ``magnetically arrested'' accretion state~\cite{narayan_03}, where the accumulation of magnetic field near the black hole is sufficient to temporarily halt the inflow of matter.  This state has been shown to produce jet efficiencies $\eta = (\dot{M} - \dot{E})/\langle\dot{M}\rangle > 1$~\cite{tchekhovskoy_11}, where
\begin{equation}
\dot{M} (t) = -\int \sqrt{-g} \rho u^r d\theta d\phi
\label{eqn:massFlux}
\end{equation}
is the mass accretion rate and
\begin{equation}
\dot{E} (t) = \int \sqrt{-g} T_t^r d\theta d\phi~
\label{eqn:energyFlux}
\end{equation}
is the energy flux, both taken at the black hole event horizon \(r_\mathrm{H}\), and the angle brackets indicate a time-averaged quantity.  This can \emph{only} happen if at least some of the energy powering the jet is being extracted from the rotational energy of the black hole itself!

\epubtkImage{}{%
\begin{figure}[htbp]
  \centerline{\includegraphics[scale=0.3]{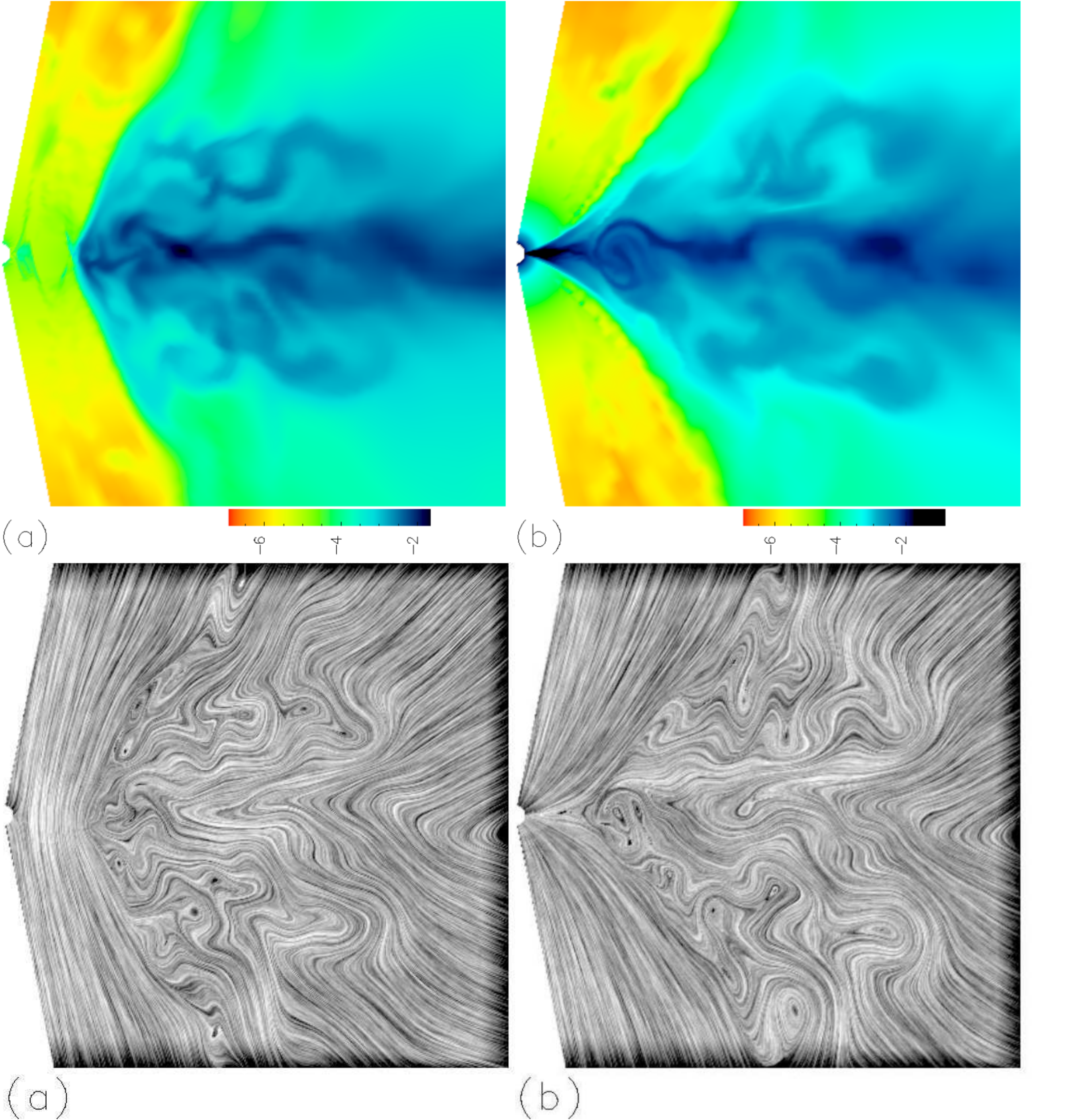}}
  \caption{\emph{Top:} Distributions of density in the meridional
    plane at different simulation times, showing a magnetically
    arrested state (\emph{left}) and a non-arrested state
    (\emph{right}). \emph{Bottom:} Snapshot of magnetic field lines at
    the same simulation times. Image reproduced by permission
    from~\cite{igumenshchev_08}, copyright by AAS.}
 \label{figure:igumenshchev08}
\end{figure}}

\newpage

\section{Selected Astrophysical Applications}
\label{section:applications}

\subsection{Measurements of black-hole mass and spin}
\label{section:spin}

Astrophysical black holes are not charged, and thus are characterized only by their mass and spin. Measurements of black hole mass are generally straightforward, requiring only the observation of an orbital companion and an application of Kepler's laws. Current mass estimates for stellar-mass black holes (in particular microquasars) are reviewed by McClintock \& Remillard~\cite{mcclintock_06} and for supermassive black holes by Kormendy \& Richstone~\cite{kormendy_95} (see also~\cite{merritt_01} for somewhat more recent data).

Nevertheless, there remain several fundamental questions connected with the mass of black holes, and some of them are directly connected to accretion disk theory. One of them is the question of ultra-luminous X-ray sources (ULXs)~\cite{colbert_99,makishima_00}. ULXs are powerful X-ray sources located outside of galactic nuclei, which have luminosities in excess of the Eddington limit for $M = 10^2\,M_{\odot}$, assuming isotropic emission. The huge luminosities of ULXs lead some to conclude that they are accreting \textit{intermediate-mass} black holes, with masses $M_\mathrm{ULX} > 10^2\,M_{\odot}$ (e.g.~\cite{krolik_04,miller_04a}). Others think that they are stellar mass black holes with $M_\mathrm{ULX} \sim 10\,M_{\odot}$, either exhibiting beamed emission~\cite{georganopoulos_02} or surrounded by disks that are somehow able to produce highly super-Eddington luminosities (e.g.~\cite{king_01,begelman_02}).  At this time, at least one ULX (ESO 243-49/HLX-1) has been convincingly demonstrated to have an intermediate mass ($\sim 500\,M_{\odot}$)~\cite{farrell_09,davis_11}.

As difficult as it is to nail down the mass on some of these systems, it is even more difficult to measure black hole spin, even though it plays a direct, and important, role in accretion disk physics. One obvious example of the role of spin is the dependence of $r_{\mathrm{ms}}$, the coordinate radius of the ISCO, on spin. In the symmetry plane of the black hole $r_{\mathrm{ms}} = 6\,r_G$ for $a_* = a/M =0$ (non-rotating), $1 r_G$ for $a_*=1$ (maximal prograde rotation), and $9\,r_G$ for $a_*=-1$ (maximal retrograde rotation) (Section~\ref{section-summary-radii-frequences}). It is believed that the inner edge of the accretion disk will be similarly affected. In addition, for rapidly rotating black holes, the Blandford--Znajek mechanism, and similar processes which depend on spin, may account for a fair share of the global energetics, comparable to that of accretion itself (see e.g.~\cite{komissarov_05, mckinney_06} and references therein). Therefore, measuring the spin of accreting black holes is integrally tied up in understanding black hole accretion generally.

Four methods for determining black hole spin have been proposed in the literature. With references to some of the earliest results, they are: 1) fitting the continuum spectra of microquasars observed in the thermally dominant state using disk emission models~\cite{davis_06b,mcclintock_06b,middleton_06,shafee_06}; 2) fitting observed relativistically broadened iron line profiles with theoretical models~\cite{karas_00,wilms_01,miller_02,miller_04b,reynolds_08}; 3) matching observed QPO frequencies to those predicted by theoretical models~\cite{cui_98,abramowicz_01,remillard_02,torok_05}; and 4) analyzing the ``shadow'' a black hole makes on the surface of an accretion disk~\cite{takahashi_04}. The first three methods have been the most commonly applied to date. Although there have been some glaring discrepancies in the spin estimates published to date (e.g. one group claiming that Cyg X-1 has a near-zero spin, $a_* = 0.05\pm0.01$~\cite{miller_09}, and then later claiming it has a near-maximal spin, $a_* > 0.95$~\cite{fabian_12}), there appears to be a settling of values in recent years and a growing confidence in the methods, particularly the continuum fitting. There are still some concerns, however. For the continuum fitting method, the main issue is that the inclination of the X-ray emitting region must be measured by some independent means. This is because the effect of the inclination on the spectrum is degenerate with the effect of spin~\cite{li_09}, so both can not be accounted for within the continuum fitting method. In cases where such an independent measure is available (e.g.~\cite{steiner_12}), the continuum fitting method appears robust.
For the relativistically-broadened iron line method, there are difficulties in properly estimating the extent of the ``red'' wing, which is most directly related to the spin of the black hole, and in modeling the hard X-ray source photons and the disk ionization, both of which strongly affect the reflection spectrum. The reviews by Remillard, McClintock, and collaborators~\cite{remillard_06,mcclintock_11} give more complete introductions to the topic of measuring black hole spin, with emphasis mainly on the continuum fitting method.

\subsection{Black hole vs. neutron star accretion disks}
\label{section:horizon_argument}

Little of what we have said so far has depended on whether the central compact object is a black hole or neutron star, provided only that the neutron star is compact enough to lie inside the inner radius of the disk $r_{\mathrm{in}}$. In this case, its presence will not be noticed by the disk except through its gravity, which will be practically the same as for a black hole (an exception would be if the neutron star is strongly magnetized~\cite{harding_06}). However, this does \emph{not} mean that accreting black hole and neutron star sources will be indistinguishable, as we have not yet fully addressed the question of what happens to energy advected past $r_{\mathrm{in}}$. For optically thick, geometrically thin Shakura--Sunyaev disks (Section~\ref{section-Shakura-Sunyaev}), a significant fraction of the gravitational energy liberated by advection is radiated by the gas prior to it passing through $r_{\mathrm{in}}$. Thus, the total luminosity of thin disks will not depend sensitively on the nature of the central object. However, this is not the case for the ADAF solution (Section~\ref{section-ADAFs}), for which much of the thermal energy gained by the gas from accretion is carried all the way in to the central object. Narayan and his collaborators~\cite{narayan_95b,narayan_97,menou_99,garcia_01} have convincingly argued that this may allow observers to distinguish between black hole and neutron star sources.

The key is that, for black hole sources, advection through the event horizon allows the excess thermal energy to be effectively absorbed without ever radiating. For neutron star sources, on the other hand, the presence of a hard surface ensures that the excess energy of accretion is released upon impact and must be radiated to infinity. This implies that for systems in the ADAF state, a black hole source should be significantly less luminous that a neutron star one with the same mass accretion rate \cite{narayan_95b}. Perhaps a more important point is that the \emph{range} of luminosities should be wider for a black hole source than for a neutron star one~\cite{narayan_97}. This is because, while the luminosity goes as $L \propto \dot{m}$ for \emph{all} neutron star states and for black holes in a high accretion state, it goes as $L \propto \dot{m}^2$ for black holes in the ADAF state for which $\dot{m} < 10^{-2}-10^{-1}$. Furthermore, since the luminosity is also proportional to the mass of the central object $L \propto M$, at the highest accretion rates a black hole source should be more luminous than a neutron star one due to its higher mass (Figure~\ref{figure:narayan97}). Another key point to this argument is that neutron stars can independently and reliably be confirmed if they display type I bursts, which are thermonuclear flashes occurring in material accumulated on the surface of the neutron star \cite{joss_84}. Thus one can compare known neutron star sources against suspected black hole sources. This has now been done in a number of otherwise similar sources and Narayan's expectations have indeed been confirmed~\cite{narayan_97,menou_99,garcia_01} (a recent example is shown in Figure~\ref{figure:narayan97}). This provides compelling observational evidence for the existence of black hole event horizons, although this falls short of being a proof~\cite{abramowicz_02}. This topic is discussed further in the review article by Narayan and McClintock~\cite{narayan_08}.

\epubtkImage{}{%
\begin{figure}[htbp]
  \centerline{\includegraphics[width=0.8\textwidth]{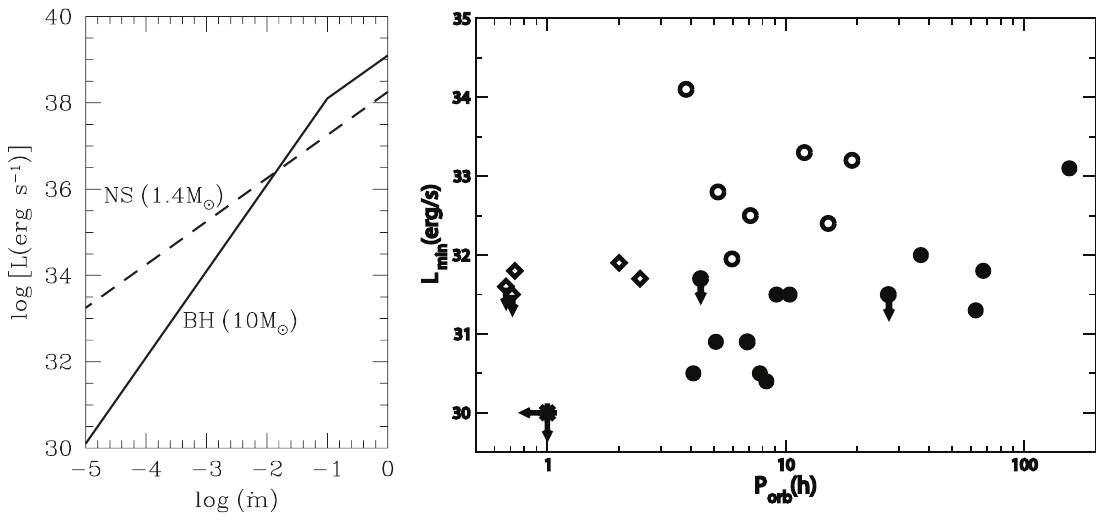}}
  \caption{\emph{Left:} Luminosity as a function of accretion rate for
    neutron star and black hole sources, illustrating that a wider
    range of luminosities are expected for black holes. Image
    reproduced by permission from~\cite{narayan_97}, copyright by
    AAS. \emph{Right:} Recent data showing that neutron star sources
    (open symbols) are systematically more luminous than black hole
    sources (filled symbols) in analogous spectral states. Image
    reproduced by permission from~\cite{lasota_08}, copyright by
    Elsevier.}
  \label{figure:narayan97}
\end{figure}}

\subsection{Black-hole accretion disk spectral states}
\label{section:SpectralStates}

Black hole accretion disks, particularly in X-ray binaries, exhibit complex spectra composed of both thermal and nonthermal components. During outbursts, the relative strengths of these components change frequently in concert with changes in luminosity and the characteristics of the radio features (i.e., jets). Astronomers have developed a set of empirical spectral classification states to broadly characterize these observations. These spectral states likely reveal important information about the underlying physical state of the system; therefore it is worth summarizing the states and their observed properties here.

Probably the easiest state to connect with a theoretical model is the ``High/Soft'' (HS) or ``Thermally Dominant'' state.  As the name implies, the spectrum in this state is dominated by the thermal component. This state is best explained as $\sim$~1~keV thermal emission from a multitemperature accretion disk, as predicted by the Shakura--Sunyaev (thin) disk model (Section~\ref{section-Shakura-Sunyaev}). However, most sources spend the majority of their lifetimes in the ``Low/Hard'' (LH) or even ``quiescent'' state. The quiescent state is characterized by exceptionally low luminosity and a hard, non-thermal spectrum (photon index $\Gamma =1.5-2.1$). As the luminosity increases the sources usually enters the Hard state. Here the 2\,--\,10~keV intensity is still comparatively low and the spectrum is still nonthermal. This spectrum is best fit with a powerlaw of photon index $\Gamma \sim 1.7$ (220~keV). In this state, the thermal component is either not detected or appears much cooler, indicating the thin disk may truncate further out than in the Thermally Dominant state, although see \cite{miller_06} for claims that the thin disk extends all the way to the ISCO even in the Hard state. Observations suggest the region interior to the thin disk may be filled with a hot (presumably thick), optically-thin plasma, which accounts for the nonthermal part of the spectrum. This is the picture suggested by the ``truncated disk model''~\cite{esin_97,esin_98,esin_01}, which pictures the Hard state as a truncated thin (Shakura--Sunyaev) disk adjoined with an inner thick (ADAF-like) flow. This model has shown tremendous phenomenological success~\cite{done_07}, although other models for the Hard state still abound~\cite{yuan_07,fragile_09a,igumenshchev_09}. The Hard state is also linked with observations of a persistent radio jet that is not seen in other states (see Section~\ref{section-jets}).  The final spectral state, which is referred to as the ``Very High'' (VH) or ``Steep Power Law'' state, is characterized by the appearance of high-frequency QPOs (Section~\ref{section-QPOs}) and the presence of both disk and powerlaw components, each of which contributes substantial luminosity. In this state, the powerlaw component is observed to be steep ($\Gamma \sim2.5$), giving the state its name. This state is sometimes associated with intermittent jets.

These states, their distinguishing observational properties, and a sampling of observations from various black hole X-ray binaries is presented in the review by McClintock and Remillard~\cite{mcclintock_06}.

\subsection{Quasi-Periodic Oscillations (QPOs)}
\label{section-QPOs}

Einstein's general theory of gravity has never been tested in its strong field limit, characteristic of the region very near black holes (or neutron stars), i.e., a few gravitational radii away from these sources. Soon VLBI measurements may be able to resolve these scales for the supermassive black hole at the center of our galaxy. However, for most sources, resolution in time seems to be a more practical approach. To zeroth order, the light curves from accreting black holes vary in a chaotic manner, resembling a loud noise. However, Fourier analysis of the light curve reveals stinkingly regular patterns buried in the noise. For Galactic black hole and neutron star sources, these quasi-periodic oscillations (QPOs) have frequencies of a few hundred Hz.

Frequencies in the range 100\,--\,1000~Hz formally correspond to orbital frequencies a few gravitational radii away from a stellar-mass object.  The focus on orbital frequencies is further motivated by the stability of observed QPO frequencies over very long periods of time. For example, QPOs of 300~Hz and 450~Hz were observed from the microquasar GRO~J1655--40 during its 1996 outburst and again almost nine years later during its 2005 outburst. This strongly suggests that the frequencies cannot depend on quantities such as magnetic field, density, temperature, or accretion rate, as these all vary greatly in time. The only parameters of a black hole accretion system that do not vary over a nine year period are the mass and the spin of the central black hole. Thus, the oscillation frequencies must only depend on these two parameters, and only frequencies connected to orbital motion have the property that they depend only on mass and spin.  Thus, the possible frequencies are: the Keplerian frequency, the two epicyclic frequencies (as originally suggested by Klu{\'z}niak and Abramowicz~\cite{abramowicz_01}), the Lense--Thirring frequency (as originally suggested by Stella and Vietri~\cite{stella_98}), and their combinations (e.g., \cite{kato_04b, kato_04c}).

In several microquasars the detected high-frequency QPOs come not as single oscillations, but as part of a pair. Furthermore, Abramowicz and Klu{\'z}niak~\cite{abramowicz_01} noticed that they are commensurable, being most often in a 2/3 ratio, as shown in Table~\ref{table-qpo}. This suggests that a resonance may be at work. Twin peak QPOs in the kilohertz range have been also detected from binaries containing accreting neutron stars. These neutron star QPOs show a similar, though less obvious, 2/3 ratio.

\begin{table}
  \caption{Frequency ratio of the ``twin peak'' QPOs in all four
    microquasars where they have been detected.}
  \label{table-qpo}
  \centering
  \begin{tabular}{c c}
    \toprule
    Microquasar  & Frequency ratio \\ 
    \midrule
    GRO J1655--40 & 300/450\,=\,0.66 \\ 
    XTE 1550--564 & 184/276\,=\,0.66 \\ 
    H 1743--322   & 166/240\,=\,0.69 \\ 
    GRS 1915+105  & 113/168\,=\,0.67 \\ 
    \bottomrule
  \end{tabular}
\end{table}

It was realized~\cite{kluzniak_01} that the behavior of the observed
QPO frequencies and amplitudes in both neutron star binaries and
microquasars is typical for a certain type of non-linear
resonance. Indeed, it may be observed when a properly tuned spring is
attached to a pendulum with a properly chosen length. Such a system
oscillates in two ``modes'': the pendulum mode and the spring
mode.%
\epubtkFootnote{It is quite remarkable that the mathematical
  theory describing resonances makes some very general predictions
  about the behavior of frequencies and amplitudes, even if the
  specific physical properties of the oscillating objects are not
  known. To some extent it is possible to accurately describe how
  things oscillate without even knowing what is oscillating.}
Mostly due to efforts of Rebusco and Hor{\'a}k~\cite{rebusco_04, horak_04,
  horak_08, horak_09}, a mathematical \textit{resonance model} was
developed to describe an arbitrary system oscillating in two modes
near a 2/3 non-linear resonance. The model's predictions for the
frequency and amplitude behaviors are strikingly similar to the ones
observed in X-ray binary QPOs, suggesting that they may indeed be
explained as a non-linear resonance of two modes of oscillation. The
model does not, however, explain what these modes are, how they are
excited, nor what energy reservoir they tap. Only when these questions
are answered satisfactorily could one say that the QPO puzzle is
solved.%
\epubtkFootnote{The mathematical foundations of the resonance model
  are described and discussed in a special issue of the
  \textit{Astronomische Nachrichten}~\cite{abramowicz_05}.}

As a final note, a crucial discovery by Barret and collaborators~\cite{barret_05a, barret_05, barret_07} concerning the behavior of the quality factor of twin peak QPOs proves that they are disk \textit{oscillations} and cannot be explained by kinematic (Doppler) effects due to the presence ``hot spots'' on the accretion disk surface. These effects are, however, important in modulating the QPO's signal \cite{bursa_04}. It is not clear, though, whether the oscillations are explained by the discoseismic modes discussed in Section~\ref{section-diskoseismology} (see, e.g., \cite{wagoner_12} for references).

\subsection{The case of Sgr A*}
\label{section-SgrA}

As already mentioned in Section~\ref{section-event-horizon}, there is a very good chance that the first direct evidence for a black hole event horizon will come from Sgr~A*, the compact, supermassive object at the center of the Milky Way. There have already been a number of strong, indirect arguments in favor of the black hole nature of Sgr~A*~\cite{broderick_06, broderick_09}, but no direct evidence yet. Sgr~A* is also of interest because it represents a unique case of black hole accretion, having by far the lowest (scaled) mass accretion rate and radiative efficiency of any known source. We anticipate greatly expanding this section in the near future as new results become available.


\section{Concluding Remarks}

Since the early 1970s, the study of black hole accretion disks has yielded to remarkable successes. And yet, as in most fields of research, each step forward has been met by new questions. In this article, we have tried to give a tour of some of the successes, such as the many disk models (thick, thin, slim, ADAF, \dots) that have given a firm foundation on which to work, the many studies of disk instabilities and oscillations that help us to understand the ways in which real disks can deviate from the simplistic models, and the numerical simulations that come as close as possible to an experimental test-bed for black hole accretion. We have also tried to indicate what we believe are some of the most pressing challenges of the day, including matching our theoretical knowledge to actual observed phenomena such as black hole spectral states, quasi-periodic oscillations, and relativistic jets. There are also observational challenges to find direct evidence of black hole event horizons and definitively constrain black hole spins. We hope, as we continue to update this Living Review, to be able to report on future discoveries in these areas, just as we expect to report new puzzles we have yet to encounter.

Clearly our tour has been incomplete.  For instance, despite their prominent role in nature, e.g., as an AGN feedback mechanism, outflows are not accounted for in any of the four main accretion models we presented, as the models all assume that the accretion rate is constant with radius. In reality, outflows may be triggered by any of three mechanisms: thermal, radiative, or centrifugal. Thermal winds are expected to result from heating of the outer regions of an accretion disk by its hot inner region. Radiative winds are driven by radiative flux acting on line opacities. Centrifugal acceleration of particles can take place along magnetic field lines which are sufficiently inclined to the disk plane.

The ADIOS (adiabatic inflow-outflow solution)~\cite{blandford_99} is a generalization of the ADAF solution that actually has much of the mass, energy, and angular momentum of the accretion ``disk'' carried away in the form of winds, rather than being advected into the black hole as in a normal ADAF.  However, the argument behind the ADIOS model is flawed~\cite{abramowicz_00}.  Blandford and Begelman~\cite{blandford_99} claim that black hole accretion flows with small radiative efficiency must necessarily experience strong outflows because the matter all has a positive Bernoulli constant. Yet a positive Bernoulli constant is only a necessary, but not sufficient, condition for outflows. For example, the classical Bondi accretion solution has ${\cal B}>0$ everywhere and yet experiences no outflows.  Furthermore, low efficiency accretion flows do not really have a positive Bernoulli constant everywhere, as boundary conditions (ignored in \cite{blandford_99}) will impose some regions of negative Bernoulli constant. A more recent look at the ADIOS solution is presented in~\cite{begelman_12}.

There are several other analytic and semi-analytic models of accretion disks. Some are closely related to the models we have already discussed. For example, the CDAF (convection-dominated accretion flow)~\cite{narayan_00, narayan_02} is another variant on the ADAF, in which long-wavelength convective instabilities transport angular momentum \emph{inward} and energy outward.  Other models relax some of the standard assumptions about accretion disks. For example, Bardeen and Petterson~\cite{bardeen_75} and others~\cite{kumar_85,scheuer_96,lubow_02} relaxed the assumption that disks are axisymmetric by considering tilted accretion disks, acted on by the Lense--Thirring precession of the central (rotating) black hole. Then there is the exact solution for stationary, axisymmetric \emph{non-circular}, accretion flows found by Klu{\'z}niak and Kita~\cite{kluzniak_00}.

On the numerical side, too, there are many interesting accretion configurations that have been identified, but are not included in this review. Some examples include quasi-spherical (low angular momentum) accretion flows~\cite{proga_03,proga_05} and convection-dominated disks~\cite{igumenshchev_03}.

For those who wish for more details or a different perspective, we can recommend several excellent text books and review articles devoted, partially or fully, to black hole accretion disks. We recall here some of the most often quoted. The oldest, but still very useful and informative, is the classic review by Pringle~\cite{pringle_81}. The most authoritative text book on accretion is \textit{Accretion Power in Astrophysics} by Frank, King and Raine~\cite{frank_02}. Two monographs devoted to black hole accretion disks are: \textit{Black-Hole Accretion Disks} by Kato, Fukue and Mineshige~\cite{kato_98}, and \textit{Theory of Black Hole Accretion Disks} by Abramowicz, Bj\"ornsson, and Pringle~\cite{abramowicz_98}. Lasota~\cite{lasota_99} also wrote an excellent non-technical \textit{Scientific American} article on black hole accretion in microquasars.  Finally, there is a nice series of lecture notes by Ogilvie available on the web~\cite{ogilvie_05}.


\section{Acknowledgements}
\label{section:acknowledgements}

MAA gratefully acknowledges supporting grants from Sweden (VR Dnr 621-2006-3288) and Poland
(UMO-2011/01/B/ST9/05439). MAA also thanks the College of Charleston for hosting him during
a portion of his work on this review. PCF gratefully acknowledges supporting grants from the
National Science Foundation under Grant No.\ NSF PHY11-25915, the College of Charleston, and
the South Carolina Space Grant Consortium. PCF also enjoyed the hospitality of NORDITA and
G\"oteborg University while working on this review. Computational support was provided under
the following NSF programs: Partnerships for Advanced Computational Infrastructure,
Distributed Terascale Facility (DTF) and Terascale Extensions: Enhancements to the
Extensible Terascale Facility.

\newpage

\bibliography{refs}

\end{document}